\documentstyle[11pt]{article}
\date{August 7th 1999}
\begin{document}
\newcommand{\Limits}
{\vspace*{-1mm}{\!\lim_{\small\begin{array}{c} N\!\!\to\!\infty \\[-1mm] n\!\to\! 0\end{array}}\!}}
\newcommand{\trace}{~{\large\rm Trace}~}
\newcommand{\s}{\varsigma}
\newcommand{\f}{\varphi}
\newcommand{\room}{\rule[-0.3cm]{0cm}{0.8cm}}
\newcommand{\smallroom}{\rule[-0.2cm]{0cm}{0.7cm}}
\newcommand{\hsp}{\hspace*{3mm}}
\newcommand{\vsp}{\vspace*{3mm}}
\newcommand{\be}{\begin{equation}}
\newcommand{\ee}{\end{equation}}
\newcommand{\bd}{\begin{displaymath}}
\newcommand{\ed}{\end{displaymath}}
\newcommand{\bdm}{\begin{displaymath}}
\newcommand{\edm}{\end{displaymath}}
\newcommand{\bea}{\begin{eqnarray}}
\newcommand{\eea}{\end{eqnarray}}
\newcommand{\sgn}{~{\rm sgn}}
\newcommand{\extr}{~{\rm extr}}
\newcommand{\Equiv}{\Longleftrightarrow}
\newcommand{\pprime}{{\prime\prime}}
\newcommand{\ppprime}{{\prime\prime\prime}}
\newcommand{\notexists}{\exists\hspace*{-2mm}/}
\newcommand{\bra}{\langle}
\newcommand{\ket}{\rangle}
\newcommand{\bigbra}{\left\langle\room}
\newcommand{\bigket}{\right\rangle\room}
\newcommand{\bras}{\langle\!\langle}
\newcommand{\kets}{\rangle\!\rangle}
\newcommand{\bigbras}{\left\langle\!\!\!\left\langle\room}
\newcommand{\bigkets}{\room\right\rangle\!\!\!\right\rangle}
\newcommand{\order}{{\cal O}}
\newcommand{\minus}{\!-\!}
\newcommand{\plus}{\!+\!}
\newcommand{\erf}{{\rm erf}}
\newcommand{\bbf}{{\mbox{\boldmath $f$}}}
\newcommand{\bk}{\mbox{\boldmath $k$}}
\newcommand{\bm}{\mbox{\boldmath $m$}}
\newcommand{\br}{\mbox{\boldmath $r$}}
\newcommand{\bq}{\mbox{\boldmath $q$}}
\newcommand{\bu}{\mbox{\boldmath $u$}}
\newcommand{\bx}{\mbox{\boldmath $x$}}
\newcommand{\bz}{\mbox{\boldmath $z$}}
\newcommand{\bA}{\mbox{\boldmath $A$}}
\newcommand{\bB}{\mbox{\boldmath $B$}}
\newcommand{\bC}{\mbox{\boldmath $C$}}
\newcommand{\bF}{\mbox{\boldmath $F$}}
\newcommand{\bH}{\mbox{\boldmath $H$}}
\newcommand{\bJ}{\mbox{\boldmath $J$}}
\newcommand{\bM}{\mbox{\boldmath $M$}}
\newcommand{\bQ}{\mbox{\boldmath $Q$}}
\newcommand{\bR}{\mbox{\boldmath $R$}}
\newcommand{\bW}{\mbox{\boldmath $W$}}
\newcommand{\hmu}{\hat{\mu}}
\newcommand{\hf}{\hat{f}}
\newcommand{\hQ}{\hat{Q}}
\newcommand{\hR}{\hat{R}}
\newcommand{\hbf}{\hat{\mbox{\boldmath $f$}}}
\newcommand{\hbh}{\hat{\mbox{\boldmath $h$}}}
\newcommand{\hbm}{\hat{\mbox{\boldmath $m$}}}
\newcommand{\hbr}{\hat{\mbox{\boldmath $r$}}}
\newcommand{\hbq}{\hat{\mbox{\boldmath $q$}}}
\newcommand{\hbD}{\hat{\mbox{\boldmath $D$}}}
\newcommand{\hbJ}{\hat{\mbox{\boldmath $J$}}}
\newcommand{\hbQ}{\hat{\mbox{\boldmath $Q$}}}
\newcommand{\hbR}{\hat{\mbox{\boldmath $R$}}}
\newcommand{\hbW}{\hat{\mbox{\boldmath $W$}}}
\newcommand{\bsigma}{\mbox{\boldmath $\sigma$}}
\newcommand{\btau}{{\mbox{\boldmath $\tau$}}}
\newcommand{\bomega}{{\mbox{\boldmath $\Omega$}}}
\newcommand{\bOmega}{{\mbox{\boldmath $\Omega$}}}
\newcommand{\bDelta}{{\mbox{\boldmath $\Delta$}}}
\newcommand{\bphi}{{\mbox{\boldmath $\Phi$}}}
\newcommand{\bpsi}{{\mbox{\boldmath $\psi$}}}
\newcommand{\bdelta}{{\mbox{\boldmath $\Delta$}}}
\newcommand{\btheta}{{\mbox{\boldmath $\theta$}}}
\newcommand{\bxi}{{\mbox{\boldmath $\xi$}}}
\newcommand{\bmu}{{\mbox{\boldmath $\mu$}}}
\newcommand{\brho}{{\mbox{\boldmath $\rho$}}}
\newcommand{\bEta}{{\mbox{\boldmath $\eta$}}}
\newcommand{\G}{{\cal G}}
\newcommand{\A}{{\cal A}}
\newcommand{\B}{{\cal B}}
\newcommand{\C}{{\cal C}}
\newcommand{\K}{{\cal K}}
\newcommand{\cA}{{\cal A}}
\newcommand{\cB}{{\cal B}}
\newcommand{\cC}{{\cal C}}
\newcommand{\cD}{{\cal D}}
\newcommand{\cE}{{\cal E}}
\newcommand{\cF}{{\cal F}}
\newcommand{\cL}{{\cal L}}
\newcommand{\cW}{{\cal W}}
\newcommand{\unity}{{\bf 1}\hspace{-1mm}{\bf I}}
\newcommand{\inn}{\!\cdot\!}
\newcommand{\set}{{\tilde{D}}}
\newcommand{\sets}{{\rm sets}}
\newcommand{\ketset}{\ket_{\!\tilde{D}}}
\newcommand{\all}{D}
\newcommand{\ketall}{\ket_{\!D}}
\newcommand{\LG}{\mbox{\normalsize ${\cal G}$}}
\newcommand{\LPsi}{\mbox{\normalsize $\Psi$}}
\newcommand{\hatildeG}{\bar{G}}
\newcommand{\ketop}{\mbox{\tiny ${\rm Q\!R\!P}$}}
\newcommand{\bsomega}{\!\!\mbox{\scriptsize\boldmath $\Omega$}}
\newcommand{\rmsets}{\!\!\mbox{\scriptsize\boldmath $\Xi$}}
\newcommand{\bnul}{\mbox{\boldmath $0$}}
\newcommand{\authorA}{A.C.C. Coolen \\[1mm]
Department of Mathematics\\
King's College London\\
Strand, London WC2R 2LS, UK}
\newcommand{\authorB}{D. Saad \\[1mm]
The Neural Computing Research Group \\
Aston University\\
Birmingham B4 7ET, UK}
\title{\bf Dynamics of Learning with Restricted Training Sets\\
I. General Theory}
\author{\authorA\and\authorB}
\maketitle

\begin{abstract}
\noindent
We study the dynamics of supervised learning in layered neural
networks, in the regime where the size $p$ of the training set is
proportional to the number $N$ of inputs. Here the local fields are no
longer described by Gaussian probability distributions and the
learning dynamics is of a spin-glass nature, with the composition of
the training set playing the role of quenched disorder.   We show how
dynamical replica theory can be used to predict the evolution of
macroscopic observables, including the two relevant performance
measures (training error and generalization error),
incorporating the old formalism developed for
complete training sets
in the limit $\alpha=p/N\to\infty$
as a special case.  For simplicity we restrict ourselves in this paper
to single-layer networks and realizable tasks.
\end{abstract}

{\small 
\vspace*{2mm}

\begin{center}{PACS: 87.10.+e, 02.50.-r, 05.20.-y}\end{center}

\vspace*{-7mm}

\tableofcontents}

\clearpage

\section{Introduction}

In the last few years much progress has been made in the
analysis of the dynamics of supervised learning in layered neural
networks, using the strategy of statistical mechanics: by
deriving from the microscopic dynamical equations of the learning process
a set of closed laws describing the evolution of suitably chosen
macroscopic observables (dynamic order parameters), in the limit of an
infinite system size
(e.g.
\cite{KinzelRujan,KinouchiCaticha,BiehlSchwarze92,BiehlSchwarze95,SaadSolla}.
A recent review and more extensive guide to the relevant references
can be found in \cite{MaceCoolen,Newton}. \cite{Newton} also contains a preliminary presentation
of some of the results in the present paper, without proofs or derivations. The main successful
procedure developed so far is built on the following four
cornerstones:
\begin{itemize}
\item
{\em The task to be learned by the network is defined by a (possibly noisy)
`teacher', which is itself a layered neural network.}
This induces a
canonical set of dynamical order parameters, typically the (rescaled)
overlaps between the various student weight vectors and the
corresponding teacher weight vectors.
\item
{\em The number of network inputs is (eventually) taken to be
infinitely large.}
This ensures that fluctuations in mean-field
observables will vanish, and creates the possibility of using the
central limit theorem.
\item
{\em The number of `hidden' neurons is finite.}
This prevents
the number of order parameters from being infinite, and ensures that
the cumulative impact of their fluctuations is insignificant.
\item
{\em The size of the training set is much larger than the number
of weight  updates made.}
Each example presented to the system is now different from
those that have already been seen, such that the local fields will
have Gaussian probability distributions, which leads to closure of the
dynamic equations.
\end{itemize}
These are not ingredients to simplify the calculations, but vital
conditions, without which the standard method fails. Although the
assumption of an infinite system size has been shown not to be
 too critical \cite{Barberetal}, the
other assumptions do
place serious restrictions on the degree of realism of the scenarios
that can be analyzed, and have thereby, to some extent, prevented
the theoretical results from being used by practitioners.

\begin{figure}[t]
\vspace*{85mm}
\hbox to \hsize{\hspace*{-5mm}\includegraphics{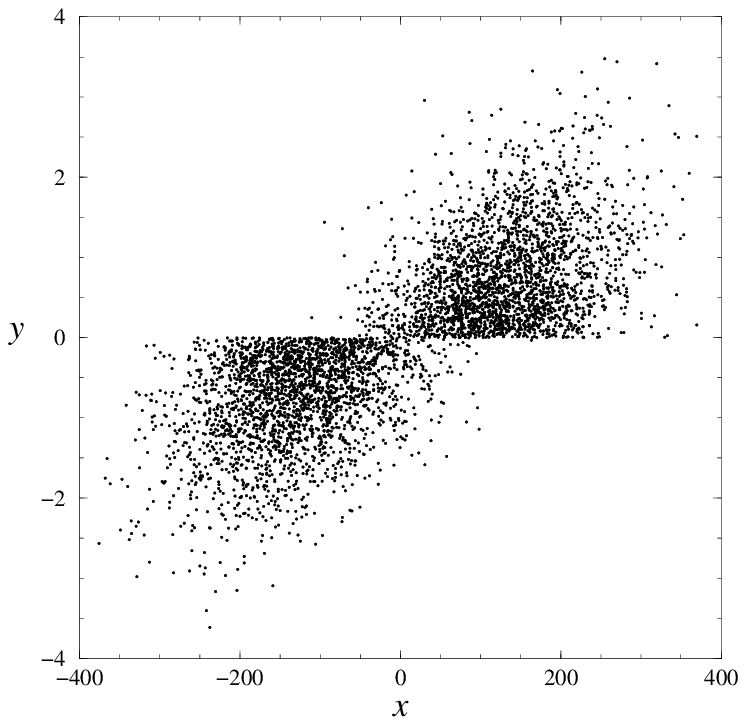}\hspace*{5mm}}
\vspace*{-4.8mm}
\hbox to
\hsize{\hspace*{80mm}\includegraphics{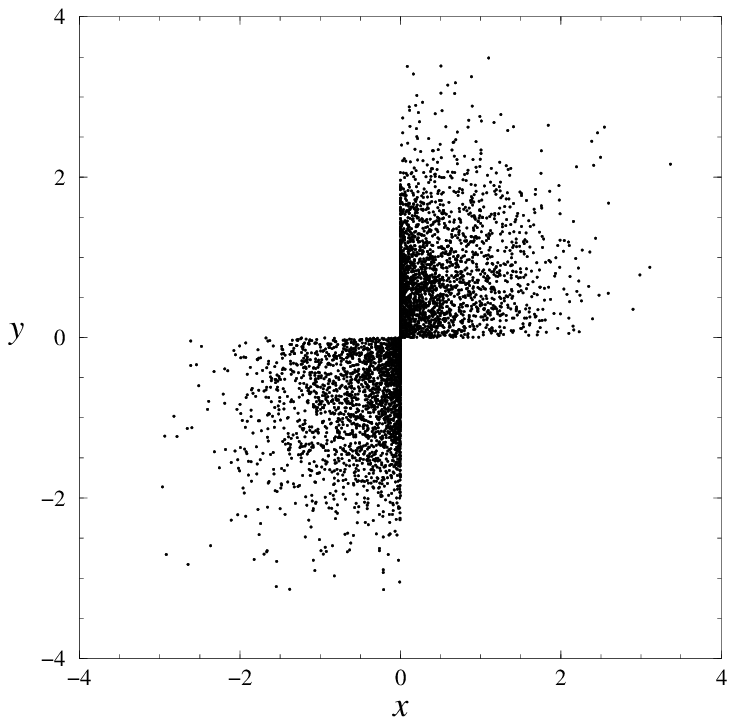}\hspace*{-80mm}}
\vspace*{-10mm}
\caption{Student and teacher fields $(x,y)=(\bJ\inn\bxi,\bB\inn\bxi)$
as observed during numerical simulations of on-line learning (learning rate $\eta=1$) in
a perceptron of size $N=10,000$ at $t=50$, using `questions' from a
restricted training set of size $p=N/2$.  Left: Hebbian
learning. Right: AdaTron learning.  Note: in the case of Gaussian
field distributions one would have found spherically shaped plots.}
\label{fig:example_scatterfields}
\end{figure}

\begin{figure}[t]
\vspace*{85mm}
\hbox to \hsize{\hspace*{-5mm}\includegraphics{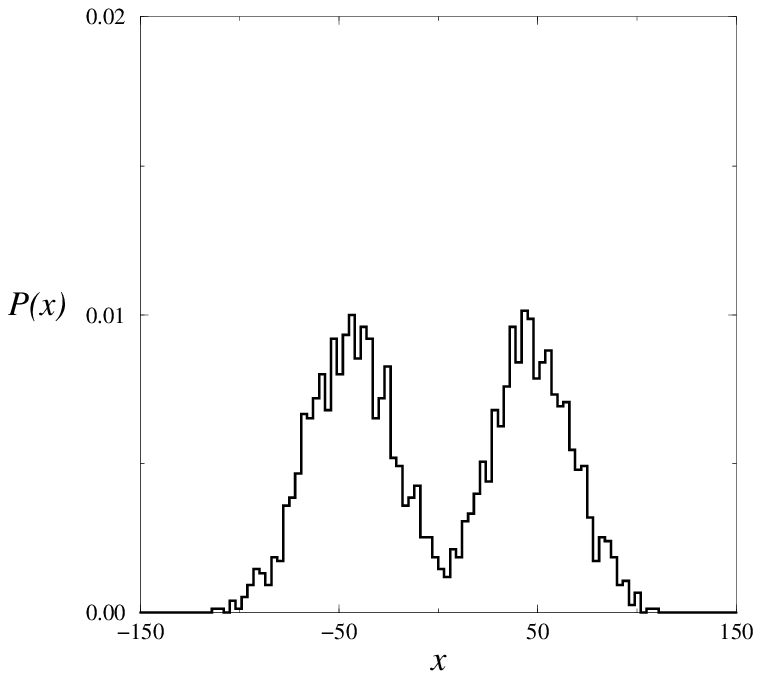}\hspace*{-5mm}}
\vspace*{-4.8mm}
\hbox to
\hsize{\hspace*{77mm}\includegraphics{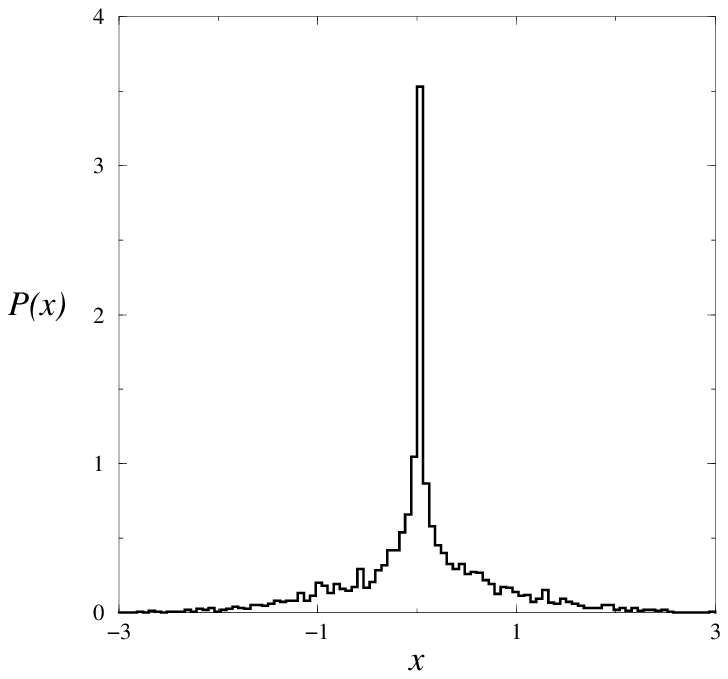}\hspace*{-77mm}}
\vspace*{-15mm}
\caption{Distribution $P(x)$ of student fields as observed
during numerical simulations of on-line learning (learning rate
$\eta=1$) in a perceptron of size $N=10,000$, using
`questions' from a restricted training set of size $p=N/4$.
Left: Hebbian learning, measured at $t=10$.  Right: AdaTron learning,
measured at $t=20$.  Note: not only are these distributions
distinctively non-Gaussian, they also appear to vary widely in their
basic characteristics, depending on the learning rule used.}
\label{fig:example_histograms}
\end{figure}
Here we study the dynamics of learning in layered neural
networks with restricted training sets, where the number $p$ of
examples (`questions' with corresponding `answers') scales linearly
with the number $N$ of inputs, i.e. $p=\alpha N$ with $0<\alpha<\infty$.
Here individual questions will
re-appear during the learning process as soon as the number of weight
updates made is of the order of the size of the training set. In the
traditional models, where the duration of an individual
update is defined as $N^{-1}$, this happens as soon as $t=\order(\alpha)$.
At that point
correlations develop between the weights and the questions in the
training set, and the dynamics is of a spin-glass type, with the
composition of the training set playing the role of `quenched
disorder'. The main consequence of this is that the central limit
theorem no longer applies to the student's local fields, which are now
indeed described by non-Gaussian distributions.  To demonstrate this we
trained (on-line) a perceptron with weights $J_{i}$
on noiseless examples generated by a teacher perceptron with
weights $B_i$, using the Hebb and AdaTron rules.  We plotted in Fig. 1
the student and teacher fields, $x = \bJ\inn\bxi$ and $y =
\bB\inn\bxi$ respectively, where $\bxi$ is the input vector, for
$p=N/2$ examples and at time $t=50$. The marginal distribution $P(x)$
for $p=N/4$, at times $t=10$ for the Hebb rule and $t=20$ for the
Adatron rule, is shown in Fig.~2.  The non-Gaussian student field
distributions observed in Figs.~1 and 2 induce a deviation between the
training- and generalization errors, which measure the network
performance on training and test examples, respectively. The former
involves averages over the non-Gaussian field distribution, whereas
the latter (which is calculated over {\em all} possible examples)
still involves Gaussian fields.

The appearance of
non-Gaussian  fields leads to a complete breakdown of the standard
formalism,  based on deriving closed equations for
a finite number of observables: the field distributions can no
longer be characterized by a few moments, and the macroscopic
laws must now be averaged over realizations of the training set.
One could still try to use Gaussian distributions as large
$\alpha$ approximations, see e.g. \cite{SollichBarber}, but it
will be clear from Figs. 1 and 2 that a systematic theory
will have to give up Gaussian distributions entirely. The
first rigorous study of the dynamics of learning with restricted
training sets in non-linear networks, via the calculation of
generating functionals, was carried out in \cite{Horner} for
perceptrons with binary weights.
The only cases where explicit and relatively simple solutions can be
obtained, even for restricted training sets, are those where linear
learning rules are used, such as \cite{KroghHertz} or \cite{Raeetal}.

In this paper we show how the formalism of dynamical replica
theory (see e.g. \cite{Coolenetal}) can be used successfully to
predict the evolution of macroscopic observables for finite $\alpha$,
incorporating the infinite training set formalism
 as a special case, for $\alpha\to\infty$.  Central to our approach is the
derivation of a diffusion equation for the joint distribution $P[x,y]$
of the
student and teacher fields, which will be found to have Gaussian
solutions only for $\alpha\to\infty$.  For simplicity and transparency
we restrict ourselves in the present paper to single-layer systems and
noise-free teachers.
Application and generalization of our methods to multi-layer systems
\cite{SaadCoolen} and learning scenarios involving `noisy'
teachers \cite{MaceCoolennew} are presently under way.

Our paper is organized as follows. In section 2 we first derive a
Fokker-Planck equation describing the evolution of arbitrary
mean-field observables for $N\to\infty$. This allows us to
identify the conditions for the latter to be described by closed
deterministic laws. We then choose as our observables the
joint field distribution $P[x,y]$, in addition to (the
traditional ones) $Q$ and $R$, and show that this set $\{Q,R,P\}$
obeys deterministic laws.  In order to close these laws
we use the tools of dynamical replica theory.  Details of the
replica calculation are given in an Appendix, so that they can
be skipped by those primarily interested in results.
In section 3 we summarize the final
replica-symmetric macroscopic theory
and its notational conventions, discuss some of its general properties,
and show how in
the limit $\alpha\to\infty$ (infinite training sets) the equations of
the conventional theory are recovered. In a subsequent paper
we will work out and apply our equations explicitly for several types
of learning rules, and compare the
predictions of our theory
with exact results (derived directly from the microscopic
equations, for Hebbian learning \cite{Raeetal}) and
with numerical simulations.


\section{From Microscopic to Macroscopic Laws}

\subsection{Definitions}

A student perceptron operates the
following rule, which is parametrised by a weight vector $\bJ\in\Re^N$:
\be
S:\{-1,1\}^N\to\{-1,1\} ~~~~~~~~~~~~~~~~~~~~
S(\bxi)=\sgn\left[\bJ\cdot\bxi\right]
\label{eq:student}
\ee
It tries to emulate the operation of a teacher perceptron, which
is assumed to operate a similar rule, characterized by a
given (fixed) weight vector $\bB\in\Re^N$:
\be
T:\{-1,1\}^N\to\{-1,1\} ~~~~~~~~~~~~~~~~~~~~
T(\bxi)=\sgn\left[\bB\cdot\bxi\right]
\label{eq:teacher}
\ee
In order to improve its performance, the student perceptron modifies
its weight vector $\bJ$ according to an iterative procedure, using
examples of input vectors (or `questions') $\bxi$, drawn at random
from a fixed training
set $\set\subseteq \all=\{-1,1\}^N$, and the corresponding values of
the teacher outputs $T(\bxi)$.

We will consider the case where the training set is a randomly composed
subset $\set\subset D$, of size $|\set|=p=\alpha N$ with $\alpha>0$:
\be \set=\{\bxi^1,\ldots,\bxi^p\}~~~~~~~~~~~~~~p=\alpha N~~~~~~~~~~~~~~
\bxi^\mu\in\all~~{\rm for~all}~\mu
\label{eq:trainingset}
\ee
We will denote averages over the training set $\set$ and averages
over the full question set $D$ in the following way:
\bd
\bra\Phi(\bxi) \ketset=\frac{1}{|\set|}\sum_{\bxi\in\set}\Phi(\bxi)
~~~~~~~{\rm and}~~~~~~~~
\bra
\Phi(\bxi)\ketall =\frac{1}{|\all|}\sum_{\bxi\in\all}\Phi(\bxi)
\ .
\ed
We will analyze the following two classes of learning rules:
\be
\begin{array}{lll}
{\rm on\!-\!line:} &&
\bJ(m\plus 1)=
\bJ(m)+
\frac{\eta}{N} \
\bxi(m) \ \G\left[\bJ(m)\inn\bxi(m), \bB\inn\bxi(m)\right]
\\[2mm]
{\rm batch:} &&
\bJ(m\plus 1)=
\bJ(m)+
\frac{\eta}{N} \
\bra \bxi~ \G\left[\bJ(m)\inn\bxi, \bB\inn\bxi\right]\ketset
\end{array}
\label{eq:weightdynamics}
\ee
In on-line learning one draws at each iteration step $m$ a question
$\bxi(m)\in\set$ at random, the dynamics is thus a stochastic process;
in batch learning one iterates a deterministic map.  The function
$\G[x,y]$ is assumed to be bounded and not to depend on $N$, other
than via its two arguments.

Our most important observables during learning are the training error
$E_{\rm t}(\bJ)$ and the generalization error $E_{\rm g}(\bJ)$,
defined as follows:
\be E_{\rm t}(\bJ)=\bra \theta[-(\bJ\inn\bxi)
(\bB\inn\bxi)]\ketset
~~~~~~~~~~~~~~~~~~~~
E_{\rm g}(\bJ)=\bra \theta[-(\bJ\inn\bxi) (\bB\inn\bxi)]\ketall \ .
\label{eq:E}
\ee
Only if the training set $\set$ is sufficiently large, and if
there are no correlations between $\bJ$ and the questions
$\bxi\in\set$, will these two errors will be identical.
\vsp

We next convert the dynamical laws (\ref{eq:weightdynamics}) into the
language of stochastic processes. We introduce the probability
$\hat{p}_m(\bJ)$ to find weight vector $\bJ$ at discrete iteration
step $m$. In terms of this microscopic probability distribution the
processes (\ref{eq:weightdynamics}) can be written in the general
Markovian form
\be \hat{p}_{m+1}(\bJ)=
\int\!d\bJ^\prime~W[\bJ;\bJ^\prime] \ \hat{p}_m(\bJ^\prime) \ ,
\label{eq:markovprocess}
\ee
with the transition probabilities
\be
\begin{array}{ll}
{\rm on\!-\!line:} &
W[\bJ;\bJ^\prime]=\bra \delta\left[
\bJ\minus \bJ^\prime\minus
\frac{\eta}{N} \ \bxi \ \G\left[\bJ^\prime\inn\bxi,\bB\inn\bxi\right]
\right]\ketset
\\[2mm]
{\rm batch:} &
W[\bJ;\bJ^\prime]=
\delta\left[\bJ\minus \bJ^\prime\minus
\frac{\eta}{N}\bra \bxi~\G\left[\bJ^\prime\inn\bxi,
\bB\inn\bxi\right]\ketset
\right]
\end{array}
\label{eq:transitionmatrix}
\ee
We make the transition to a description involving real-valued time
labels by choosing the duration of each iteration step to be a
real-valued random number, such that the probability that at time $t$
precisely $m$ steps have been made is given by the Poisson expression
\be \pi_m(t)=\frac{1}{m!}(Nt)^me^{-Nt} \ .
\label{eq:poisson}
\ee
For times $t\gg N^{-1}$ we find $t=m/N+\order(N^{-\frac{1}{2}})$,
the usual time unit.  Due to the random durations of the iteration
steps we have to switch to the following microscopic probability
distribution: \be p_t(\bJ)= \sum_{m\geq 0}\pi_m(t) \ \hat{p}_m(\bJ) \
.
\label{eq:conttime}
\ee
This distribution obeys a simple differential equation, which
immediately follows from the pleasant properties of (\ref{eq:poisson})
under temporal differentiation:
\be \frac{d}{dt} \ p_t(\bJ) =
N\int\!d\bJ^\prime~\left\{ W[\bJ;\bJ^\prime]-\delta[\bJ\minus
\bJ^\prime]\right\} \ p_t(\bJ^\prime) \ .
\label{eq:conttimemarkov}
\ee
So far no approximations have been made, equation
(\ref{eq:conttimemarkov}) is exact for any $N$. It is the equivalent
of the master equation often introduced to define the dynamics of spin
systems.

\subsection{Derivation of Macroscopic Fokker-Planck Equation}

We now wish to investigate the dynamics of a number of as yet
arbitrary {\em
macroscopic} observables
$\bOmega[\bJ]=(\Omega_1[\bJ],\ldots,\Omega_k[\bJ])$. To do so we
introduce a macroscopic probability distribution
\be
P_t(\bOmega)=
\int\!d\bJ~p_t(\bJ)\delta\left[\bOmega-\bOmega[\bJ]\right]
\ee
Its time derivative immediately follows from that in (\ref{eq:conttimemarkov}):
\bd
\frac{d}{dt}P_t(\bOmega)=N\int\!d\bJ
d\bJ^\prime~\delta\left[\bOmega\minus \bOmega[\bJ]\right]
\left\{
W[\bJ;\bJ^\prime]\minus \delta[\bJ\minus \bJ^\prime]\right\}
p_t(\bJ^\prime)
\ed
\bd
=N\int\!d\bOmega^\prime
\int\!d\bJ d\bJ^\prime~\delta\left[\bOmega\minus \bOmega[\bJ]\right]
\delta\left[\bOmega^\prime\minus \bOmega[\bJ^\prime]\right]
\left\{
W[\bJ;\bJ^\prime]\minus \delta[\bJ\minus \bJ^\prime]\right\}
p_t(\bJ^\prime)
\ed
This then can be written in the standard form
\be
\frac{d}{dt}P_t(\bOmega)
=
\int\!d\bOmega^\prime~ {\cal W}_t[\bOmega;\bOmega^\prime]
P_t(\bOmega^\prime)
\label{eq:macrodynamics}
\ee
where
\bd
{\cal W}_t[\bOmega;\bOmega^\prime]=
\frac{\int\!d\bJ^\prime~p_t(\bJ^\prime)
\delta\left[\bOmega^\prime\minus \bOmega[\bJ^\prime]\right]
\int\!d\bJ~
\delta\left[\bOmega\minus \bOmega[\bJ]\right]
N\left\{
W[\bJ;\bJ^\prime]\minus \delta[\bJ\minus \bJ^\prime]\right\}
}
{\int\!d\bJ^\prime~p_t(\bJ^\prime)\delta\left[\bOmega^\prime\minus
\bOmega[\bJ^\prime]\right]}
\ed
If we now insert the relevant expressions
(\ref{eq:transitionmatrix}) for $W[\bJ;\bJ^\prime]$ we can perform the
$\bJ$-integrations, and obtain results given in
terms of so-called sub-shell averages, which are defined as
\bd
\bra f(\bJ)\ket_{\bOmega;t}=
\frac{\int\!d\bJ~p_t(\bJ)\delta\left[\bOmega\minus
\bOmega[\bJ]\right]f(\bJ)}
{\int\!d\bJ~p_t(\bJ)\delta\left[\bOmega\minus
\bOmega[\bJ]\right]}
\ed
For the two classes of learning rules at hand we obtain:
\bd
{\cal W}^{\rm onl}_t[\bOmega;\bOmega^\prime]=N\bigbra\bra
\delta\left[
\bOmega\minus
\bOmega[\bJ\plus\frac{\eta}{N}\bxi\G[\bJ\cdot\bxi,\bB\cdot\bxi]]
\right]\ket_\set
\minus
\delta\left[\bOmega\minus \bOmega[\bJ]\right]
\bigket_{\!\bOmega^\prime;t}
\ed
\bd
{\cal W}^{\rm bat}_t[\bOmega;\bOmega^\prime]=
N\bigbra
\delta\left[\bOmega\minus
\bOmega[\bJ
\plus\frac{\eta}{N}\bra\bxi\G[\bJ\cdot\bxi,\bB\cdot\bxi]\ket_\set
]
\right]
\minus
\delta\left[\bOmega\minus \bOmega[\bJ]\right]
\bigket_{\!\bOmega^\prime;t}
\ed
We now insert integral representations for the $\delta$-distributions.
The observables $\bOmega[\bJ]\in \Re^k$ are assumed to be $\order(1)$ each, and
finite in number (i.e. $k\ll N$):
\be
\delta[\bOmega\minus
\bQ]=\int\!\frac{d\hat{\bOmega}}{(2\pi)^k}e^{i\hat{\bOmega}\cdot[\bOmega-\bQ]}
\label{eq:delta_function}
\ee
which gives for our two learning scenario's:
\be
{\cal W}^{\rm onl}_t[\bOmega;\bOmega^\prime]=
\int\!\frac{d\hat{\bOmega}}{(2\pi)^k}e^{i\hat{\bOmega}\cdot\bOmega}~
N\bigbra\bra
e^{-i\hat{\bOmega}\cdot
\bOmega[\bJ
\plus\frac{\eta}{N}\bxi\G[\bJ\cdot\bxi,\bB\cdot\bxi]
]}
\ket_\set
\minus
e^{-i\hat{\bOmega}\cdot\bOmega[\bJ]}
\bigket_{\!\bOmega^\prime;t}
\label{eq:macromap1}
\ee
\be
{\cal W}^{\rm bat}_t[\bOmega;\bOmega^\prime]=
\int\!\frac{d\hat{\bOmega}}{(2\pi)^k}e^{i\hat{\bOmega}\cdot\bOmega}~
N\bigbra
e^{-i\hat{\bOmega}\cdot\bOmega
[\bJ
\plus\frac{\eta}{N}\bra\bxi\G[\bJ\cdot\bxi,\bB\cdot\bxi]\ket_\set]}
\minus
e^{-i\hat{\bOmega}\cdot\bOmega[\bJ]}
\bigket_{\!\bOmega^\prime;t}
\label{eq:macromap2}
\ee
Still no approximations have been made. The above two expressions
differ only in at which stage the averaging over the training set
occurs.
\vsp

In expanding equations (\ref{eq:macromap1},\ref{eq:macromap2}) for
large $N$ and finite $t$ we have to be careful, since the system size
$N$ enters both as a small parameter to control the magnitude of the
modification of individual components of the weight vector, but also
determines the dimensions and lengths of various vectors that occur.
We therefore inspect more closely the usual Taylor expansions:
\bd
F[\bJ\plus\bk] -F[\bJ]=\sum_{\ell\geq 1}\frac{1}{\ell
!}\sum_{i_1=1}^N\cdots\sum_{i_\ell=1}^N k_{i_1}\cdots k_{i_\ell}
\frac{\partial^\ell F[\bJ]}{\partial J_{i_1}\cdots\partial J_{i_\ell}} \ .
\ed
If
we assess how derivatives with respect to individual components $J_i$
scale for mean-field observables such as
$Q[\bJ]=\bJ^2$ and $R[\bJ]=\bB\inn\bJ$, we find the
following scaling property which we will choose as our definition
of {\em simple} mean-field observables:
\be
F[\bJ]=\order(N^0),~~~~~~~~~~
\frac{\partial^\ell F[\bJ]}{\partial
J_{i_1}\cdots\partial J_{i_\ell}}=\order\left(|\bJ|^{-\ell} N^{\frac{1}{2}\ell-d}
\!\right)
~~~~~~(N\to\infty)
\label{eq:simplemeanfield}
\ee
in which $d$ is the number of different elements in the set
$\{i_1,\ldots,i_\ell\}$. For simple mean-field observables
we can now estimate the scaling of the
various terms in the Taylor expansion.
However, we will find that for restricted training sets not all relevant
observables will have the properties (\ref{eq:simplemeanfield}). In
particular,
the joint distribution of student and teacher fields will, for on-line
learning, have a contribution for which
all terms in the Taylor series
will have to be summed, giving rise to an additional term
$\Delta[\bJ;\bk]$ \footnote{We are grateful to Dr. Yuan-sheng Xiong
for alerting us to this important point.}.
The latter type of more {\em general} mean-field observables will
have to be defined via the identities
\be
F[\bJ\plus\bk]- F[\bJ]
=\Delta[\bJ;\bk]
+\sum_i k_i\frac{\partial F[\bJ]}{\partial
J_i} +\frac{1}{2}\sum_{ij}k_i k_j\frac{\partial^2 F[\bJ]}{\partial J_i
\partial J_j} +\sum_{\ell\geq 3}
\order\left(\!
\left[\frac{|\bk|}{|\bJ|}\right]^\ell\!\right)
\label{eq:meanfield}
\ee
\be
F[\bJ]=\order(N^0),~~~~~~~~~~
\Delta[\bJ;\bk]=\order\left(\!
|\bk|^2/|\bJ|^2\!\right)
\label{eq:Deltaterm}
\ee
(in the assessment of the order of the remainder terms of
(\ref{eq:meanfield}) we have used $\sum_i k_i=\order(\sqrt{N}|\bk|)$).
Simple mean-field observables correspond to $\Delta[\bJ;\bk]=0$.

We expand our macroscopic equations
(\ref{eq:macromap1},\ref{eq:macromap2}) for large $N$ and finite
times,
restricting ourselves from
now on to mean-field observables in the
sense of (\ref{eq:meanfield},\ref{eq:Deltaterm}).
One of our observables we choose to be
$\bJ^2$. In the present problem
the shifts $\bk$, being either
$\frac{\eta}{N}\bxi~ G[\bJ\cdot\bxi,\bB\cdot\bxi]$
or
$\frac{\eta}{N}\bra\bxi~
\G[\bJ\cdot\bxi,\bB\cdot\bxi]\ketset$, scale as
$|\bk|=\order(N^{-\frac{1}{2}})$. Consequently:
\bd
e^{-i\hat{\bOmega} \cdot \bOmega\left[\bJ\plus\bk\right]} =
e^{-i\hat{\bOmega} \cdot \bOmega\left[\bJ\right]
}\left\{ 1-i\hat{\bOmega} \cdot \bDelta\left[\bJ;\bk\right]
- i\sum_i
k_i\frac{\partial}{\partial J_i}(\hat{\bOmega}\cdot\bOmega[\bJ])
- \frac{i}{2}\sum_{ij}k_i k_j\frac{\partial^2}{\partial J_i
\partial J_j} (\hat{\bOmega}\cdot\bOmega[\bJ])
\right.
\ed
\bd
\left.
-
\frac{1}{2}\left[ \sum_i k_i\frac{\partial}{\partial
J_i}(\hat{\bOmega}\cdot\bOmega[\bJ]) \right]^2\right\}
+\order(N^{-\frac{3}{2}}) \ .
\ed
This, in turn, gives
\bd
\int\!\frac{d\hat{\bOmega}}{(2\pi)^k}e^{i\hat{\bOmega}\cdot\bOmega}~
N\left[e^{-i\hat{\bOmega}  \cdot  \bOmega\left[\bJ\plus\bk\right]} \minus
e^{-i\hat{\bOmega}  \cdot  \bOmega\left[\bJ\right]}\right]
~~~~~~~~~~~~~~~~~~~~~~~~~~~~~~~~~~~~~~~~~~~~~~~~~~~~~~~~~~~~~~~~~~~~~~
\ed
\bd
=-N \left\{ \sum_{\mu}\frac{\partial}{\partial\Omega_\mu}
\left[\Delta_\mu[\bJ;\bk]+\sum_i k_i\frac{\partial \Omega_\mu[\bJ]}{\partial J_i} \plus
\frac{1}{2}\sum_{ij}k_i k_j\frac{\partial^2\Omega_\mu[\bJ]}{\partial
J_i \partial J_j}\right] \right. ~~~~~~~~~~~~~~~~~~~~~~~~
\ed
\bd
\left.  ~~~~~~~~~~~~~~
-\frac{1}{2}\sum_{\mu\nu}\frac{\partial^2}{\partial\Omega_\mu
\partial\Omega_\nu} \sum_{ij}
k_ik_j\frac{\partial\Omega_\mu[\bJ]}{\partial J_i}
\frac{\partial\Omega_\nu[\bJ]}{\partial J_j}\right\}
\delta\left[\bOmega-\bOmega[\bJ]\right] +\order(N^{-\frac{1}{2}}) \ .
\ed
It is now evident, in view of
(\ref{eq:macromap1},\ref{eq:macromap2}),
that both types of dynamics are described by macroscopic laws with
transition probability densities of the general form
\bd
{\cal
W}^{\rm\star\star\star}_t[\bOmega;\bOmega^\prime]= \left\{
-\sum_{\mu}F_\mu[\bOmega^\prime;t]\frac{\partial}{\partial\Omega_\mu}
+ \frac{1}{2}\sum_{\mu\nu}G_{\mu\nu}[\bOmega^\prime;t]
\frac{\partial^2}{\partial\Omega_\mu\partial\Omega_\nu}
\right\}\delta\left[\bOmega\minus \bOmega^\prime\right]~
+~\order(N^{-\frac{1}{2}})
\ed
which, due to (\ref{eq:macrodynamics}) and for $N\to\infty$ and finite times,
leads to a Fokker-Planck equation:
\be
\frac{d}{dt}P_t(\bOmega)
=
-\sum_{\mu=1}^k\frac{\partial}{\partial\Omega_\mu}
\left\{F_\mu[\bOmega;t]P_t(\bOmega)\right\}
+\frac{1}{2}\sum_{\mu\nu=1}^k
\frac{\partial^2}{\partial\Omega_\mu\partial\Omega_\nu}
\left\{ G_{\mu\nu}[\bOmega;t]P_t(\bOmega)\right\} \ .
\label{eq:fokkerplanck}
\ee
The differences between the two types of dynamics are in the explicit
expressions for the flow- and diffusion terms:
\bd
F^{\rm onl}_\mu[\bOmega;t]=
\lim_{N\to\infty}\bigbra ~
N \bra \Delta_\mu[\bJ;\frac{\eta}{N}\bxi~G[\bJ\inn\bxi,\bB\inn\bxi]]
\ketset
+
\eta \sum_i \bra \xi_i \G[\bJ\inn\bxi,\bB\inn\bxi] \ketset
\frac{\partial \Omega_\mu[\bJ]}{\partial
J_i}
~~~~~~~~~~~
\right.
\ed
\bd
\left.
~~~~~~~~~~~~~~~~~~~~~~~~~~~~~~~~~~~~~~~~~~~~~~~~~~~~~~~~~~~~~~
+ \frac{\eta^2}{2N}\sum_{ij}
\bra \xi_i\xi_j \G^2[\bJ\inn\bxi,\bB\inn\bxi] \ketset
\frac{\partial^2\Omega_\mu[\bJ]}{\partial J_i \partial J_j}
~\bigket_{\bsomega;t}
\ed
\bd
G^{\rm onl}_{\mu\nu}[\bOmega;t]=\lim_{N\to\infty}
\frac{\eta^2}{N}  \bigbra~ \sum_{ij}
\bra \xi_i\xi_j \G^2[\bJ\inn\bxi,\bB\inn\bxi]\ketset
\left[\frac{\partial\Omega_\mu[\bJ]}{\partial J_i}\right]
\left[\frac{\partial\Omega_\nu[\bJ]}{\partial J_j}\right]
 ~\bigket_{\bsomega;t}
~~~~~~~~~~~~~~~~~~~~~~~~
\ed
\bd
F^{\rm bat}_\mu[\bOmega;t]=
\lim_{N\to\infty}\bigbra~
N \Delta_\mu[\bJ;\frac{\eta}{N}\bra \bxi~G[\bJ\cdot\bxi;\bB\cdot\bxi]\ketset]
+
\eta \sum_i
\bra\xi_i \ \G[\bJ\cdot\bxi,\bB\cdot\bxi]  \ketset
\frac{\partial \Omega_\mu[\bJ]}{\partial J_i}
~~~~~~~
\right.
\ed
\bd
\left.
~~~~~~~~~~~~~~~~~~~~~~~~~~~~~~~~~~~~~~~~~
+ \frac{\eta^2}{2N} \sum_{ij}
\bra\xi_i \ \G[\bJ\cdot\bxi,\bB\cdot\bxi]\ketset
\bra\xi_j \ \G[\bJ\cdot\bxi,\bB\cdot\bxi]\ketset
\frac{\partial^2\Omega_\mu[\bJ]}{\partial J_i \partial J_j}
~
\bigket_{\bsomega;t}
\ed
\bd
G^{\rm bat}_{\mu\nu}[\bOmega;t] =\lim_{N\to\infty}
\frac{\eta^2}{N} \bigbra ~ \sum_{ij}
\bra\xi_i\G[\bJ\!\cdot\!\bxi,\bB\!\cdot\!\bxi]
\ketset  \
\bra\xi_j\G[\bJ\!\cdot\!\bxi,\bB\!\cdot\!\bxi]
\ketset
\left[\frac{\partial\Omega_\mu[\bJ]}{\partial J_i}\right] \!
\left[\frac{\partial\Omega_\nu[\bJ]}{\partial J_j}\right]
~\bigket_{\bsomega;t}
~~~~~
\ed
Equation (\ref{eq:fokkerplanck}) allows us to define the goal of our
exercise in more explicit form. If we wish to arrive at closed
deterministic macroscopic equations, we have to choose our observables
such that
\begin{enumerate}
\item $\lim_{N\to\infty}G_{\mu\nu}[\bOmega;t]=0$ ~~~~~~~~(this ensures
determinism)
\item $\lim_{N\to\infty}\frac{\partial}{\partial
t}F_\mu[\bOmega;t]=0$ ~~~~~~~(this ensures closure)
\end{enumerate}
In the case of having time-dependent global parameters, such as learning
rates or decay rates, the latter condition relaxes to the requirement
that any explicit time-dependence of $F_\mu[\bOmega;t]$ is restricted
to these global parameters.

\subsection{Choice and Properties of Canonical Observables}

We next apply the general results obtained so far to a specific set of
observables, $\bOmega\to\{Q,R,P\}$,
which are tailored to the problem at hand (note that
we restrict ourselves to
$\bJ^2=\order(1)$ and $\bB^2=1$):
\be
Q[\bJ]=\bJ^2,~~~~~~~~~~~
R[\bJ]=\bJ\inn\bB,~~~~~~~~~~~
P[x,y;\bJ]=\bra\delta[x\minus \bJ\inn\bxi] \
\delta[y\minus\bB\inn\bxi]\ketset
\label{eq:learningobservables}
\ee
with $x,y\in \Re$.
This choice is motivated by the following considerations: (i) in order
to incorporate the standard theory in the limit $\alpha\to\infty$ we
need at least $Q[\bJ]$ and $R[\bJ]$, (ii) we need to be able to
calculate the training error, which involves field statistics
calculated over the training set $\set$, as described by $P[x,y;\bJ]$,
and (iii) for finite $\alpha$ one cannot expect closed macroscopic
equations for just a finite number of order parameters, the present
choice (involving the order parameter {\em function} $P[x,y;\bJ]$)
represents effectively an infinite number \footnote{A simple rule of
thumb is the following: if a process requires replica theory for its
stationary state analysis, as does learning with restricted training
sets, its dynamics is of a spin-glass type and cannot be described by
a finite set of closed dynamic equations.}.  In subsequent
calculations we will, however, assume the number of arguments $(x,y)$
for which $P[x,y;\bJ]$ is to be evaluated (and thus our number of
order parameters) to go to infinity only after the limit $N\to\infty$
has been taken. This will eliminate many technical subtleties and will
allow us to use the Fokker-Planck equation (\ref{eq:fokkerplanck}).

The observables (\ref{eq:learningobservables}) are indeed of the
general mean-field type in the sense of
(\ref{eq:meanfield},\ref{eq:Deltaterm}).
Insertion into the stronger condition
(\ref{eq:simplemeanfield}) immediately shows this to be true for the scalar
observables $Q[\bJ]$ and $R[\bJ]$ (they are simple mean field
observables, for which the term (\ref{eq:Deltaterm}) is absent).
Verification of  (\ref{eq:meanfield},\ref{eq:Deltaterm}) for
the function $P[x,y;\bJ]$ is less trivial.
We denote
with $\cal{I}$ the set of all {\em different} indices in the list
$(i_1,\ldots,i_\ell)$, with $n_k$ giving the number of times a number
$k$ occurs, and with ${\cal I}^\pm\subseteq{\cal I}$ defined as the
set of all indices $k\in{\cal I}$ for which $n_k$ is even ($+$), or
odd ($-$). Note that with these definitions
$\ell=\sum_{k\in{\cal I}^+} n_k+\sum_{k\in{\cal I}^{-}}n_k\geq 2|{\cal
I}^+| +|{\cal I}^-|$.
We then have:
\bd
\frac{\partial^\ell P[x,y;\bJ]}{\partial
J_{i_1}\!\ldots\!\partial J_{i_\ell}}
=(\minus 1)^\ell\frac{\partial^\ell}{\partial x^\ell}
\int\frac{d\hat{x} \ d\hat{y}}{(2\pi)^2}
e^{i[x\hat{x}+y\hat{y}]}
\bigbra
\left[\prod_{k\in {\cal I}}\xi_k^{n_k}e^{-i\xi_k
[\hat{x}J_k+\hat{y}B_k]}\right]
\left[\prod_{k\notin {\cal I}}e^{-i\xi_k
[\hat{x}J_k+\hat{y}B_k]}\right]
\bigket_{\!\set}
\ed
Upon writing averaging over {\em all} training sets of size $p=\alpha N$ (where
each realization of $\set$ has equal probability) as $\bra
\ldots\ket_\sets$, this allows us to conclude
\bd
\bigbra\frac{\partial^\ell P[x,y;\bJ]}{\partial
J_{i_1}\!\ldots\!\partial J_{i_\ell}}\bigket_{\!\sets}\!=
\order\left(N^{-\frac{1}{2}|{\cal I}^-|}\right)
\ed
Since $\frac{1}{2}\ell \minus |{\cal I}|\plus \frac{1}{2}|{\cal I}^-|=
\frac{1}{2}[\ell\minus |{\cal I}^-| \minus 2|{\cal I}^+|]\geq 0$, the
{\em average over all training sets} of the
function $P[x,y;\bJ]$
 is found to be a simple mean-field observable in the sense of
(\ref{eq:simplemeanfield}).

The scaling
properties of expansions or derivations of $P[x,y;\bJ]$ for a given
training set $\set$, however,
need not be identical to those of its average over all training sets
$\bra P[x,y;\bJ]\ket_\sets$.
Here we have to use the
fact  that $\set$ has been composed in a random manner, as well as
the specific form of the shifts $\bk$ in $P[x,y;\bJ\plus\bk]$ that
occur for the two types of dynamics under consideration:
\bd
P[x,y;\bJ\plus \bk]-P[x,y;\bJ]=
\int\frac{d\hat{x} \ d\hat{y}}{(2\pi)^2}
e^{i[x\hat{x}+y\hat{y}]}
\frac{1}{p} \sum_{\mu=1}^p e^{-i\hat{x}\bJ\cdot\bxi^\mu
-i\hat{y}\bB\cdot\bxi^\mu}\left[
e^{-i\hat{x}\bk\cdot\bxi^\mu}-1\right]
\ed
All complications are caused by the dependence of $\bk$ on the
composition of the training set $\set$, and would therefore have been
absent in the $\alpha\rightarrow\infty$ case.
This dependence will turn out to be harmless in the case of batch
learning, where  $\bk=\frac{\eta}{N}\bra\bxi
\G[\bJ\inn\bxi,\bB\inn\bxi]\ketset$  is an average over $\set$,
but will
have a considerable impact
 in the case of on-line learning, where
$\bk=\frac{\eta}{N}\bxi \G[\bJ\inn\bxi,\bB\inn\bxi]$
is proportional to an individual member of $\set$.
Working out the relevant expression for on-line learning gives
\bd
P[x,y;\bJ\plus \bk^{\rm onl}]-P[x,y;\bJ]=
\int\frac{d\hat{x} \ d\hat{y}}{(2\pi)^2}
e^{i[x\hat{x}+y\hat{y}]}\frac{1}{p} \sum_{\mu=1}^p e^{-i\hat{x}\bJ\cdot\bxi^\mu
-i\hat{y}\bB\cdot\bxi^\mu}
\left\{\room
\delta_{\bxi^\mu \bxi} \left[e^{-i\eta \hat{x}
\G[\bJ\cdot\bxi,\bB\cdot\bxi]}-1\right]
\right.
\ed
\bd
\left.
-
[1\minus \delta_{\bxi^\mu,\bxi}]
\left[\frac{i\eta\hat{x}}{N}(\bxi\cdot\bxi^\mu)\G[\bJ\inn\bxi,\bB\inn\bxi]
+\frac{\eta^2\hat{x}^2}{2N^2}(\bxi\cdot\bxi^\mu)^2
\G^2[\bJ\inn\bxi,\bB\inn\bxi]
+\order(N^{-\frac{3}{2}})
\right]\right\}
\ed
\bd
=\frac{1}{p}\int\frac{d\hat{x} d\hat{y}}{(2\pi)^2}
e^{i[x\hat{x}+y\hat{y}]}
 e^{-i\hat{x}\bJ\cdot\bxi-i\hat{y}\bB\cdot\bxi}
 \left\{
\left[e^{-i\eta \hat{x}\G[\bJ\cdot\bxi,\bB\cdot\bxi]}-1\right]
+i\eta\hat{x}\G[\bJ\inn\bxi,\bB\inn\bxi]
+\frac{1}{2}\eta^2\hat{x}^2\G^2[\bJ\inn\bxi,\bB\inn\bxi]
\right\}
\ed
\bd
+\sum_i k_i^{\rm onl}\frac{\partial}{\partial J_i}P[x,y;\bJ]
+\frac{1}{2}\sum_{ij} k_i^{\rm onl}k_j^{\rm onl}
\frac{\partial^2}{\partial J_i\partial J_j} P[x,y;\bJ]
+\order(N^{-\frac{3}{2}})
\ed
We conclude that, at least for the purpose of the expansions relevant
to on-line learning, $P[x,y;\bJ]$ is a mean field
observable in the sense of (\ref{eq:meanfield},\ref{eq:Deltaterm}),
with the non-trivial contribution of (\ref{eq:Deltaterm}) given by
\bd
\Delta[\bJ;\bk^{\rm onl}]
=
\frac{1}{p}\left\{\room
\delta[x\minus \bJ\inn\bxi\minus \eta
\G[\bJ\inn\bxi,\bB\inn\bxi]]\delta[y\minus\bB\inn\bxi]
-\delta[x\minus \bJ\inn\bxi]\delta[y\minus\bB\inn\bxi]
~~~~~~~~~~~~~~~~~~~~~~~~~
\right.
\ed
\be
\left.
~~~~~~
+\eta\frac{\partial}{\partial x}\left[\G[x,y]
\delta[x\minus \bJ\inn\bxi]\delta[y\minus\bB\inn\bxi]\right]
-\frac{1}{2}\eta^2\frac{\partial^2}{\partial x^2}\left[\G^2[x,y]
\delta[x\minus \bJ\inn\bxi]\delta[y\minus\bB\inn\bxi]\right]
\room\right\}
\label{eq:deltaterm_online}
\ee
Note that $\lim_{N\to\infty}N\Delta[\bJ;\bk^{\rm
onl}]=\order(\eta^3/\alpha)$, so that for small learning rates or
large training sets this non-trivial term will vanish.
Working out the relevant expression for batch learning, on the other
hand,
gives
\bd
P[x,y;\bJ\plus \bk^{\rm bat}]-P[x,y;\bJ]=
\int\frac{d\hat{x} \ d\hat{y}}{(2\pi)^2}
e^{i[x\hat{x}+y\hat{y}]}
\frac{1}{p} \sum_{\mu=1}^p e^{-i\hat{x}\bJ\cdot\bxi^\mu
-i\hat{y}\bB\cdot\bxi^\mu}~~~~~~~~~~~~~~~~~~~~~
\ed
\bd
\times\left\{\room\left[
1-\frac{i\eta\hat{x}}{p}\G[\bJ\inn\bxi^\mu,\bB\inn\bxi^\mu]
+~\order(N^{-\frac{3}{2}})\right]-1\right\}
\ed
\bd
=\sum_i k_i^{\rm bat}\frac{\partial}{\partial J_i}P[x,y;\bJ]
+\frac{1}{2}\sum_{ij} k_i^{\rm bat}k_j^{\rm bat}
\frac{\partial^2}{\partial J_i\partial J_j} P[x,y;\bJ]
+\order(N^{-\frac{3}{2}})
\ed
Here the term $\Delta[\bJ;\bk^{\rm bat}]$ is absent.
In fact also the quadratic contribution $\sum_{ij}k_i^{\rm bat}
k_j^{\rm bat}\ldots$ in the above expansion will turn out to
be of insignificant
order in $N$.
For the purpose of the expansions relevant
to batch learning, $P[x,y;\bJ]$ is apparently a simple mean field
observable in the sense of (\ref{eq:simplemeanfield}).
This could have been anticipated, since one should ultimately obtain the batch
learning equations upon expanding those of on-line learning for small
learning rate $\eta$, and retaining only the leading order $\eta^1$ in
this expansion.

\subsection{Derivation of Deterministic Dynamical Laws}

Having defined our order parameters $Q$, $R$ and $\{P[x,y]\}$,
from this stage onwards
the notation $\bra\cdots \ket_{\ketop;t}$ will be used to
denote sub-shell  averages defined
with respect to these order parameters, at time
$t$. With a modest amount of foresight
we define the complementary Kronecker delta
$\overline{\delta}_{ab}=1\minus \delta_{ab}$,
and the following key functions:
\be
\A[x,y;x^\prime,y^\prime]
=\lim_{N\to\infty}
\bigbra
\bra\bra~
\overline{\delta}_{\bxi\bxi^\prime}(\bxi\inn\bxi^\prime)
\delta[x\minus\bJ\inn\bxi]\delta[y\minus\bB\inn\bxi]
\delta[x^\prime\!\minus\bJ\inn\bxi^\prime]\delta[y^\prime\!\minus\bB\inn\bxi^\prime]
~\ketset\ketset
\bigket_{\!\!\ketop;t}
\label{eq:greensfunction}
\ee
\be
\B[x,y;x^\prime,y^\prime]
=\lim_{N\to\infty}
\bigbra\frac{1}{N}\sum_{i\neq j}
\bra\bra~
\overline{\delta}_{\bxi\bxi^\prime}(\xi_i\xi_j\xi^\prime_i\xi^\prime_j)
\delta[x\minus\bJ\inn\bxi]\delta[y\minus\bB\inn\bxi]
\delta[x^\prime\!\minus\bJ\inn\bxi^\prime]\delta[y^\prime\!\minus\bB\inn\bxi^\prime]
~\ketset\ketset
\bigket_{\!\!\ketop;t}
\label{eq:zerogreensfunction}
\ee
\bd
\C[x,y;x^\prime,y^\prime;x^\pprime,y^\pprime]
=\lim_{N\to\infty}
~~~~~~~~~~~~~~~~~~~~~~~~~~~~~~~~~~~~~~~~~~~~~~~~~~~~~~~~~~~~~~~~~~~~~~~~~~~~~~~~~~~~~~~~~
\ed
\be
\bigbra
\bra\bra\bra~
\overline{\delta}_{\bxi\bxi^\pprime}\overline{\delta}_{\bxi^\prime\bxi^\pprime}
\frac{(\bxi\inn\bxi^\pprime)(\bxi^\prime\inn\bxi^\pprime)}{N}
\delta[x\minus\bJ\inn\bxi]\delta[y\minus\bB\inn\bxi]
\delta[x^\prime\!\minus\bJ\inn\bxi^\prime]\delta[y^\prime\!\minus\bB\inn\bxi^\prime]
\delta[x^\pprime\!\minus\bJ\inn\bxi^\pprime]\delta[y^\pprime\!\minus\bB\inn\bxi^\pprime]
~\ketset\ketset\ketset
\bigket_{\!\!\ketop;t}
\label{eq:zerodiffusion}
\ee
We will eventually show in a subsequent section
that (\ref{eq:zerogreensfunction}) and
(\ref{eq:zerodiffusion}) are zero. The function
(\ref{eq:greensfunction}), on the other hand, will contain all the
interesting physics of the learning process,
and its calculation will turn out to be our central problem.

We next show that for the observables (\ref{eq:learningobservables})
the diffusion matrix elements $G_{\mu\nu}^{\star\star\star}$ in the
Fokker-Planck equation (\ref{eq:fokkerplanck}) vanish for
$N\to\infty$. Our observables will consequently obey deterministic
dynamical laws.
Calculating diffusion terms associated with $Q[\bJ]$ and
$R[\bJ]$ is trivial:
\bd
\left[\!\!\begin{array}{ll}
G^{\rm onl}_{QQ}[\ldots]\\[2mm]
G^{\rm onl}_{QR}[\ldots]\\[2mm]
G^{\rm onl}_{RR}[\ldots]
\end{array}\!\!\right]
=\lim_{N\to\infty}
\frac{\eta^2}{N}\int\!dxdy~P[x,y]~\G^2[x,y]
\left[\!\!\begin{array}{c}
4 x^2
\\
2xy
\\
y^2
\end{array}\!\!\right]=0
\ed
\bd
\left[\!\!\begin{array}{ll}
G^{\rm bat}_{QQ}[\ldots]\\[2mm]
G^{\rm bat}_{QR}[\ldots]\\[2mm]
G^{\rm bat}_{RR}[\ldots]
\end{array}\!\!\right]
=\lim_{N\to\infty}
\frac{\eta^2}{N}
\left[\!\!\begin{array}{c}
4\left\{\room\int\!dxdy~P[x,y]~x \G[x,y]
\right\}^2
\\
2
\left\{\room \int\!dxdy~P[x;y]~ x\G[x,y]
\right\}
\left\{\room \int\!dxdy~P[x;y] ~ y \G[x,y]
\right\}
\\
\left\{\room \int\!dx dy~P[x;y] ~ y  \G[x,y]
\right\}^2
\end{array}\!\!\right]=0
\ed
We next turn to diffusion
terms with one occurrence of $P[x,y;\bJ]$.
Here we repeatedly
build on the cornerstone assumption that all fields $\bJ\inn\bxi$ and
$\bB\inn\bxi$ are of order unity (which is clear from
numerical simulations, and will be supported
self-consistently by the equations resulting from our theory),
in combination with
two simple scaling consequences  of the random composition of
$\set$, as $N\to\infty$:
\be
\bxi\in\set:~~~
\frac{1}{p}\sum_{\bxi^\prime\in\set}\delta_{\bxi\bxi^\prime}=p^{-1}+\order(p^{-2})
~~~~~~~~~~~~~~
\frac{1}{p^2}\sum_{\bxi\in\set}\sum_{\bxi^\prime\in\set}[1\minus\delta_{\bxi\bxi^\prime}]~|\bxi\inn\bxi^\prime|=\order(N^{\frac{1}{2}})
\label{eq:simplescalingrelations}
\ee
For on-line learning we find:
\bd
\left[\!\!\begin{array}{ll}
G^{\rm onl}_{Q,P[x,y]}[\ldots]\\[2mm]
G^{\rm onl}_{R,P[x,y]}[\ldots]
\end{array}\!\!\right]
=
-\lim_{N\to\infty}\frac{\eta^2}{N}\frac{\partial}{\partial x}
 \bigbra
\bra\bra~ \G^2[\bJ\inn\bxi,\bB\inn\bxi]
\left[\!\begin{array}{c}2\bJ\inn\bxi\\ \bB\inn\bxi
\end{array}\!\right](\bxi\inn\bxi^\prime)
\delta[x\minus\bJ\inn\bxi^\prime]\delta[y\minus\bB\inn\bxi^\prime]
~\ketset\ketset
\bigket_{\!\!\ketop;t}
\ed
\bd
=
-\eta^2\frac{\partial}{\partial x}
\lim_{N\to\infty}
 \bigbra~
\frac{1}{N}
\bra\bra~
[1\minus \delta_{\bxi\bxi^\prime}] \G^2[\bJ\inn\bxi,\bB\inn\bxi]
\left[\!\!\begin{array}{c}2\bJ\inn\bxi\\ \bB\inn\bxi
\end{array}\!\!\right](\bxi\inn\bxi^\prime)
\delta[x^\prime\minus\bJ\inn\bxi^\prime]
\delta[y^\prime\minus\bB\inn\bxi^\prime]
~\ketset\ketset
\right.
\ed
\bd
\left.
~~~~~~~~~~~~~~~~~~~~~~~~~~~~~~~~~~~~~~~~~~~~~~~~~~~~~~~~~~~~~
+~
\G^2[x,y]\left[\!\!\begin{array}{c}2x \\ y\end{array}\!\!\right]
\bra\bra~ \delta_{\bxi\bxi^\prime}
\delta[x\minus\bJ\inn\bxi]\delta[y\minus\bB\inn\bxi]
~\ketset\ketset~
\bigket_{\!\!\ketop;t}
\ed
\bd
=-\eta^2\frac{\partial}{\partial x}
\lim_{N\to\infty}\bigbra \order(N^{-\frac{1}{2}})+\order(N^{-1})
\bigket_{\!\!\ketop;t}=0
\ed
For batch learning we find:
\bd
\left[\!\begin{array}{ll}
G^{\rm bat}_{Q,P[x,y]}[\ldots]\\[2mm]
G^{\rm bat}_{R,P[x,y]}[\ldots]
\end{array}\!\right]
=
-\lim_{N\to\infty}
\frac{\eta^2}{N} \frac{\partial}{\partial x}
\int\!dx^\prime dy^\prime P[x^\prime,y^\prime]\G[x^\prime,y^\prime]
\left[\!\!\begin{array}{c}2x^\prime \\ y^\prime\end{array}\!\!\right]
~~~~~~~~~~~~~~~~~~~~~~~~~~~~~~~~~~~~~~~~
\ed
\bd
~~~~~~~~~~~~~~~~~~~~~~~~~~~~~~~~~~~~~~~~
\times~
\bigbra \bra\bra~
\G[\bJ\inn\bxi,\bB\inn\bxi](\bxi\inn\bxi^\prime)
\delta[x\minus\bJ\inn\bxi^\prime]
\delta[y\minus\bB\inn\bxi^\prime]
~\ketset\ketset
\bigket_{\!\!\ketop;t}
\ed
\bd
= -\eta^2\frac{\partial}{\partial x}
\int\!dx^\prime dy^\prime P[x^\prime,y^\prime]\G[x^\prime,y^\prime]
\left[\!\!\begin{array}{c}2x^\prime \\ y^\prime\end{array}\!\!\right]
\lim_{N\to\infty}\bigbra~
\G[x,y] \bra\bra~\delta_{\bxi\bxi^\prime}
\delta[x\minus\bJ\inn\bxi]
\delta[y\minus\bB\inn\bxi]
~\ketset\ketset
\right.
~~~~~~~~~~
\ed
\bd
\left.
~~~~~~~~~~~~~~~~~~~~~~~~~~~~~~~~~~~~~~~~~~~~~~~
+\frac{1}{N}
 \bra\bra~[1\minus\delta_{\bxi\bxi^\prime}]
\G[\bJ\inn\bxi,\bB\inn\bxi](\bxi\inn\bxi^\prime)
\delta[x\minus\bJ\inn\bxi^\prime]
\delta[y\minus\bB\inn\bxi^\prime]
~\ketset\ketset
~\bigket_{\!\!\ketop;t}
\ed
\bd
=-\eta^2\frac{\partial}{\partial x}
\lim_{N\to\infty}\bigbra \order(N^{-1})+\order(N^{-\frac{1}{2}})
\bigket_{\!\!\ketop;t}=0
\ed
The difficult terms are those where two derivatives of the order
parameter function $P[x,y;\bJ]$ come into play. Here we have to
deal separately with four distinct contributions,
defined according to which of the vectors
from the trio $\{\bxi,\bxi^\prime,\bxi^\pprime\}$
are identical.
For on-line learning we find:
\bd
G^{\rm onl}_{P[x,y],P[x^\prime,y^\prime]}[\ldots]=
\lim_{N\to\infty}\frac{\eta^2}{N}
\frac{\partial^2}{\partial x\partial x^\prime}
~~~~~~~~~~~~~~~~~~~~~~~~~~~~~~~~~~~~~~~~~~~~~~~~~~~~~~~~~~~~~~~~~~~~~~~~~~~~~~~
\ed
\bd
\bigbra
\bra\bra\bra~
\G^2[\bJ\inn\bxi^\pprime,\bB\inn\bxi^\pprime]
(\bxi\inn\bxi^\pprime)(\bxi^\prime\inn\bxi^\pprime)
\delta[x\minus\bJ\inn\bxi]\delta[y\minus\bB\inn\bxi]
\delta[x^\prime\minus\bJ\inn\bxi^\prime]\delta[y^\prime\minus\bB\inn\bxi^\prime]
~\ketset\ketset\ketset
\bigket_{\!\!\ketop;t}
\ed
\bd
=\eta^2\frac{\partial^2}{\partial x\partial x^\prime}
\lim_{N\to\infty}
\bigbra~~
N\G^2[x,y]\delta[x^\prime\minus x]\delta[y^\prime\minus y]~
\bra\bra\bra~
\delta_{\bxi\bxi^\pprime}\delta_{\bxi^\prime\bxi^\pprime}
\delta[x\minus\bJ\inn\bxi]\delta[y\minus\bB\inn\bxi]
~\ketset\ketset\ketset
\right.
\ed
\bd
\left.\room
+~\G^2[x^\prime,y^\prime]~
\bra\bra\bra~
\overline{\delta}_{\bxi\bxi^\pprime}\delta_{\bxi^\prime\bxi^\pprime}
(\bxi\inn\bxi^\prime)
\delta[x\minus\bJ\inn\bxi]\delta[y\minus\bB\inn\bxi]
\delta[x^\prime\minus\bJ\inn\bxi^\prime]\delta[y^\prime\minus\bB\inn\bxi^\prime]
~\ketset\ketset\ketset
\right.
\ed
\bd
\left.\room
+~\G^2[x,y]~
\bra\bra\bra~
\delta_{\bxi\bxi^\pprime}\overline{\delta}_{\bxi^\prime\bxi^\pprime}
(\bxi\inn\bxi^\prime)
\delta[x\minus\bJ\inn\bxi]\delta[y\minus\bB\inn\bxi]
\delta[x^\prime\minus\bJ\inn\bxi^\prime]\delta[y^\prime\minus\bB\inn\bxi^\prime]
~\ketset\ketset\ketset
\right.
\ed
\bd
\left.
+
\bra\bra\bra~
\overline{\delta}_{\bxi\bxi^\pprime}\overline{\delta}_{\bxi^\prime\bxi^\pprime}
\G^2[\bJ\inn\bxi^\pprime\!,\bB\inn\bxi^\pprime]
\frac{(\bxi\inn\bxi^\pprime)(\bxi^\prime\inn\bxi^\pprime)}{N}
\delta[x\minus\bJ\inn\bxi]\delta[y\minus\bB\inn\bxi]
\delta[x^\prime\!\minus\bJ\inn\bxi^\prime]\delta[y^\prime\!\minus\bB\inn\bxi^\prime]
~\ketset\ketset\ketset
~\bigket_{\!\!\ketop;t}
\ed
\bd
=
\eta^2\frac{\partial^2}{\partial x\partial x^\prime}\left\{
\lim_{N\to\infty}
 \bigbra
\order(N^{-1})+\order(N^{-\frac{1}{2}})+\order(N^{-\frac{1}{2}})
\!\bigket_{\!\!\ketop;t}
+\int\!dx^\pprime dy^\pprime
\G^2[x^\pprime,y^\pprime]~\C[x,y;x^\prime,y^\prime;x^\pprime,y^\pprime]
\right\}
\ed
\bd
=
\eta^2
\int\!dx^\pprime dy^\pprime
\G^2[x^\pprime,y^\pprime]~
\frac{\partial^2}{\partial x\partial x^\prime}
\C[x,y;x^\prime,y^\prime;x^\pprime,y^\pprime]
\ed
Similarly:
\bd
G^{\rm bat}_{P[x,y],P[x^\prime,y^\prime]}[\ldots]=
\lim_{N\to\infty}
\frac{\eta^2}{N}\frac{\partial^2}{\partial x\partial x^\prime}
~~~~~~~~~~~~~~~~~~~~~~~~~~~~~~~~~~~~~~~~~~~~~~~~~~~~~~~~~~~~~~~~~~~~~~~~~~~~~~~
\ed
\bd
\bigbra
\bra\bra\G[\bJ\inn\bxi^\prime,\bB\inn\bxi^\prime](\bxi\inn\bxi^\prime)
\delta[x\minus\bJ\inn\bxi]\delta[y\minus\bB\inn\bxi]\ketset\ketset~
\bra\bra\G[\bJ\inn\bxi,\bB\inn\bxi](\bxi\inn\bxi^\prime)
\delta[x^\prime\minus\bJ\inn\bxi^\prime]\delta[y^\prime\minus\bB\inn\bxi^\prime]\ketset\ketset
\bigket_{\!\!\ketop;t}
\ed
\bd
=\eta^2\frac{\partial^2}{\partial x\partial x^\prime}\lim_{N\to\infty}
\bigbra
\left\{
\G[x,y]\bra\bra \delta_{\bxi\bxi^\prime}
\delta[x\minus\bJ\inn\bxi]\delta[y\minus\bB\inn\bxi]\ketset\ketset
+
\bra\bra \overline{\delta}_{\bxi\bxi^\prime}
\G[\bJ\inn\bxi^\prime,\bB\inn\bxi^\prime]\frac{\bxi\inn\bxi^\prime}{N}
\delta[x\minus\bJ\inn\bxi]\delta[y\minus\bB\inn\bxi]\ketset\ketset
\right\}
\right.
\ed
\bd
\left.
\times\left\{
\G[x^\prime,y^\prime]\bra\bra \delta_{\bxi\bxi^\prime}
\delta[x^\prime\minus\bJ\inn\bxi]\delta[y^\prime\minus\bB\inn\bxi]\ketset\ketset
+
\bra\bra \overline{\delta}_{\bxi\bxi^\prime}
\G[\bJ\inn\bxi,\bB\inn\bxi]\frac{\bxi\inn\bxi^\prime}{N}
\delta[x^\prime\minus\bJ\inn\bxi^\prime]\delta[y^\prime\minus\bB\inn\bxi^\prime]
\ketset\ketset
\right\}
\bigket_{\!\!\ketop;t}
\ed
\bd
=\eta^2\frac{\partial^2}{\partial x\partial
x^\prime}\lim_{N\to\infty}\bigbra
\left\{\order(N^{-1})+\order(N^{-\frac{1}{2}})\right\}
\left\{\order(N^{-1})+\order(N^{-\frac{1}{2}})\right\}
\bigket_{\!\!\ketop;t} = 0
\ed
For batch learning all diffusion matrix elements of
(\ref{eq:fokkerplanck})
vanish in a straightforward manner. For on-line learning
all diffusion terms vanish provided we can prove that the function
$\C[\ldots]$ of (\ref{eq:zerodiffusion}) is zero. This is indeed the
case within the present theory, as will be verified
in the Appendix.  The
Fokker-Planck equation (\ref{eq:fokkerplanck}) now
reduces to the
Liouville equation
$\frac{d}{dt}P_t(\bOmega)=-\sum_\mu\frac{\partial}{\partial
\Omega_\mu}[F_\mu[\bOmega;t]P_t(\bOmega)]$, describing deterministic
evolution for our macroscopic observables:
$\frac{d}{dt}\bOmega=\bF[\bOmega;t]$.  These deterministic equations
we will now work out explicitly.

\subsubsection*{On-Line Learning}

First we deal with the scalar observables
$Q$ and $R$:
\bd
\frac{d}{dt}Q = \lim_{N\to\infty}\left\{
2\eta  \bigbra\bra(\bJ\inn\bxi)
\G[\bJ\inn\bxi,\bB\inn\bxi]\ketset\bigket_{\!\!\ketop;t}
+ \eta^2\bigbra\bra \G^2[\bJ\inn\bxi,\bB\inn\bxi]
\ketset\bigket_{\!\!\ketop;t}
\right\}
\ed
\bd
=2\eta \int\!dx dy~P[x,y]~ x~ \G[x,y]
+ \eta^2\int\!dx dy~P[x,y] ~ \G^2[x,y]
\ed
\bd
\frac{d}{dt}R =
\lim_{N\to \infty}
\eta \bigbra\bra(\bB\inn\bxi)  \G[\bJ\inn\bxi,\bB\inn\bxi]
\ketset\bigket_{\!\!\ketop;t}
=
\eta  \int\!dx dy~P[x,y]~ y ~ \G[x,y]
\ed
These equations are identical
to those found in the $\alpha \rightarrow \infty$ formalism.  The
difference is in the function to be substituted for $P[x,y]$, which
here is the solution of
\bd
\frac{\partial }{\partial t}P[x,y]
= \lim_{N\to\infty}\left\{
- \eta\frac{\partial}{\partial x}
\bigbra \bra\bra \G[\bJ\inn\bxi^\prime,\bB\inn\bxi^\prime](\bxi\inn\bxi^\prime)
\delta[x\minus \bJ\inn\bxi]
\delta[y\minus \bB\inn\bxi]\ketset\ketset
\bigket_{\!\!\ketop;t}
\right.
\ed
\bd
\left.
+ \frac{\eta^2}{2N}\frac{\partial^2}{\partial x^2}\bigbra
\bra\bra\G^2[\bJ\inn\bxi^\prime,\bB\inn\bxi^\prime](\bxi\inn\bxi^\prime)^2
\delta[x\minus \bJ\inn\bxi] \delta[y\minus \bB\inn\bxi]\ketset\ketset
\bigket_{\!\!\ketop;t}
\right.
\ed
\bd
\left.
+
\frac{1}{\alpha}\bigbra
\bra\delta[x\minus \bJ\inn\bxi\minus \eta
\G[\bJ\inn\bxi,\bB\inn\bxi]]\delta[y\minus\bB\inn\bxi]
-\delta[x\minus \bJ\inn\bxi]\delta[y\minus\bB\inn\bxi]
\ketset
\right.
\right.
\ed
\bd
\left.\left.
+\eta\frac{\partial}{\partial x}\left[\G[x,y]
\bra\delta[x\minus \bJ\inn\bxi]\delta[y\minus\bB\inn\bxi]\ketset \right]
-\frac{1}{2}\eta^2\frac{\partial^2}{\partial x^2}\left[\G^2[x,y]
\bra\delta[x\minus \bJ\inn\bxi]\delta[y\minus\bB\inn\bxi]\right]\ketset
\bigket_{\!\!\ketop;t}
\room\right\}
\ed
(where we have inserted (\ref{eq:deltaterm_online}))
\bd
=\frac{1}{\alpha}\left\{\int\!dx^\prime~P[x^\prime,y]\delta[x\minus
x^\prime\minus \eta\G[x^\prime,y]]-P[x,y]\right\}
\ed
\bd
- \eta\frac{\partial}{\partial x}\int\!dx^\prime
dy^\prime~\cA[x,y;x^\prime,y^\prime]\G[x^\prime,y^\prime]
+ \frac{1}{2}\eta^2\int\!dx^\prime
dy^\prime~P[x^\prime,y^\prime]\G^2[x^\prime,y^\prime]
\frac{\partial^2}{\partial x^2}P[x,y]
\ed
\bd
+ \frac{1}{2}\eta^2\frac{\partial^2}{\partial x^2}\int\!dx^\prime
dy^\prime~
\B[x,y;x^\prime,y^\prime]\G^2[x^\prime,y^\prime]
\ed
Anticipating the term $\B[\ldots]$ to be zero (as shown
 in the Appendix) we thus arrive at the following
set of coupled  deterministic
macroscopic equations
\be
\frac{d}{dt}Q=2\eta \int\!dx dy~P[x,y] ~x~ \G[x,y]
+ \eta^2\int\!dx dy~P[x,y] ~ \G^2[x,y]
\label{eq:onlinedQdt}
\ee
\be
\frac{d}{dt}R =
\eta  \int\!dx dy~P[x,y]~ y~  \G[x,y]
\label{eq:onlinedRdt}
\ee
\bd
\frac{d}{dt}P[x,y]=\frac{1}{\alpha}\left\{\int\!dx^\prime~P[x^\prime,y]\delta[x\minus
x^\prime\minus \eta\G[x^\prime,y]]-P[x,y]\right\}
\ed
\be
- \eta\frac{\partial}{\partial x}\int\!dx^\prime
dy^\prime~\cA[x,y;x^\prime,y^\prime]~\G[x^\prime,y^\prime]
+ \frac{1}{2}\eta^2\int\!dx^\prime
dy^\prime~P[x^\prime,y^\prime]\G^2[x^\prime,y^\prime]
\frac{\partial^2}{\partial x^2}P[x,y]
\label{eq:onlinedPdt}
\ee

\subsubsection*{Batch Learning}

For $Q$ and $R$ one again finds simple equations:
\bd
\frac{d}{dt}Q=\lim_{N\to \infty}\left\{
2\eta  \bigbra\bra(\bJ\inn\bxi)  \G[\bJ\inn\bxi,\bB\inn\bxi]\ketset
\bigket_{\!\!\ketop;t}
+ \frac{\eta^2}{N}\bigbra\sum_{i}
\bra\xi_i  \G[\bJ\inn\bxi,\bB\inn\bxi]
\ket_{\ketset}^2
\bigket_{\!\!\ketop;t}
\right\}
\ed
\bd
=
2\eta \int\!dx dy~P[x,y] ~ x~ \G[x;y]
\ed
\bd
\frac{d}{dt}R=\lim_{N\to\infty}
\eta  \bigbra
\bra(\bB\inn\bxi) \G[\bJ\inn\bxi,\bB\inn\bxi]\ketset
\bigket_{\!\!\ketop;t}
=\eta\int\!dx dy~ P[x,y] ~ y~  \G[x;y]
\ed
Finally we calculate the temporal derivative of the joint field
distribution:
\bd
\frac{\partial }{\partial t}P[x,y]
= \lim_{N\to\infty}\left\{\room
- \eta \frac{\partial}{\partial x}
\bigbra
\bra\bra  \G[\bJ\inn\bxi^\prime,\bB\inn\bxi^\prime]
(\bxi\inn\bxi^\prime)  \delta[x\minus\bJ\inn\bxi]
\delta[y\minus\bB\inn\bxi]
\ketset\ketset
\bigket_{\!\!\ketop;t}
\right.
\ed
\bd
\left.
+\frac{\eta^2}{2N}\frac{\partial^2}{\partial x^2}
\bigbra
\bra\bra\bra  \G[\bJ\inn\bxi^\prime,\bB\inn\bxi^\prime]
\G[\bJ\inn\bxi^\pprime,\bB\inn\bxi^\pprime](\bxi\inn\bxi^\prime)(\bxi\inn\bxi^\pprime)
\delta[x\minus\bJ\inn\!\bxi]  \delta[y\minus\bB\inn\bxi]
\ketset\ketset\ketset
\bigket_{\!\!\ketop;t}\room
\right\}
\ed
\bd
=
- \frac{\eta}{\alpha} \frac{\partial}{\partial x}
\left[\G[x,y]P[x,y]\right]
- \eta\frac{\partial}{\partial x}\int\!dx^\prime
dy^\prime~\cA[x,y;x^\prime,y^\prime]\G[x^\prime,y^\prime]
\ed
\bd
+\frac{1}{2}\eta^2\frac{\partial^2}{\partial x^2}
\int\!dx^\prime dy^\prime dx^\pprime dy^\pprime \C[x,y;x^\prime,y^\prime;x^\pprime,y^\pprime]\G[x^\prime,y^\prime]\G[x^\pprime,y^\pprime]
\ed
Anticipating the term $\C[\ldots]$ to be zero (to be
demonstrated in the Appendix) we thus arrive at the following
coupled deterministic macroscopic equations:
\be
\frac{d}{dt}Q=
2\eta \int\!dx dy~P[x,y] ~x ~\G[x;y]
\label{eq:batchdQdt}
\ee
\be
\frac{d}{dt}R
=\eta\int\!dx dy~ P[x,y] ~ y~  \G[x;y]
\label{eq:batchdRdt}
\ee
\be
\frac{d}{dt}P[x,y]=
- \frac{\eta}{\alpha} \frac{\partial}{\partial x}
\left[\G[x,y]P[x,y]\right]
- \eta\frac{\partial}{\partial x}\int\!dx^\prime
dy^\prime~\cA[x,y;x^\prime,y^\prime]~\G[x^\prime,y^\prime]
\label{eq:batchdPdt}
\ee
The difference between the macroscopic equations for batch and on-line
learning is merely the presence (on-line) or absence (batch) of those
terms which are not linear in the learning rate $\eta$ (i.e. of order
$\eta^2$ or higher).

\subsection{Closure of Macroscopic Dynamical Laws}

The complexity of the problem is fully concentrated in
the Green's function $\cA[x,y;x^\prime,y^\prime]$ defined in
(\ref{eq:greensfunction}).
Our macroscopic laws are exact for
$N\to\infty$ but not yet closed due to the appearance of the
microscopic probability density $p_t(\bJ)$ in the sub-shell average of
(\ref{eq:greensfunction}).
We now close our macroscopic laws by making, for $N\to\infty$, the two
key assumptions underlying dynamical replica theories:
\begin{enumerate}
\item
Our macroscopic observables $\{Q,R,P\}$  obey {\em closed} dynamic equations.
\item
These macroscopic equations are self-averaging with respect to the
disorder, i.e. the microscopic realisation of the training set $\set$.
\end{enumerate}
Assumption 1 implies that all microscopic probability variations
within the $\{Q,R,P\}$ sub-shells of the $\bJ$-ensemble are either
absent or irrelevant to the evolution of $\{Q,R,P\}$. We may
consequently make the simplest self-consistent choice for $p_t(\bJ)$
in evaluating the macroscopic laws, i.e. in (\ref{eq:greensfunction}):
microscopic probability equipartitioning in the $\{Q,R,P\}$-subshells
of the ensemble, or
\be
p_t(\bJ)~~~\to ~~~w(\bJ)~\sim~
\delta[Q\minus Q[\bJ]] \
\delta[R\minus R[\bJ]] \ \prod_{xy}\delta[P[x,y]\minus P[x,y;\bJ]]
\label{eq:equipart}
\ee
This new microscopic distribution $w(\bJ)$ depends on time via the
order parameters
$\{Q,R,P\}$.
Note that (\ref{eq:equipart})
leads to exact macroscopic laws if our observables $\{Q,R,P\}$
for $N\to\infty$ indeed obey closed equations, and is true
in equilibrium for detailed balance models in which the  Hamiltonian
can be written in terms of $\{Q,R,P\}$.
It is an approximation if our observables do not obey closed
equations. Assumption 2 allows us to average the macroscopic laws over the
disorder; for mean-field models it is
usually convincingly supported  by numerical simulations, and
can be proven using the path integral formalism (see e.g.
\cite{Horner}).
We write averages over all training
sets $\set\subseteq \{-1,1\}^N$, with $|\set|=p$, as $\bra
\ldots\ket_{ \ \rmsets}$.
Our assumptions result in the closure of the two sets
(\ref{eq:onlinedQdt},\ref{eq:onlinedRdt},\ref{eq:onlinedPdt})
and (\ref{eq:batchdQdt},\ref{eq:batchdRdt},\ref{eq:batchdPdt}),
since now the function $\A[x,y;x^\prime,y^\prime]$ is expressed fully in
terms of $\{Q,R,P\}$:
\bd
\A[x,y;x^\prime,y^\prime]=
\lim_{N\to\infty}
\bigbra\!
\frac{\int\!d\bJ~w(\bJ)~
\bra\!\bra\delta[x\minus\bJ\inn\bxi]~ \delta[y\minus\bB\inn\bxi]~
(\bxi\inn\bxi^\prime)~ \overline{\delta}_{\bxi\bxi^\prime} ~
\delta[x^\prime\minus\bJ\inn\bxi^\prime] ~
\delta[y^\prime\minus\bB\inn\bxi^\prime]
\ket_{\ketset}\ket_{\ketset}}
{\int\!d\bJ~w(\bJ)}\!\bigket_{\!\rmsets}
\ed
The final ingredient of dynamical replica theory is the realization
that averages of fractions can be calculated with the replica identity
\bd
\bigbra \frac{\int\!d\bJ~W[\bJ,\bz]G[\bJ,\bz]}
{\int\!d\bJ~W[\bJ,\bz]}\bigket_{\!\!\bz}=
\lim_{n\to 0} \int\!d\bJ^1\cdots d\bJ^n~\bra G[\bJ^1,\bz]
\prod_{\alpha=1}^n W[\bJ^\alpha,\bz]\ket_{\bz}
\ed
Since each weight component scales as $J^\alpha_i=
\order(N^{-\frac{1}{2}})$ we transform variables in such a way that
our calculations will involve $\order(1)$ objects:
\bd
(\forall i)(\forall\alpha):~~~~~~
J^\alpha_i=(Q/N)^{\frac{1}{2}}\sigma^\alpha_i,~~~~
B_i=N^{-\frac{1}{2}}\tau_i
\ed
This ensures
$\sigma_i^\alpha=\order(1)$, $\tau_i=\order(1)$, and reduces various
constraints to ordinary spherical ones: $(\bsigma^\alpha)^2=\btau^2=N$
for all $\alpha$.  Overall prefactors generated by these
transformations always vanish due to $n\to 0$. We find a new effective
measure: $\prod_{\alpha=1}^n w(\bJ^\alpha) \ d\bJ^\alpha \rightarrow
\prod_{\alpha=1}^n \tilde{w}(\bsigma^\alpha) \ d\bsigma^\alpha$, with
\be
\tilde{w}(\bsigma)~ \sim ~\delta\left[N\minus \bsigma^2\right]
\delta\left[NRQ^{-\frac{1}{2}}\minus\btau\inn \bsigma\right]
\prod_{xy}\delta\left[P[x,y]\minus
P[x,y;(Q/N)^{\frac{1}{2}}\bsigma]\right]
\label{eq:newmeasure}
\ee
We thus arrive at
\bd
\A[x,y;x^\prime,y^\prime]=\Limits
\int\!\prod_{\alpha=1}^n \tilde{w}(\bsigma^\alpha)  d\bsigma^\alpha
~~~~~~~~~~~~~~~~~~~~~~~~~~~~~~~~~~~~~~~~~~~~~~~~~~~~~~~~~~~~~~~~~~~~~~~~~~~~~~~~
\ed
\be
\bigbra ~
\bigbras
(\bxi^\prime\inn\bxi)\overline{\delta}_{\bxi\bxi^\prime}
\delta\left[x\minus \frac{\sqrt{Q}\bsigma^1\inn\bxi}{\sqrt{N}}\right]
\delta\left[y\minus \frac{\btau\inn\bxi}{\sqrt{N}}\right]
\delta\left[x^\prime\minus
\frac{\sqrt{Q}\bsigma^1\inn\bxi^\prime}{\sqrt{N}}
\right]
\delta\left[y^\prime\minus \frac{\btau\inn\bxi^\prime}{\sqrt{N}}\right]
\bigket_{\!\!\!\set}\bigket_{\!\!\!\set}
~\bigket_{\rmsets}
\label{eq:Ainreplicas}
\ee
In the same fashion one can also express $P[x,y]$ in replica form
(which will prove useful for normalization purposes and for
self-consistency tests):
\be
P[x,y]
=\Limits
\int\!\prod_{\alpha=1}^n \tilde{w}(\bsigma^\alpha)d\bsigma^\alpha
\bigbra\bigbra
\delta\left[x\minus \frac{\sqrt{Q}\bsigma^1\inn\bxi}{\sqrt{N}}\right]
\delta\left[y\minus \frac{\btau\inn\bxi}{\sqrt{N}}\right]
\bigket_{\!\!\!\set}
\bigket_{\rmsets}
\label{eq:Pinreplicas}
\ee
Finally we will have to demonstrate that the two functions $\B[\ldots]$ and
$\C[\ldots]$, as defined in
(\ref{eq:zerogreensfunction},\ref{eq:zerodiffusion}), do indeed
vanish self-consistently, as claimed. To achieve this we again express
them in replica form:
\bd
\B[x,y;x^\prime,y^\prime]
=\Limits
\int\!\prod_{\alpha=1}^n \tilde{w}(\bsigma^\alpha)d\bsigma^\alpha
~~~~~~~~~~~~~~~~~~~~~~~~~~~~~~~~~~~~~~~~~~~~~~~~~~~~~~~~~~~~~~~~~~~~~~~~~~~~~~~~~~
\ed
\be
\bigbra~
\bigbras\overline{\delta}_{\bxi\bxi^\prime}
\left[\frac{1}{N}\sum_{i\neq j}
\xi_i\xi_j\xi^\prime_i\xi^\prime_j\right]
\delta\left[x\minus \frac{\sqrt{Q}\bsigma^1\inn\bxi}{\sqrt{N}}\right]
\delta\left[y\minus \frac{\btau\inn\bxi}{\sqrt{N}}\right]
\delta\left[x^\prime\minus \frac{\sqrt{Q}\bsigma^1\inn\bxi^\prime}{\sqrt{N}}\right]
\delta\left[y^\prime\minus \frac{\btau\inn\bxi^\prime}{\sqrt{N}}\right]
\bigket_{\!\!\!\set}\bigket_{\!\!\!\set}
~\bigket_{\rmsets}
\label{eq:Binreplicas}
\ee
and
\bd
\C[x,y;x^\prime,y^\prime;x^\pprime,y^\pprime]
=\Limits
\int\!\prod_{\alpha=1}^n \tilde{w}(\bsigma^\alpha)d\bsigma^\alpha
\bigbra~
\bigbras\!\!\!\bigbra
\overline{\delta}_{\bxi\bxi^\pprime}\overline{\delta}_{\bxi^\prime\bxi^\pprime}
\frac{(\bxi\inn\bxi^\pprime)(\bxi^\prime\inn\bxi^\pprime)}{N}
\delta\left[x\minus \frac{\sqrt{Q}\bsigma^1\inn\bxi}{\sqrt{N}}\right]
\delta\left[y\minus \frac{\btau\inn\bxi}{\sqrt{N}}\right]
\right.\right.\right.\right.
\ed
\be
\left.\left.\left.\left.
\times~
\delta\left[x^\prime\minus \frac{\sqrt{Q}\bsigma^1\inn\bxi^\prime}{\sqrt{N}}\right]
\delta\left[y^\prime\minus\frac{\btau\inn\bxi^\prime}{\sqrt{N}}\right]
\delta\left[x^\pprime\minus \frac{\sqrt{Q}\bsigma^1\inn\bxi^\pprime}{\sqrt{N}}\right]
\delta\left[y^\pprime\minus \frac{\btau\inn\bxi^\pprime}{\sqrt{N}}\right]
\bigket_{\!\!\!\set}\bigket_{\!\!\!\set}\bigket_{\!\!\!\set}
~\bigket_{\rmsets}
\label{eq:Cinreplicas}
\ee
At this stage the physics is over, what remains is to perform the
summations and integrations in
(\ref{eq:Ainreplicas},\ref{eq:Pinreplicas},\ref{eq:Binreplicas},\ref{eq:Cinreplicas})
in the limit $N\to \infty$.  Full details of this exercise are given
in Appendix A, where we show that (\ref{eq:Binreplicas}) and
(\ref{eq:Cinreplicas}) are indeed zero, and where we derive, in
replica symmetric ansatz, an expression for the Green's function
(\ref{eq:Ainreplicas}). It turns out that to calculate this Green's
function $\cA[\ldots]$ one has to solve two coupled saddle-point
equations at each time-step, one scalar equation relating to a
spin-glass order parameter $q$, and one functional saddle-point
equation relating to an effective single-spin measure.


\section{Summary of the Theory and Connection with
$\alpha\!\rightarrow\!\infty$ Formalism}

In this section we summarize the results obtained so far (including the
replica calculation in Appendix A),
and we show that
our general theory has the satisfactory property that it
incorporates the standard formalism developed for
infinite training sets (with Gaussian joint field distributions
$P[x,y]$ at any time) as a special case, recovered in the limit
$\alpha\to\infty$. In addition we provide
a proof of the uniqueness of the RS functional saddle-point equation
and show that it can be found as the fixed-point of an iterative
map.

\subsection{Summary of the Theory}

Our theory can be summarized in the following compact way:

\subsubsection*{Dynamic Equations for Observables}

Our observables are  $Q=\bJ^2$,
$R=\bJ\cdot\bB$,
and the joint distribution of student and teacher fields
$P[x,y]=\bra\delta[x\minus \bJ\inn\bxi]\delta[y\minus
\bB\inn\bxi]\ketset$. For $N\to\infty$ these quantities obey
closed,
deterministic, and self-averaging macroscopic dynamic
equations. One always has $P[x,y]\!=\!P[x|y]P[y]$ with
$P[y]\!=\!(2\pi)^{-{\frac{1}{2}}}e^{-\frac{1}{2}y^2}$.
We define $\bra f[x,y]\ket \!=\!\int\!dx Dy ~P[x|y]f[x,y]$,
with the familiar short-hand
$Dy=(2\pi)^{-\frac{1}{2}}e^{-\frac{1}{2}y^2}dy$, and the
following four averages
(the function $\Phi[x,y]$ will be given below):
\be
U=\bra \Phi[x,y]\G[x,y]\ket~~~~~~
V=\bra x \G[x,y]\ket~~~~~~
W=\bra y \G[x,y]\ket~~~~~~
Z=\bra \G^2[x,y]\ket
\label{eq:fouraverages}
\ee
For on-line learning our macroscopic laws are
\be
\frac{d}{dt}Q=2\eta V +\eta^2 Z
~~~~~~~~~~~~~~~~
\frac{d}{dt}R =
\eta W
\label{eq:summaryonlinedQdtdRdt}
\ee
\bd
\frac{d}{dt}P[x|y]=\frac{1}{\alpha}\int\!dx^\prime P[x^\prime|y]\left[
\delta[x\minus
x^\prime\minus \eta\G[x^\prime\!,y]]\minus \delta[x\minus
x^\prime]\right]
- \eta\frac{\partial}{\partial x} \left\{\room
P[x|y]\left[ U(x\minus Ry) \plus Wy\right]\right\}
\ed
\be
+ \frac{1}{2}\eta^2 Z
\frac{\partial^2}{\partial x^2}P[x|y]
- \eta\left[V\! \minus RW\! \minus (Q\minus R^2)U \right]
\frac{\partial}{\partial x}\left\{\room
P[x|y]\Phi[x,y]
\right\}
\label{eq:summaryonlinedPdt}
\ee
For batch learning one has:
\be
\frac{d}{dt}Q=
2\eta V
~~~~~~~~~~~~~~~
\frac{d}{dt}R
=\eta W
\label{eq:summarybatchdQdtdRdt}
\ee
\bd
\frac{d}{dt}P[x|y]=
- \frac{\eta}{\alpha}\frac{\partial}{\partial x}\left[
P[x|y]\G[x,y]\right]
- \eta\frac{\partial}{\partial x} \left\{\room
P[x|y]\left[ U(x\minus Ry) \plus Wy\right]\right\}
\ed
\be
- \eta\left[V\! \minus RW\! \minus (Q\minus R^2)U \right]
\frac{\partial}{\partial x}\left\{\room
P[x|y]\Phi[x,y]
\right\}
\label{eq:summarybatchdPdt}
\ee
Note that
the batch equations follow from the on-line ones by
retaining only terms which are linear in the
learning rate.
From the solution of the above equations follow, in turn,
the training- and generalization
errors:
\be
E_{\rm t}=\bra \theta[-xy]\ket
~~~~~~~~~~~~~~~~
E_{\rm g}=\frac{1}{\pi}\arccos[R/\sqrt{Q}]
\label{eq:errors}
\ee

\subsubsection*{Saddle-Point Equations and the function $\Phi$}


The function $\Phi[x,y]$ appearing in the above equations (generated
by the Green's function $\cA[\ldots]$) is expressed in terms of
auxiliary order parameters.  These come about in the replica
calculation of Appendix A, where the order parameters are defined
through Dirac $\delta$ functions in their integral representation
(similar to equation~\ref{eq:delta_function}).  The first auxiliary
order parameter is a spin-glass type order parameter $q=\bra\bra
\bJ\ket^2 \ketset/Q$, with $R^2/Q \leq q\leq 1$. The second, defined
similarly for the joint probability $P[x,y]$ is the function
$\chi[x,y]$ (for details see the Appendix). The latter is not
necessarily normalised and in what follows it is useful to consider
the effective measure $M[x,y]$ which is related to $\chi[x,y]$ through
a simple transformation (equation~\ref{eq:chi_M}). The measure
$M[x,y]$ is non-negative and can be always normalized such that
$\int\!dx~M[x,y]=1$ for all $y\in\Re$, as emphasized in our notation
by writing $M[x,y]\to M[x|y]$.  The auxiliary order parameters are
calculated at each time-step by solving the following two coupled
saddle-point equations:
\be
\bra (x\minus Ry)^2\ket
+(qQ\minus R^2)(1\minus \frac{1}{\alpha})
=\left[\frac{1\plus q\minus 2R^2/Q}{1\minus q}\right]
\int\!DyDz~\left[\bra x^2\ket_\star-\bra x\ket_\star^2\right]
\label{eq:summarysaddle_q}
\ee
\be
P[X|y]
=\int\! Dz~ \bra \delta[X\minus x]\ket_\star
\label{eq:summarysaddle_M}
\ee
in which
\be
\bra f[x,y,z]\ket_\star =
 \frac{\int\!dx~M[x|y]e^{Bxz}f[x,y,z]}
{\int\!dx~M[x|y]e^{Bxz}}
~~~~~~~~~~~~~
B=\frac{\sqrt{qQ\minus R^2}}{Q(1\minus q)}
\label{eq:notation}
\ee
After $q$ and $M[x|y]$ have been determined,
the key function $\Phi[x,y]$ in
(\ref{eq:fouraverages},\ref{eq:summaryonlinedPdt},\ref{eq:summarybatchdPdt})
is calculated as
\be
\Phi[X,y]=
\left\{\room Q(1\minus q)P[X|y]\right\}^{-1}\!\int\!Dz
\bra X\minus x \ket_\star
\bra \delta[X\minus x]\ket_\star
\label{eq:defPhi}
\ee
or, equivalently:
\be
\Phi[X,y]=
\left\{\room \sqrt{qQ\minus R^2} P[X|y]\right\}^{-1}\!\int\!Dz
~z~
\bra \delta[X\minus x]\ket_\star
\label{eq:defPhi2}
\ee
Finding a saddle-point problem for an order parameter function, rather
than a finite number of scalar order parameters, introduces the
possibility of a proliferation of saddle-points. In the next section
 we will show that
this does not happen:  the solution of the
functional saddle-point problem is unique, and can even by found iteratively
by executing a specific non-linear mapping.

\subsection{Uniqueness and Iterative Calculation
of the Functional Saddle-Point}

The uniqueness proof is more easily
set up in terms of the original order parameter function $\chi[x,y]$,
rather than the new (normalised) measure $M[x|y]$ (see the Appendix).
For a given state $\{Q,R,P\}$ and a given value for $q\in[R^2/Q,1]$ we
have to find the functional saddle-points of the functional
$\Psi[\chi]$, defined as:
\be
\Psi[\chi]=\alpha \int\!DyDz~\log
\int\!dx~e^{-\frac{x^2}{2Q(1-q)}+x[Ay+Bz]+\alpha^{-1}\chi[x,y]}
-\int\!Dy dx P[x|y]\chi[x,y]
\label{eq:functional}
\ee
Our proof will carry the existence of the various
integrals as an implicit condition for validity.
To reduce
notational ballast we define
\bd
w(x,y,z)=
\frac{e^{-\frac{x^2}{2Q(1-q)}+x[Ay+Bz]+\alpha^{-1}\chi[x,y]}}
{\int\!dx^\prime~e^{-\frac{x^{\prime 2}}{2Q(1-q)}+x^\prime[Ay+Bz]+\alpha^{-1}\chi[x^\prime,y]}},~~~~~~~~~~~~\bra f[x,y,z]\ket_\star=\int\!dx~w(x,y,z)f[x,y,z]
\ed
Note: $w(x,y,z)=M[x|y]e^{Bxz}/\int\!dx^\prime~M[x^\prime|y]e^{Bx^\prime z}$.
The function $w(u,v,z)$ obeys
\bd
\frac{\delta
w(u,v,z)}{\delta\chi[u^\prime,v^\prime]}=
\alpha^{-1}\delta[v\minus
v^\prime]
\left[
\delta[u\minus u^\prime]w(u,v,z)
- w(u,v,z)w(u^\prime,v,z)\right]
\ed
The functional saddle-point equation is obtained by requiring the first
functional derivative of $\Psi[\chi]$ with respect to $\chi[u,v]$ to
be zero for all $u,v\in\Re$, where
\be
\left.
\frac{\delta\Psi}{\delta\chi[u,v]}
\right|_{\chi}=\frac{e^{-\frac{1}{2}v^2}}{\sqrt{2\pi}}
\left\{
 \int\!Dz~w(u,v,z)
-P[u|v]
\right\}
\label{eq:firstderivative}
\ee
Clearly, if
the function $\chi[x,y]$ is a saddle-point, then also the function
$\chi[x,y]+\rho(y)$ for any
$\rho(y)$. This degree of freedom is
irrelevant because such terms $\rho(y)$ will drop out of the measure
$\bra \ldots\ket_\star$. Furthermore, one immediately verifies that
transformations of the form $\chi[x,y]\to\chi[x,y]+\rho(y)$ leave
the functional $\Psi[\ldots]$ (\ref{eq:functional}) invariant.
Next we calculate the Hessian (or curvature) operator
$H[u,v;u^\prime,v^\prime;\chi]$, using (\ref{eq:firstderivative}):
\bd
H[u,v;u^\prime,v^\prime;\chi]
=\left.\frac{\delta^2\Psi}
{\delta\chi[u,v]\delta\chi[u^\prime,v^\prime]}\right|_{\chi}
=
\frac{e^{-\frac{1}{2}v^2}}{\sqrt{2\pi}}
 \int\!Dz ~\frac{\delta w(u,v,z)}{\delta\chi[u^\prime,v^\prime]}
~~~~~~~~~~
\ed
\be
~~~~~~~~~~
=
\delta[v\minus
v^\prime]
\frac{e^{-\frac{1}{2}v^2}}{\alpha\sqrt{2\pi}}
 \int\!Dz
\left[
\delta[u\minus u^\prime]w(u,v,z)
- w(u,v,z)w(u^\prime,v,z)\right]
\label{eq:secondderivative}
\ee
$H[u,v;u^\prime,v^\prime;\chi]$ is
non-negative definite for each $\chi$, and thus
the functional $\Psi$ is convex, since for any function $\phi[u,v]$ for
which the relevant integrals exist we find
\bd
\int\!dudvdu^\prime dv^\prime~\phi[u,v]
H[u,v;u^\prime,v^\prime;\chi] \phi[u^\prime,v^\prime]
=
\frac{1}{\alpha}\int\!DvDz~\left[\bra \phi^2[u,v]\ket_\star -\bra \phi[u,v]\ket_\star^2
\right]\geq 0
\ed
The kernel of $H[u,v;u^\prime,v^\prime;\chi]$, for a given `point'
$\chi$ in $\chi$-space, is determined by requiring {\rm equality} in the above
inequality, i.e.
\bd
{\rm for~each}~~v,z\in\Re:~~~~~~~~
\bra \left[\phi[u,v]-\bra \phi[u,v]\ket_\star\right]^2\ket_\star= 0
~~~~{\rm so}~~~~
\frac{\partial}{\partial u}\phi[u,v]=0
\ed
For each $\chi$ the kernel of the second functional derivative
$H[x,y;x^\prime,y^\prime;\chi]$ thus consists of the set of all
(integrable) functions $\phi[x,y]$
which depend on $y$ only.
\vsp

We now find that, if $\chi_0[x,y]$ and
$\chi_1[x,y]$ are both functional saddle-points of $\Psi[\chi]$, then
$\chi_1[x,y]-\chi_0[x,y]=\rho(y)$ for some function $\rho(y)$.  In
other words: apart from the aforementioned irrelevant degree of
freedom, the solution of the functional saddle-point equation
(\ref{eq:summarysaddle_M})
is unique.
To show this, consider two functions $\chi_0[x,y]$ and $\chi_1[x,y]$ which are both
functional saddle-points of $\Psi$, i.e. corresponding to solutions of
(\ref{eq:summarysaddle_M}).
Define a path $\{\chi_t\}$ through $\chi$-space,
connecting these two functions:
\bd
\chi_t[x,y]=\chi_0[x,y]+
t\left\{\chi_1[x,y]-\chi_0[x,y]\right\}
,~~~~~~t\in[0,1]
\ed
Integration along this path will bring us from $\chi_0$ to $\chi_1$.
Thus for any functional $L[\chi]$ one has
\bd
L[\chi_1]-L[\chi_0]=\int_{\chi_0}^{\chi_1}\!dL[\chi]=
\int\!dudv \int_{\chi_0}^{\chi_1}\!d\chi[u,v] \frac{\delta
L}{\delta\chi[u,v]}
\ed
\bd
=\int\!dudv \left[\chi_1[u,v]\minus \chi_0[u,v]\right]
\int_{0}^{1}\!dt \left.\frac{\delta L}{\delta\chi[u,v]} \right|_{\chi_t}
\ed
For the functional $L[\chi]$ we now choose a functional first
derivative of $\Psi[\chi]$,
i.e. $L[\chi]=\delta\Psi/\delta\chi[x,y]$ for some $x,y\in\Re$.
Since both $\chi_0$ and $\chi_1$ are saddle-points one finds
$L[\chi_0]=L[\chi_1]=0$. Thus
\bd
\int\!dudv \left[\chi_1[u,v]\minus \chi_0[u,v]\right]
\int_{0}^{1}\!dt \left.\frac{\delta^2
\Psi}{\delta\chi[u,v]\delta\chi[x,y]}
\right|_{\chi_t}=0
\ed
Multiply both sides by $\chi_1[x,y]\minus \chi_0[x,y]$ and
integrate the result over $x,y\in\Re$:
\bd
\int_{0}^{1}\!dt\int\!dudvdxdy \left[\chi_1[u,v]\minus \chi_0[u,v]\right]
H[u,v;x,y;\chi_t]
\left[\chi_1[x,y]\minus \chi_0[x,y]\right]
=0
\ed
One concludes (since the Hessian is a symmetric non-negative operator):
\bd
{\rm for~all}~~t\in[0,1],~u,v\in\Re:~~~~\int\!dxdy~
H[u,v;x,y;\chi_t]\left[\chi_1[x,y]\minus \chi_0[x,y]\right]=0
\ed
The function $\chi_1[x,y]\minus\chi_0[x,y]$ is in the kernel of
$H|_{\chi_t}$ for any $t\in[0,1]$. The kernel of $H$
was already determined to be the set of all integrable
functions which depend on $y$ only, whatever the point $\chi$ where
one chooses to evaluate $H$.
Hence $\chi_1[x,y]\minus
\chi_0[x,y]=\rho(y)$ for some function $\rho(y)$.
Finally, the remaining freedom in choosing a function $\rho$ is
eliminated by our normalisation $\int\!dx~M[x|y]=1$ (for each
$y$), so that the solution $M[x|y]$ is indeed truly unique.
\vsp

Next we will show how for any given value of the scalar order
parameter $q$ and the observables $\{Q,R,P\}$ (and thus of $B$), for
which the relevant integrals exist, the unique solution $M[x|y]$ of the
functional saddle-point equation (\ref{eq:summarysaddle_M}) can be
constructed as the stable fixed-point of the following functional map:
\vspace*{-3mm}
\be
{\rm for~each}~y\in\Re:~~~~~~ M_{\ell+1}[x|y]= \frac{P[x|y]\left\{
\int\!Dz\left[\int\!dx^\prime~e^{Bz(x^\prime-x)}M_{\ell}[x^\prime|y]\right]^{-1}
\right\}^{-1}} {\int\!du~P[u|y]\left\{
\int\!Dz\left[\int\!dx^\prime~e^{Bz(x^\prime-u)}M_{\ell}[x^\prime|y]\right]^{-1}
\right\}^{-1}}
\label{eq:iteration}
\ee
Clearly all fixed-points of this map correspond to normalised
solutions $M[x|y]$ of a functional saddle-point equation
(\ref{eq:summarysaddle_M}), of which there can be only one.
Thus we only need to verify the convergence of
(\ref{eq:iteration}), which can be done most efficiently using an
appropriate Lyapunov functional. Note that the functional
(\ref{eq:functional}) can be written as
\bd
\Psi[M|y]=\alpha \int\!Dy~\tilde{\Psi}[M|y]~+~ {\rm terms
~independent~of}~M[\ldots]
\ed
with
\be
\tilde{\Psi}[M|y]=\int\!Dz~\log
\int\!dx~M[x|y]e^{Bxz}
-\int\!dx P[x|y]\log M[x|y]
\label{eq:Lyapunov}
\ee
For any given $y\in\Re$ we will show (\ref{eq:Lyapunov}) to be a
Lyapunov functional for the mapping (\ref{eq:iteration}),
i.e. $\tilde{\Psi}[M|y]$ is bounded from below and monotonically
increasing during the iteration of (\ref{eq:iteration})
with stationarity obtained only when $M[\ldots]$ is the (unique)
fixed-point of (\ref{eq:iteration}).
First we prove that a lower bound for $\tilde{\Psi}$ is given by
the entropy of the conditional distribution $P[x|y]$:
\be
{\rm for ~any}~M[\ldots]~{\rm and ~any}~y\in\Re:~~~~~~~~
\tilde{\Psi}[M|y]~\geq~ -\int\!dx~P[x|y]\log P[x|y]
\label{eq:boundedness}
\ee
The proof is elementary (using Jenssen's inequality):
\bd
\tilde{\Psi}[M|y]=\int\!Dz~\log\left\{\int\!dx~P[x|y] e^{Bxz+\log
M[x|y]-\log P[x|y]}\right\}-\int\!dx~P[x|y]\log M[x|y]
\ed
\bd
\geq
\int\!Dz\int\!dx~P[x|y]\left\{Bxz+\log
M[x|y]-\log P[x|y]\right\}-\int\!dx~P[x|y]\log M[x|y]
\ed
\bd
=-\int\!dx~P[x|y]\log P[x|y]
\ed
Secondly we show that (\ref{eq:Lyapunov}) indeed decreases
monotonically under (\ref{eq:iteration}) until the fixed-point
of (\ref{eq:iteration}) is reached. To do so we introduce
the short-hand notations
 $\lambda_\ell(x,y,z)=Bxz+\log M_\ell[x|y]-\log P[x|y]$,
$\bra f[x]\ket=\int\!dx~P[x|y]f[x]$, and
\bd
v_\ell(x,y)=
\left\{
\int\!Dz~e^{\lambda_\ell(x,y,z)}\bra e^{\lambda_\ell(x^\prime,y,z)}\ket^{-1}
\right\}^{-1}
\ed
 The iterative map can now
be written as
\bd
M_{\ell+1}[x|y]=
\frac{M_\ell[x|y] v_\ell(x,y)}
{\int\!du~M_\ell[u|y] v_\ell(u,y)}
\ed
This gives for the change in $\tilde{\Psi}[\ldots]$ during one
iteration of the mapping, again with Jenssen's inequality:
\bd
\tilde{\Psi}[M_{\ell+1}|y]-\tilde{\Psi}[M_\ell|y]=
\int\!Dz~\log\left\{\frac{\int\!dx~M_{\ell+1}[x|y]e^{Bxz}}
{\int\!dx~M_\ell[x|y]e^{Bxz}}\right\}
-\int\!dx P[x|y]\log \left\{\frac{M_{\ell+1}[x|y]}{M_\ell[x|y]}
\right\}
\ed
\bd
=\int\!Dz\left\{\log \frac{
\bra e^{\lambda_\ell(x,y,z)}v_\ell(x,y)\ket}
{\bra e^{\lambda_\ell(x,y,z)}\ket}
\right\}
-\bra \log v_\ell(x,y)\ket
\ed
\bd
\leq~\log \left\{
\bra v_\ell(x,y)\int\!Dz\left[ e^{\lambda_\ell(x,y,z)}
\bra e^{\lambda_\ell(x^\prime,y,z)}\ket^{-1}
\right]\ket
\right\}
-\bra \log v_\ell(x,y)\ket
\ed
\bd
=
-\bra \log v_\ell(x,y)\ket
=
\bra \log
\int\!Dz~e^{\lambda_\ell(x,y,z)}\bra
e^{\lambda_\ell(x^\prime,y,z)}\ket^{-1}\ket
\ed
\bd
\leq~
 \log
\int\!Dz~\bra e^{\lambda_\ell(x,y,z)}\ket
\bra e^{\lambda_\ell(x^\prime,y,z)}\ket^{-1}
=0
\ed
Finally we round off our argument by inspecting the implications
of having strict equality in the above inequality. Equality can
only occur if at both instances where Jenssen's inequality was
used in replacements of the form $\bra \log (X)\ket\leq \log\bra X\ket$
the relevant stochastic variable $X$ was a constant. In our
problem this gives the two conditions
\bd
\frac{\partial}{\partial z}
\frac{
\bra e^{\lambda_\ell(x,y,z)}v_\ell(x,y)\ket}
{\bra e^{\lambda_\ell(x,y,z)}\ket}=0,
~~~~~~~~~~
\frac{\partial}{\partial x}v_\ell(x,y)=0
\ed
If the second condition is met, the first immediately follows.
Working out the second condition gives, in combination with
the property that $P[x|y]$ is normalised:
\bd
\int\!Dz\frac{M_\ell[x|y]e^{Bxz}}
{\int\!dx^\prime~M_\ell[x^\prime|y]e^{Bx^\prime z}}=P[x|y]
\ed
Thus we have confirmed that $\tilde{\Psi}[M_{\ell+1}|y]=
\tilde{\Psi}[M_\ell|y]$ if and only if $M_{\ell}[\ldots]$ is the
(unique) fixed-point of (\ref{eq:iteration}).
\vsp

As a consequence of the above we may now write the normalised
solution of our functional
saddle-point equation (\ref{eq:summarysaddle_M}) in terms of
repeated execution of the mapping (\ref{eq:iteration}) following an
an in principle arbitrary initialisation:
\bd
{\rm for~all}~y\in\Re:~~~~~~~~M[x|y]=\lim_{\ell\to\infty}M_\ell[x|y],~~~~~~~~
M_0[x|y]=P[x|y]
\ed
This property simplifies the numerical solution of our equations
drastically.

\subsection{Fourier Representation and Conditionally-Gaussian  Solutions}

There are two potential  advantages of rewriting our equations in
Fourier representation. Firstly, after a Fourier transform
the functional saddle-point equation
(\ref{eq:summarysaddle_M})
will acquire a much
simpler form. Secondly, in those cases where we expect $P[x|y]$
to be of a Gaussian shape in $x$ this would simplify solution of the
diffusion equations
(\ref{eq:summaryonlinedPdt},\ref{eq:summarybatchdPdt}).
Clearly, $P[x,y]$ being Gaussian in $(x,y)$ is not
equivalent to $P[x|y]$ being Gaussian in $x$ only. The former
requires
\bd
\frac{\partial^2}{\partial y^2}\int\!dx~xP[x|y]=
\frac{\partial}{\partial y}\left\{\int\!dx~x^2 P[x|y]
-\left[\int\!dx~x P[x|y]\right]^2\right\}=0,
\ed
which  only will turn out to happen
for $\alpha\to\infty$.
A Gaussian $P[x|y]$ with moments
which depend  on $y$ in a non-trivial way, on the other hand,
is found to occur also
for $\alpha<\infty$, provided we consider simple learning rules
and small $\eta$ (see \cite{CoolenSaad2}). To avoid ambiguity we
will call solutions of the latter type `conditionally-Gaussian'.

We introduce the Fourier transforms
\be
\hat{P}[k|y]=\int\!dx~e^{-ikx}P[x|y]
~~~~~~~~~~~~~~
\hat{M}[k|y]=\int\!dx~e^{-ikx}M[x|y]
\label{eq:Fouriertransforms}
\ee
The transformed functional saddle-point equation thereby acquires
a very simple form
\be
\hat{P}[k|y]=\int\!Dz ~\frac{\hat{M}[k\plus iBz|y]}{\hat{M}[iBz|y]}
\label{eq:FourierSPE}
\ee
Note that, in contrast to the original equation
(\ref{eq:summarysaddle_M}), the transformed equation
(\ref{eq:FourierSPE}) need not have a unique solution
(it could allow for solutions corresponding to non-integrable
functions in the original problem).
 Consider, for
instance, the transformation
\bd
\hat{M}[k|y]~\to~\dot{M}[k|y]=\frac{e^{\frac{1}{2}k^2/B^2}}{\hat{M}[-k|y]}
\ed
with the property (verified by a simple transformation of variables):
\bd
\int\!Dz~\frac{\dot{M}[k\plus iBz|y]}{\dot{M}[iBz|y]}=
\int_{ik/B-\infty}^{ik/B+\infty}\!Dz~\frac{\hat{M}[k\plus
iBz|y]}{\hat{M}[iBz|y]}
\ed
If $\hat{M}[k]$, which by definition cannot have poles,
is sufficiently well behaved, a simple deformation of the integration
path (via contour integration)  leads to the statement that if
$\hat{M}[k|y]$ is a solution of (\ref{eq:FourierSPE}), then so is
$\dot{M}[k|y]$.

Transformation of the dynamical on-line equation
(\ref{eq:summaryonlinedPdt}) for $P[x|y]$
(from the which the batch equation (\ref{eq:summarybatchdPdt}) can be
obtained
by expansion in $\eta$)
gives:
\bd
\frac{d}{dt}\log\hat{P}[k|y]~=~
\frac{1}{\alpha}
\left\{\int\!dk^\prime~\frac{\hat{P}[k^\prime|y]}{\hat{P}[k|y]}
\int\!\frac{dx^\prime}{2\pi}e^{ix^\prime(k^\prime-k)
-i\eta k\G[x^\prime,y]}
-1\right\}
- i\eta k (W\!\minus UR)y
\ed
\be
+~\eta k U\frac{\partial}{\partial k}\log\hat{P}[k|y]
- \frac{1}{2}\eta^2 k^2 Z
~-~ i\eta k
\left[\frac{V\!\minus RW\!\minus (Q\minus R^2)U}{\sqrt{qQ\minus R^2}\hat{P}[k|y]}\right]
\int\!Dz ~z~\frac{\hat{M}[k+iBz|y]}{\hat{M}[iBz]}
\label{eq:FourierdPdt}
\ee
\vsp

We now determine the conditions for equation (\ref{eq:FourierdPdt}) to
have conditionally-Gaussian solutions. If $P[x|y]$ is Gaussian in $x$ we can
solve the functional saddle-point equation (\ref{eq:summarysaddle_M})
(whose solution is unique), and find the resulting pair of measures
\be
P[x|y]=\frac{e^{-\frac{1}{2}[x-\overline{x}(y)]^2/\Delta^2(y)}}{\Delta(y)\sqrt{2\pi}}
~~~~~~~~~~~~~~~~~~
M[x|y]=\frac{e^{-\frac{1}{2}[x-\overline{x}(y)]^2/\sigma^2(y)}}{\sigma(y)\sqrt{2\pi}}
\label{eq:locallygaussian}
\ee
\be
\Delta^2(y)=\sigma^2(y)+B^2\sigma^4(y)
\label{eq:twovariances}
\ee
with their Fourier transforms
$\hat{P}[k|y]=\exp\left[\minus
ik\overline{x}(y)\minus \frac{1}{2}k^2\Delta^2(y)\right]$
and
$\hat{M}[k|y]=\exp\left[\minus
ik\overline{x}(y)\minus \frac{1}{2}k^2\sigma^2(y)\right]$.
Insertion of these expressions as an Ansatz into
(\ref{eq:FourierdPdt}), using the identity
\bd
\int\!Dz ~z~\frac{\hat{M}[k+iBz|y]}{\hat{M}[iBz]}=ikB\sigma^2(y)\hat{P}[k|y]
\ed
and performing some simple manipulations,
gives the following simplified equation:
\bd
-ik\frac{d}{dt}\overline{x}(y)-\frac{1}{2}k^2\frac{d}{dt}\Delta^2(y)
~=~
\frac{1}{\alpha}\left\{
\int\!\frac{du}{\sqrt{2\pi}}
e^{-\frac{1}{2}[u-ik\Delta(y)]^2-ik\eta \G[\overline{x}(y)+u\Delta(y),y]}
-1\right\}
- i\eta k\left\{ Wy +U[\overline{x}(y)\minus Ry]\right\}
\ed
\be
- \frac{1}{2}k^2 \left\{\eta^2  Z +2 \eta U\Delta^2(y)
~+ 2\eta\sigma^2(y)
\left[\frac{V\!\minus RW\!\minus (Q\minus R^2)U}
{Q(1\minus q)}\right]\right\}
\label{eq:FourierGaussiandPdt}
\ee
From this it follows that conditionally-Gaussian solutions can occur in two
situations only:
\be
\alpha\rightarrow\infty~~~~~~~~
{\rm or} ~~~~~~~~
\frac{\partial^3}{\partial k^3}
\int\!\frac{du}{\sqrt{2\pi}}~
e^{-\frac{1}{2}[u-ik\Delta(y)]^2-ik\eta \G[\overline{x}(y)+u\Delta(y),y]}
 =0
\label{eq:Gaussianconditions}
\ee
The first case corresponds to the familiar theory of infinite training
sets (see next section). The second case occurs for sufficiently
simple learning rules $\G[x,y]$, in combination either with batch
execution (so that of (\ref{eq:Gaussianconditions}) we retain only the
term linear in $\eta$) or with on-line execution for small $\eta$
(retaining in (\ref{eq:Gaussianconditions}) only $\eta$ and $\eta^2$
terms). The latter cases will be dealt with in more detail in \cite{CoolenSaad2}.

\subsection{Link with the Formalism for Complete Training Sets}

The very least we should require of our theory is that it reduces to
the simple $(Q,R)$ formalism of complete training sets
\cite{KinouchiCaticha,BiehlSchwarze92} in the limit
$\alpha\to\infty$. Here we will show that this indeed happens. In the
previous section we have seen that for
$\alpha\to\infty$ our driven diffusion equation for the conditional
distribution
$P[x|y]$ has conditionally-Gaussian solutions, with $\int\!dx
~xP[x|y]=\overline{x}(y)$ and $\int\!dx~[x-\overline{x}(y)]^2
P[x|y]=\Delta^2(y)$. Note that for such solutions we can calculate
objects such as $\bra x\ket_\star$ and the
function $\Phi[x,y]$ (\ref{eq:defPhi}) directly, giving
\bd
\bra x\ket_\star = \overline{x}(y)+zB\sigma^2(y)
~~~~~~~~~~~~~~~~~~
\Phi[x,y]=
\frac{x-\overline{x}(y)}{Q(1\minus q)[1\plus B^2\sigma^2(y)]}
\ed
with
$\Delta^2(y)=\sigma^2(y)\plus B^2\sigma^4(y)$
and $B=\sqrt{qQ\minus R^2}/Q(1\minus q)$.
The remaining dynamical equations to be solved
are those for $Q$ and $R$, in combination with dynamical equations for
the $y$-dependent cumulants $\overline{x}(y)$ and $\Delta^2(y)$.
These equations reduce to:
\be
\frac{d}{dt}Q=
\left\{\begin{array}{lc} 2\eta  \bra x\G[x,y]\ket
+\eta^2 \bra \G^2[x,y]\ket & ({\rm on\!-\! line})\\[2mm]
2\eta  \bra x\G[x,y]\ket & ({\rm batch})
\end{array}
~~~~~~~~~~~~~~~~
\frac{d}{dt}R =
\eta \bra y\G[x,y]\ket
\right.
\label{eq:QdtdRdtcomplete}
\ee
\be
\frac{1}{\eta}\frac{d}{dt}\left[\room \overline{x}(y)\minus Ry\right]=
[\overline{x}(y)\minus Ry]
\bra \Phi[x^\prime,y^\prime]\G[x^\prime,y^\prime]\ket
\label{eq:dxdtcomplete}
\ee
\be
\frac{1}{2\eta}\frac{d}{dt}\left[\room \Delta^2(y)\minus Q\plus R^2\right]=
\bra (x^\prime\minus Ry^\prime)\G[x^\prime,y^\prime]\ket
\left[\frac{\sigma^2(y)}{Q(1\minus q)}\minus 1\right]
+\bra \Phi[x^\prime,y^\prime]\G[x^\prime,y^\prime]\ket
\left[\Delta^2(y)\minus \frac{Q\minus R^2}{Q(1\minus q)\sigma^2(y)}\right]
\label{eq:ddeltadtcomplete}
\ee
with one remaining saddle-point equation to determine $q$,
obtained upon working out (\ref{eq:summarysaddle_q}) for
conditionally-Gaussian 
solutions:
\be
\int\!Dy \left\{[\overline{x}(y)\minus Ry]^2+ \Delta^2(y)\right\}
+qQ\minus  R^2
=\left[2\frac{qQ\minus R^2}{Q(1\minus q)}\plus 1\right]
\int\!Dy~\sigma^2(y)
\label{eq:saddlecomplete}
\ee
We now make the Ansatz that $\overline{x}(y)=Ry$ and
$\Delta^2(y)=Q\minus R^2$, i.e.
\be
P[x|y]=\frac{e^{-\frac{1}{2}[x-Ry]^2/(Q-R^2)}}{\sqrt{2\pi(Q-R^2)}},
\label{eq:ansatz}
\ee
Insertion into the dynamical equations shows that
(\ref{eq:dxdtcomplete}) is now immediately satisfied, that
(\ref{eq:ddeltadtcomplete}) reduces to $\sigma^2(y)=Q(1\minus q)$, and
that as a result the saddle-point equation (\ref{eq:saddlecomplete}) is
automatically satisfied. Since (\ref{eq:ansatz}) is parametrized by
$Q$ and $R$ only, this leaves us with the closed equations
\be
\frac{d}{dt}Q=
\left\{\begin{array}{ccc} 2\eta  \bra x\G[x,y]\ket
+\eta^2 \bra \G^2[x,y]\ket && ({\rm on\!-\! line})\\[2mm]
2\eta  \bra x\G[x,y]\ket && ({\rm batch})
\end{array}
~~~~~~~~~~~~~~~~
\frac{d}{dt}R=\eta \bra y\G[x,y]\ket
\right.
\label{eq:standardtheory}
\ee
These are the equations found in e.g.
\cite{KinouchiCaticha,BiehlSchwarze92}.
From our general theory for restricted training sets we thus
indeed recover
in the limit
$\alpha\to\infty$ the standard formalism
(\ref{eq:ansatz},\ref{eq:standardtheory}) describing learning with complete training
sets,
as claimed.

\section{Discussion}

In this paper we have shown how the formalism of dynamical replica
theory (see e.g. \cite{Coolenetal}) can be successfully employed to
construct a general theory which enables one to predict the evolution
of the relevant macroscopic performance measures for supervised
(on-line and batch) learning in layered neural networks, with randomly
chosen but restricted training sets, i.e. for finite $\alpha=p/N$
where weight updates are carried out by sampling with repetition. In
this case the student nodes local fields are no longer described by
(multivariate) Gaussian distributions and the traditional and
familiar statistical mechanical formalism consequently breaks down.
For simplicity and transparency we have restricted ourselves to
single-layer systems and realizable tasks.

In our approach the joint field distribution $P[x,y]$ for the student
and teacher local fields  is itself taken to
be a dynamical order parameter, in addition to the conventional
observables $Q$ and $R$ representing overlaps between the
student-student and student-teacher vectors respectively. The new
order parameter set $\{Q,R,P\}$, in turn, enables one to monitor the
generalization error $E_{\rm g}$ as well as the training error $E_{\rm
t}$.  This then results, following the prescriptions of dynamical
replica theory\footnote{The reason why the replica formalism is
inevitable (unless we are willing to pay the price of having
observables with two time arguments, and turn to path integrals) is
the necessity, for finite $\alpha$, to average the macroscopic
equations over all possible realizations of the training set.}, in a
diffusion equation for $P[x,y]$, which we have evaluated by making the
replica-symmetric ansatz in the saddle-point equations. This diffusion
equation is generally found to have Gaussian solutions only for
$\alpha\to\infty$; in the latter case we indeed recover correctly from
our theory the more familiar formalism of infinite training sets (in
the $N\!\to\!\infty$ limit), providing closed equations for $Q$ and
$R$ only. For finite $\alpha$ our theory is by construction exact if
for $N\!\to\!\infty$ the dynamical order parameters $\{Q,R,P\}$ obey
closed deterministic equations, which are self-averaging (i.e.
independent of the microscopic realization of the training set).
If this is not the case, our theory can be interpreted as employing
a maximum entropy approximation.
In a sequel paper \cite{CoolenSaad2} we will work out our equations explicitly for
various choices of learning rules, and compare our theoretical
predictions both to exact solutions, derived for special cases directly
from the microscopic equations \footnote{Such exact results can only
be obtained for Hebbian-type rules, where the dependence of the
updates $\Delta\bJ(t)$ on the weights $\bJ(t)$ is trivial or even
absent (a decay term at most), whereas our present theory generates
macroscopic equations for arbitrary learning rules.}, and with
numerical simulations. We will also construct a number of simple
but effective approximations to our full equations. As it will turn out, our theory describes the
various learning processes examined highly accurately.

The present study represents only a first step in understanding
on-line learning with restricted training sets. It opens up many
extensions, applications and generalizations that can be carried out
(some of which are already under way). Firstly, our theory would
simplify significantly if one could find a more explicit solution of
the functional saddle-point equation (\ref{eq:RSsaddle1}), enabling us
to express the function $\Phi[x,y]$ directly in terms of our order
parameters.  The benefits of such a solution will become even greater
as we apply our theory to more sophisticated learning rules, such as
to perceptron or AdaTron learning, or to learning in multi-layer
networks (which run the risk of requiring a serious amount of CPU
time). Secondly, this theory opens up new possibilities for
considering unrealizable learning scenarios, either due to structural
limitations or due to noise, which require some sort of
regularization. The examination of regularization techniques in such
scenarios, which is of great practical significance, was out of reach
so far as they come into effect only where the error-surface is
fixed by having a fixed example set. Thirdly, at a more fundamental
level one could explore the effects of (dynamic) replica symmetry
breaking (by calculating the AT-surface, signalling instability of the
replica symmetric solution with respect to replicon fluctuations), or
one could improve the built-in accuracy of our theory by adding new
observables to the present set (such as the Green's function
$\cA[x,y;x^\prime,y^\prime]$ itself). Finally it would be interesting
to see the connection between the present formalism and a suitable
adaptation of the generating functional methods,
as applied in \cite{Horner} to networks with binary weights, to
the learning processes studied in this paper.

\subsection*{Acknowledgements}

It is our pleasure to thank Yuan-sheng Xiong and Charles Mace for
valuable discussions. DS acknowledges support by EPSRC Grant GR/L52093
and the British Council grant: British-German Academic Research
Collaboration Programme project 1037.


\clearpage\appendix
\section{Replica Calculation of the Green's Function}

The main objective of
this Appendix is to calculate the Green's function $\A[\ldots]$,
with which we obtain our macroscopic dynamic equations in explicit
form.
We first carry out the disorder averages, leading to an effective
single-spin problem. The integrations are done by steepest descent,
giving a saddle-point problem for replicated order parameters at each
time step. In the saddle point equations we then make the
replica symmetry (RS) ansatz, so that the limit $n\to 0$ can be taken.
In addition we show that the two functions $\B[\ldots]$ and
$\C[\ldots]$ do indeed vanish, as claimed.

\subsection{Disorder Averaging}

The fundamental quantities $\A[x,y;x^\prime,y^\prime]$,
$\B[x,y;x^\prime,y^\prime]$, $\C[x,y;x^\prime,y^\prime;x^\pprime,y^\pprime]$,
 and $P[x,y]$,
which control the macroscopic equations
can be written as
\bd
\left\{\!\!
\begin{array}{c}
P[x,y]\\[2mm]
\A[x,y;x^\prime,y^\prime]\\[2mm]
\B[x,y;x^\prime,y^\prime]\\[2mm]
\C[x,y;x^\prime,y^\prime;x^\pprime,y^\pprime]
\end{array}\!\!
\right\}
=\!
\Limits\bigbra~ \bigbras\!\!\!\!\bigbra
~~~~~~~~~~~~~~~~~~~~~~~~~~~~~~~~~~~~~~~~~~~~~~~~~~~~~~~~~~~~~~~~~~~~~~~~~~~~~~~~~~~~~~~~~
\right.\right.\right.\right.
\ed
\bd
\left.\left.\left.\left.
\int\!
\prod_{\alpha} \left\{
\delta\left[N\minus (\bsigma^\alpha)^2\right]
\delta\left[\frac{NR}{\sqrt{Q}}\minus\btau\inn \bsigma^\alpha\right]
d\bsigma^\alpha
\prod_{x_\alpha y_\alpha}\!\delta\left[P[x_\alpha,y_\alpha]\minus
P[x_\alpha,y_\alpha;\frac{\sqrt{Q}\bsigma^\alpha}{\sqrt{N}}]\right]
\right\}
\delta\left[x\minus \frac{\sqrt{Q}\bsigma^1\inn\bxi}{\sqrt{N}}\right]
\delta\left[y\minus \frac{\btau\inn\bxi}{\sqrt{N}}\right]
\right.\right.\right.\right.
\ed
\bd
\left.\left.\left.\left.
\left\{\!
\begin{array}{c}1\\[3mm]
(\bxi^\prime\inn\bxi)\overline{\delta}_{\bxi\bxi^\prime}
\delta\left[x^\prime\minus\frac{\sqrt{Q}\bsigma^1\cdot\bxi^\prime}{\sqrt{N}}\right]
\delta\left[y^\prime\minus\frac{\btau\cdot\bxi^\prime}{\sqrt{N}}\right]
\\[3mm]
\left[\frac{1}{N}\sum_{i\neq j}\xi_i\xi_j\xi^\prime_i\xi^\prime_j\right]
\overline{\delta}_{\bxi\bxi^\prime}
\delta\left[x^\prime\minus\frac{\sqrt{Q}\bsigma^1\cdot\bxi^\prime}{\sqrt{N}}\right]
\delta\left[y^\prime\minus\frac{\btau\cdot\bxi^\prime}{\sqrt{N}}\right]
\\[3mm]
\frac{1}{N}(\bxi\inn\bxi^\pprime)(\bxi^\prime\inn\bxi^\pprime)
\overline{\delta}_{\bxi\bxi^\pprime}\overline{\delta}_{\bxi^\prime\bxi^\pprime}
\delta\left[x^\prime\minus\frac{\sqrt{Q}\bsigma^1\cdot\bxi^\prime}{\sqrt{N}}\right]
\delta\left[y^\prime\minus\frac{\btau\cdot\bxi^\prime}{\sqrt{N}}\right]
\delta\left[x^\pprime\minus\frac{\sqrt{Q}\bsigma^1\cdot\bxi^\pprime}{\sqrt{N}}\right]
\delta\left[y^\pprime\minus\frac{\btau\cdot\bxi^\pprime}{\sqrt{N}}\right]
\end{array}
\!
\right\}~
\bigket_{\!\!\!\set}\bigket_{\!\!\!\set}\bigket_{\!\!\!\set}
~\bigket_{\rmsets}
\vspace*{2mm}
\ed
We next use the definition of $P[x,y;\bJ]$, introduce
integral representations for the $\delta$-distributions involving
$P[x,y]$,  and obtain
\bd
\left\{\!\!
\begin{array}{c}
P[x,y]\\[2mm]
\A[x,y;x^\prime,y^\prime]\\[2mm]
\B[x,y;x^\prime,y^\prime]\\[2mm]
\C[x,y;x^\prime,y^\prime;x^\pprime,y^\pprime]
\end{array}\!\!
\right\}
=\!\Limits\!\int\!
\prod_{\alpha} \left\{
\delta\left[N\minus (\bsigma^\alpha)^2\right]
\delta\left[\frac{NR}{\sqrt{Q}}\minus\btau\inn \bsigma^\alpha\right]
d\bsigma^\alpha\!
\prod_{x_\alpha
y_\alpha}e^{iN\pi_\alpha[x_\alpha,y_\alpha]P[x_\alpha,y_\alpha]}
d\pi(x_\alpha,y_\alpha)
\right\}
\ed
\bd
\times~\bigbra~
\bigbras\!\!\!\!\bigbra
e^{-iN\sum_\alpha\sum_{x_\alpha y_\alpha}\pi_\alpha[x_\alpha,y_\alpha]
\bra\delta[x_\alpha -
\frac{\sqrt{Q}\bsigma^\alpha\cdot\bxi^\ppprime}{\sqrt{N}}]\delta[y_\alpha - \frac{\btau\cdot\bxi^\ppprime}{\sqrt{N}}]\ketset}
\delta\left[x\minus \frac{\sqrt{Q}\bsigma^1\inn\bxi}{\sqrt{N}}\right]
\delta\left[y\minus \frac{\btau\inn\bxi}{\sqrt{N}}\right]
\right.\right.\right.\right.
~~~~~~~~~~
\ed
\bd
\left.\left.\left.\left.
\times~\left\{\!\!
\begin{array}{c}1\\[3mm]
(\bxi^\prime\inn\bxi)\overline{\delta}_{\bxi\bxi^\prime}
\delta\left[x^\prime\minus\frac{\sqrt{Q}\bsigma^1\cdot\bxi^\prime}{\sqrt{N}}\right]
\delta\left[y^\prime\minus\frac{\btau\cdot\bxi^\prime}{\sqrt{N}}\right]
\\[3mm]
\left[\frac{1}{N}\sum_{i\neq j}\xi_i\xi_j\xi^\prime_i\xi^\prime_j\right]
\overline{\delta}_{\bxi\bxi^\prime}
\delta\left[x^\prime\minus\frac{\sqrt{Q}\bsigma^1\cdot\bxi^\prime}{\sqrt{N}}\right]
\delta\left[y^\prime\minus\frac{\btau\cdot\bxi^\prime}{\sqrt{N}}\right]
\\[3mm]
\frac{1}{N}(\bxi\inn\bxi^\pprime)(\bxi^\prime\inn\bxi^\pprime)
\overline{\delta}_{\bxi\bxi^\pprime}\overline{\delta}_{\bxi^\prime\bxi^\pprime}
\delta\left[x^\prime\minus\frac{\sqrt{Q}\bsigma^1\cdot\bxi^\prime}{\sqrt{N}}\right]
\delta\left[y^\prime\minus\frac{\btau\cdot\bxi^\prime}{\sqrt{N}}\right]
\delta\left[x^\pprime\minus\frac{\sqrt{Q}\bsigma^1\cdot\bxi^\pprime}{\sqrt{N}}\right]
\delta\left[y^\pprime\minus\frac{\btau\cdot\bxi^\pprime}{\sqrt{N}}\right]
\end{array}
\!\!\right\}
\bigket_{\!\!\!\set}\bigket_{\!\!\!\set}\bigket_{\!\!\!\set}
~\bigket_{\rmsets}
\vspace*{2mm}
\ed
The summations involving $(x_\alpha,y_\alpha)$ automatically lead to
integrals, which can be performed due to the $\delta$-distributions
involved. We define new conjugate functions $\hat{P}_\alpha[x,y]$ via
\bd
\sum_{x_\alpha y_\alpha}\pi_\alpha[x_\alpha,y_\alpha]f[x_\alpha,y_\alpha]~
\rightarrow~
\int\!dx^\pprime
dy^\pprime~\hat{P}_\alpha[x^\pprime,y^\pprime]f[x^\pprime,y^\pprime]
\ed
We write averages over the training set explicitly in terms of
the $p=\alpha N$ constituent vectors $\{\bxi^\mu\}$.
Finally we introduce integrals representations for the remaining
delta-distributions, and obtain the following expressions
(at this stage we will have to separate the various structurally
different cases):
\bd
P[x,y]
=\int\!\frac{d\hat{x} d\hat{y}}{(2\pi)^2}e^{i[x\hat{x}+y\hat{y}]}
\Limits
~~~~~~~~~~~~~~~~~~~~~~~~~~~~~~~~~~~~~~~~~~~~~~~~~~~~~~~~~~~~~~~~~~~~~~~~~~~~~~~~~~~~~~~~~
\ed
\bd
\int\!
\prod_{\alpha} \left\{
\delta\left[N\minus (\bsigma^\alpha)^2\right]
\delta\left[\frac{NR}{\sqrt{Q}}\minus\btau\cdot \bsigma^\alpha\right]
d\bsigma^\alpha
e^{iN\int dx^\pprime
dy^\pprime~\hat{P}_\alpha[x^\pprime,y^\pprime]P_t[x^\pprime,y^\pprime]}
\prod_{x^\pprime y^\pprime}d\hat{P}_\alpha[x^\pprime,y^\pprime]
\right\}
\frac{1}{p}
\sum_{\mu=1}^p
\ed
\be
\bigbra
e^{-\frac{i}{\alpha}\sum_\alpha\sum_{\lambda}
\hat{P}_\alpha(\frac{\sqrt{Q}\bsigma^\alpha\cdot\bxi^\lambda}{\sqrt{N}},
\frac{\btau\cdot\bxi^\lambda}{\sqrt{N}})
-i[\hat{x}\sqrt{Q}\bsigma^1\cdot\bxi^\mu
+\hat{y}\btau\cdot\bxi^\mu]/\sqrt{N}}
\bigket_{\rmsets}
\label{eq:intermediateP}
\ee
\bd
\left\{\!\begin{array}{c}
\A[x,y;x^\prime,y^\prime]\\[3mm]
\B[x,y;x^\prime,y^\prime]
\end{array}\!\right\}
=\int\!\frac{d\hat{x}d\hat{x}^\prime d\hat{y}
d\hat{y}^\prime}{(2\pi)^4}e^{i[x\hat{x}+x^\prime\hat{x}^\prime+y\hat{y}+y\hat{y}^\prime]}
\Limits
~~~~~~~~~~~~~~~~~~~~~~~~~~~~~~~~~~~~~~~~~~~~~~~~~~~~~~~~~~~~
\ed
\bd
\int\!
\prod_{\alpha} \left\{
\delta\left[N\minus (\bsigma^\alpha)^2\right]
\delta\left[\frac{NR}{\sqrt{Q}}\minus\btau\cdot \bsigma^\alpha\right]
d\bsigma^\alpha
e^{iN\int dx^\pprime
dy^\pprime~\hat{P}_\alpha[x^\pprime,y^\pprime]P[x^\pprime,y^\pprime]}
\prod_{x^\pprime y^\pprime}d\hat{P}_\alpha(x^\pprime,y^\pprime)
\right\}
 \frac{1}{p^2}\!
\sum_{\mu\neq \nu=1}^p
\ed
\be
\bigbra
\left\{\!\!\begin{array}{c}(\bxi^\mu\cdot\bxi^\nu)\\[3mm]
\frac{1}{N}\sum_{i\neq j}\xi_i^\mu\xi_i^\nu\xi_j^\mu\xi_j^\nu
\end{array}\!\!\right\}
e^{-\frac{i}{\alpha}\sum_\alpha\sum_{\lambda}
\hat{P}_\alpha[\frac{\sqrt{Q}\bsigma^\alpha\cdot\bxi^\lambda}{\sqrt{N}},
\frac{\btau\cdot\bxi^\lambda}{\sqrt{N}}]
-i[\hat{x}\sqrt{Q}\bsigma^1\cdot\bxi^\mu
+\hat{y}\btau\cdot\bxi^\mu
+\hat{x}^\prime\sqrt{Q}\bsigma^1\cdot\bxi^\nu
+\hat{y}^\prime\btau\cdot\bxi^\nu]/\sqrt{N}}
\bigket_{\rmsets}
\label{eq:intermediateAB}
\ee
\bd
\C[x,y;x^\prime,y^\prime;x^\pprime,y^\pprime]
=\int\!\frac{d\hat{x}d\hat{x}^\prime d\hat{x}^\pprime
d\hat{y}d\hat{y}^\prime d\hat{y}^\pprime}{(2\pi)^6}
e^{i[x\hat{x}+x^\prime\hat{x}^\prime+x^\pprime\hat{x}^\pprime
+y\hat{y}+y^\prime\hat{y}^\prime+y^\pprime\hat{y}^\pprime]}
\Limits
~~~~~~~~~~~~~~~~~~~~~~~~~~~~~~~~~~~~~~~~~~~~~~~~~~~~~~~~~~~~
\ed
\bd
\int\!
\prod_{\alpha} \left\{
\delta\left[N\minus (\bsigma^\alpha)^2\right]
\delta\left[\frac{NR}{\sqrt{Q}}\minus\btau\cdot \bsigma^\alpha\right]
d\bsigma^\alpha
e^{iN\int dx^\pprime
dy^\pprime~\hat{P}_\alpha[x^\pprime,y^\pprime]P[x^\pprime,y^\pprime]}
\prod_{x^\pprime y^\pprime}d\hat{P}_\alpha(x^\pprime,y^\pprime)
\right\}
 \frac{1}{p^3}\!
\sum_{\mu\nu\rho=1}^p\overline{\delta}_{\mu\rho}\overline{\delta}_{\nu\rho}
\ed
\be
\bigbra
(\bxi^\mu\inn\bxi^\rho)(\bxi^\nu\inn\bxi^\rho)
e^{-\frac{i}{\alpha}\sum_\alpha\sum_{\lambda}
\hat{P}_\alpha[\frac{\sqrt{Q}\bsigma^\alpha\cdot\bxi^\lambda}{\sqrt{N}},
\frac{\btau\cdot\bxi^\lambda}{\sqrt{N}}]
-i[\hat{x}\sqrt{Q}\bsigma^1\cdot\bxi^\mu
+\hat{y}\btau\cdot\bxi^\mu
+\hat{x}^\prime\sqrt{Q}\bsigma^1\cdot\bxi^\nu
+\hat{y}^\prime\btau\cdot\bxi^\nu
+\hat{x}^\pprime\sqrt{Q}\bsigma^1\cdot\bxi^\rho
+\hat{y}^\pprime\btau\cdot\bxi^\rho
]/\sqrt{N}}
\bigket_{\rmsets}
\label{eq:intermediateC}
\ee
The averages over the training sets $\bra \ldots\ket_{\rmsets}$
 in (\ref{eq:intermediateP},\ref{eq:intermediateAB},\ref{eq:intermediateC})
will now
be done separately.
First we define some relevant objects:
\be
\cD[u,v]=
\bigbra
e^{-\frac{i}{\alpha}\sum_\alpha
\hat{P}_\alpha(\frac{\sqrt{Q}\bsigma^\alpha\cdot\bxi}{\sqrt{N}},
\frac{\btau\cdot\bxi}{\sqrt{N}})
-i[u\sqrt{Q}\bsigma^1\cdot\bxi
+v\btau\cdot\bxi]/\sqrt{N}}
\bigket_{\bxi}
\label{eq:defineD}
\ee
\be
\cE_j[u,v]=
\bigbra \sqrt{N}\xi_j~
e^{-\frac{i}{\alpha}\sum_\alpha
\hat{P}_\alpha(\frac{\sqrt{Q}\bsigma^\alpha\cdot\bxi}{\sqrt{N}},
\frac{\btau\cdot\bxi}{\sqrt{N}})
-i[u\sqrt{Q}\bsigma^1\cdot\bxi
+v\btau\cdot\bxi]/\sqrt{N}}
\bigket_{\bxi}
\label{eq:defineE}
\ee
\be
\cE_{ij}[u,v]=
\bigbra N\xi_i\xi_j~
e^{-\frac{i}{\alpha}\sum_\alpha
\hat{P}_\alpha(\frac{\sqrt{Q}\bsigma^\alpha\cdot\bxi}{\sqrt{N}},
\frac{\btau\cdot\bxi}{\sqrt{N}})
-i[u\sqrt{Q}\bsigma^1\cdot\bxi
+v\btau\cdot\bxi]/\sqrt{N}}
\bigket_{\bxi}~~~~(i\neq j)
\label{eq:defineanotherE}
\ee
As we will see, all are of order $\order(N^0)$ as
$N\to\infty$.
We next use the permutation
invariance of our integrations and summations with respect to pattern labels.
First we calculate the first training sets average occurring in
(\ref{eq:intermediateP}):
\bd
\frac{1}{p}\sum_{\mu=1}^p
\bigbra
e^{\ldots}
\bigket_{\rmsets}
=
\bigbra
e^{-\frac{i}{\alpha}\sum_\alpha
\hat{P}_\alpha(\frac{\sqrt{Q}\bsigma^\alpha\cdot\bxi}{\sqrt{N}},
\frac{\btau\cdot\bxi}{\sqrt{N}})}
\bigket_{\bxi}^{p-1}
\bigbra
e^{-\frac{i}{\alpha}\sum_\alpha
\hat{P}_\alpha(\frac{\sqrt{Q}\bsigma^\alpha\cdot\bxi}{\sqrt{N}},
\frac{\btau\cdot\bxi}{\sqrt{N}})
-i[\hat{x}\sqrt{Q}\bsigma^1\cdot\bxi
+\hat{y}\btau\cdot\bxi]/\sqrt{N}}\bigket_{\bxi}
\ed
\be
=e^{p\log \cD[0,0]}~
\frac{\cD[\hat{x},\hat{y}]}
{\cD[0,0]}
\label{eq:Pworkedout}
\ee
The prefactor $e^{p\log \cD[0,0]}$,  will turn out to take care of
appropriate normalisation, and will drop out of the final result for
all four functions
$P[x,y]$, $\cA[x,y;x^\prime,y^\prime]$, $\cB[x,y;x^\prime,y^\prime]$
and  $\cC[x,y;x^\prime,y^\prime;x^\pprime,y^\pprime]$.
Secondly we evaluate
the training sets average of the expression for
$\A[\ldots]$ in (\ref{eq:intermediateAB}):
\bd
 \frac{1}{p^2}
\sum_{\mu\neq \nu}^p
\bigbra
(\bxi^\mu\inn\bxi^\nu)~
e^{\ldots}
\bigket_{\rmsets}
=
\frac{p\minus 1}{p}\bigbra
(\bxi^1\cdot\bxi^2)~
e^{\ldots}
\bigket_{\rmsets}
~~~~~~~~~~~~~~~~~~~~~~~~~~~~~~~~~~~~~~~~~~~~~~~~~~~~~~~~~~~~~~~~~~~~~~~~~~
\ed
\bd
=\frac{p-1}{p}\sum_j
\bigbra
e^{-\frac{i}{\alpha}\sum_\alpha
\hat{P}_\alpha(\frac{\sqrt{Q}\bsigma^\alpha\cdot\bxi}{\sqrt{N}},
\frac{\btau\cdot\bxi}{\sqrt{N}})}
\bigket_{\bxi}^{p-2}
\bigbra
\xi_j
e^{-\frac{i}{\alpha}\sum_\alpha
\hat{P}_\alpha(\frac{\sqrt{Q}\bsigma^\alpha\cdot\bxi}{\sqrt{N}},
\frac{\btau\cdot\bxi}{\sqrt{N}})
-i[\hat{x}\sqrt{Q}\bsigma^1\cdot\bxi
+\hat{y}\btau\cdot\bxi]/\sqrt{N}}
\bigket_{\bxi}
\ed
\bd
~~~~~~~~~~~~~~~~~~~~~~~~~~~~~~~~~~~~~~~~~~~~~~~~~~~~~~~
\times
\bigbra
\xi_j
e^{-\frac{i}{\alpha}\sum_\alpha
\hat{P}_\alpha(\frac{\sqrt{Q}\bsigma^\alpha\cdot\bxi}{\sqrt{N}},
\frac{\btau\cdot\bxi}{\sqrt{N}})
-i[\hat{x}^\prime\sqrt{Q}\bsigma^1\cdot\bxi
+\hat{y}^\prime\btau\cdot\bxi]/\sqrt{N}}
\bigket_{\bxi}
\ed
\be
=e^{p\log \cD[0,0]}\left\{\frac{1}{N}\sum_{j=1}^N
\frac{\cE_j[\hat{x},\hat{y}]\cE_j[\hat{x}^\prime,\hat{y}^\prime]}
{\cD^{2}[0,0]}
+\order(N^{-1})\right\}
\label{eq:Aworkedout}
\ee
(provided we indeed show that $\cE_j[u,v]=\order(N^0)$ as $N\to\infty$).
Secondly, the training sets average of the expression for
$\B[\ldots]$ in (\ref{eq:intermediateAB}) is given by:
\bd
 \frac{1}{p^2}
\sum_{\mu\neq \nu}^p
\bigbra
\frac{1}{N}\sum_{i\neq j}\xi_i^\mu\xi_i^\nu\xi_j^\mu\xi_j^\nu
e^{\ldots}
\bigket_{\rmsets}
=
\frac{p\minus 1}{pN}\sum_{i\neq j}\bigbra (\xi_i^1\xi_j^1)(\xi_i^2\xi_j^2)~
e^{\ldots}
\bigket_{\rmsets}
~~~~~~~~~~~~~~~~~~~~~~~~~~~~~~~~~~~~~~~~~~~~~~~~~~~~~~~~~~~~~~~
\ed
\bd
=\frac{p-1}{pN}\sum_{i\neq j}
\bigbra
e^{-\frac{i}{\alpha}\sum_\alpha
\hat{P}_\alpha(\frac{\sqrt{Q}\bsigma^\alpha\cdot\bxi}{\sqrt{N}},
\frac{\btau\cdot\bxi}{\sqrt{N}})}
\bigket_{\bxi}^{p-2}
\bigbra
\xi_i\xi_j
e^{-\frac{i}{\alpha}\sum_\alpha
\hat{P}_\alpha(\frac{\sqrt{Q}\bsigma^\alpha\cdot\bxi}{\sqrt{N}},
\frac{\btau\cdot\bxi}{\sqrt{N}})
-i[\hat{x}\sqrt{Q}\bsigma^1\cdot\bxi
+\hat{y}\btau\cdot\bxi]/\sqrt{N}}
\bigket_{\bxi}
\ed
\bd
\times
\bigbra
\xi_i\xi_j
e^{-\frac{i}{\alpha}\sum_\alpha
\hat{P}_\alpha(\frac{\sqrt{Q}\bsigma^\alpha\cdot\bxi}{\sqrt{N}},
\frac{\btau\cdot\bxi}{\sqrt{N}})
-i[\hat{x}^\prime\sqrt{Q}\bsigma^1\cdot\bxi
+\hat{y}^\prime\btau\cdot\bxi]/\sqrt{N}}
\bigket_{\bxi}
\ed
\be
=e^{p\log \cD[0,0]}\left\{\frac{1}{N^3}\sum_{i\neq j=1}^N
\frac{\cE_{ij}[\hat{x},\hat{y}]\cE_{ij}[\hat{x}^\prime,\hat{y}^\prime]}
{\cD^{2}[0,0]}
+\order(N^{-\frac{3}{2}})\right\}=e^{p\log \cD[0,0]}\left\{\order(N^{-1})\right\}
\label{eq:Bworkedout}
\ee
(provided we indeed show that $\cE_{ij}[u,v]=\order(N^0)$ as $N\to\infty$).
Finally we also obtain for the training sets average in
(\ref{eq:intermediateC}), in a similar fashion:
\bd
 \frac{1}{p^3}\sum_{\rho=1}^p
\sum_{\mu,\nu\neq  \rho}^p
\bigbra
\frac{1}{N}(\bxi^\mu\inn\bxi^\rho)(\bxi^\nu\inn\bxi^\rho)~
e^{\ldots}
\bigket_{\rmsets}
=
\frac{p\minus 1}{p^2N}\sum_{ij}
\bigbra \xi_i^1\xi_j^1\xi_i^2\xi_j^2~ e^{\ldots} \bigket_{\rmsets}+
\frac{(p\minus 1)(p\minus 2)}{p^2N}\sum_{ij}\bigbra
\xi_i^1\xi_j^2\xi^3_i\xi_j^3~
e^{\ldots}
\bigket_{\rmsets}
~~~~~~~~~~
\ed
\bd
=\sum_{i\neq j}
\bigbra \xi_i^1\xi_j^1\xi_i^2\xi_j^2 ~
e^{\ldots} \bigket_{\rmsets}\!.\order(N^{-2})
+\bigbra e^{\ldots} \bigket_{\rmsets}\!.\order(N^{-1})
+\sum_{i\neq j}\bigbra \xi_i^1\xi_j^2\xi^3_i\xi_j^3~
e^{\ldots}\bigket_{\rmsets}\!.\order(N^{-1})
+\sum_{i}\bigbra
\xi_i^1\xi_i^2~e^{\ldots}\bigket_{\rmsets}\!.\order(N^{-1})
\ed
\bd
=\cD[0,0]^{p}\left\{
\sum_{i\neq j}\cD[\hat{x}^\pprime,\hat{y}^\pprime]
\cE_{ij}[\hat{x},\hat{y}]\cE_{ij}[\hat{x}^\prime,\hat{y}^\prime].
\order(N^{-4})+\order(N^{-1})
+\sum_{i\neq j}
\cE_i[\hat{x},\hat{y}]\cE_j[\hat{x}^\prime,\hat{y}^\prime]
\cE_{ij}[\hat{x}^\pprime,\hat{y}^\pprime].\order(N^{-3})
\right.
\ed
\bd
\left.
~~~~~~~~~~~~~~~~~~~~~~~~~~~~~~~~~~~~~~~~~~~~~~~~~~~~~~~~~~+
\sum_i\cD[\hat{x}^\pprime,\hat{y}^\pprime]
\cE_j[\hat{x},\hat{y}]\cE_j[\hat{x}^\prime,\hat{y}^\prime]
.\order(N^{-2})
\right\}
\ed
\be
=e^{p\log \cD[0,0]}\left\{\order(N^{-1})\right\}
\label{eq:Cworkedout}
\ee
We now work out
(\ref{eq:defineE})
and we show that it is of order $N^0$.
 This is achieved by separating in the
exponent the terms with site label $i=j$ from those with site labels
$i\neq j$, followed by expansion in powers of the (relatively small)
$i=j$ terms,
and  will involve the following two functions:
\be
\cF_1^\alpha[u,v]=
\bigbra
\partial_x\hat{P}_\alpha
(\frac{\sqrt{Q}\bsigma^\alpha\!\cdot\!\bxi}{\sqrt{N}},
\frac{\btau\!\cdot\!\bxi}{\sqrt{N}})~
e^{
-\frac{i}{\alpha}\sum_\alpha
\hat{P}_\alpha(\frac{\sqrt{Q}\bsigma^\alpha\cdot\bxi}{\sqrt{N}},
\frac{\btau\cdot\bxi}{\sqrt{N}})
-i[u\sqrt{Q}\bsigma^1\cdot\bxi
+v\btau\cdot\bxi]/\sqrt{N}}
\bigket_{\bxi}
\label{eq:defineF1}
\ee
\be
\cF_2^\alpha[u,v]=
\bigbra
\partial_y\hat{P}_\alpha
(\frac{\sqrt{Q}\bsigma^\alpha\!\cdot\!\bxi}{\sqrt{N}},
\frac{\btau\!\cdot\!\bxi}{\sqrt{N}})~
e^{
-\frac{i}{\alpha}\sum_\alpha
\hat{P}_\alpha(\frac{\sqrt{Q}\bsigma^\alpha\cdot\bxi}{\sqrt{N}},
\frac{\btau\cdot\bxi}{\sqrt{N}})
-i[u\sqrt{Q}\bsigma\cdot\bxi
+v\btau\cdot\bxi]/\sqrt{N}}
\bigket_{\bxi}
\label{eq:defineF2}
\ee
Note that there is no need to calculate the auxiliary functions
(\ref{eq:defineanotherE}); we only need to
verify their magnitude  to scale as $\order(N^0)$ for $N\to\infty$.
\bd
\cE_j[u,v]=
\bigbra \sqrt{N}\xi_j
e^{-\frac{i}{\alpha}\sum_\alpha
\hat{P}_\alpha(\frac{\sqrt{Q}}{\sqrt{N}}[\sum_{i\neq j}\sigma_i^\alpha\xi_i
+ \sigma_j^\alpha\xi_j],
\frac{1}{\sqrt{N}}[\sum_{i\neq j}\tau_i\xi_i+ \tau_j\xi_j])
-i[\frac{u\sqrt{Q}}{\sqrt{N}}[\sum_{i\neq j}\sigma^1_i\xi_i+ \sigma^1_j\xi_j]
+\frac{v}{\sqrt{N}}[\sum_{i\neq j}\tau_i\xi_i+  \tau_j\xi_j]]}
\bigket_{\bxi}
\ed
\bd
=
\bigbra \sqrt{N}\xi_j
e^{
-\frac{i}{\alpha}\sum_\alpha
\hat{P}_\alpha(\frac{\sqrt{Q}}{\sqrt{N}}\sum_{i\neq j}\sigma_i^\alpha\xi_i,
\frac{1}{\sqrt{N}}\sum_{i\neq j}\tau_i\xi_i)
-i[\frac{u\sqrt{Q}}{\sqrt{N}}\sum_{i\neq j}\sigma^1_i\xi_i
+\frac{v}{\sqrt{N}}\sum_{i\neq j}\tau_i\xi_i]}
\times~~~~~~~~~~
\right.
\ed
\bd
\left.
 e^{-\frac{i\sqrt{Q}}{\alpha\sqrt{N}}\sum_\alpha\sigma_j^\alpha\xi_j
\partial_x\hat{P}_\alpha
(\frac{\sqrt{Q}}{\sqrt{N}}\sum_{i\neq j}\sigma_i^\alpha\xi_i,
\frac{1}{\sqrt{N}}\sum_{i\neq j}\tau_i\xi_i)
- \frac{i}{\alpha\sqrt{N}}\tau_j\xi_j\sum_\alpha \partial_y
\hat{P}_\alpha(\frac{\sqrt{Q}}{\sqrt{N}}\sum_{i\neq j}\sigma_i^\alpha\xi_i,
\frac{1}{\sqrt{N}}\sum_{i\neq j}\tau_i\xi_i)}
\right.
\ed
\bd
\left.
\times
e^{-i[\frac{u\sqrt{Q}}{\sqrt{N}}\sigma_j^1\xi_j
+\frac{v}{\sqrt{N}}\tau_j\xi_j]+\order(N^{-1})}
\bigket_{\bxi}
\ed
\bd
=
\bigbra
e^{
-\frac{i}{\alpha}\sum_\alpha
\hat{P}_\alpha(\frac{\sqrt{Q}}{\sqrt{N}}\sum_{i\neq j}\sigma_i^\alpha\xi_i,
\frac{1}{\sqrt{N}}\sum_{i\neq j}\tau_i\xi_i)
-i[\frac{u\sqrt{Q}}{\sqrt{N}}\sum_{i\neq j}\sigma_i^1\xi_i
+\frac{v}{\sqrt{N}}\sum_{i\neq j}\tau_i\xi_i]}
\times~~~~~~~~~~~~~~~~~~
\right.
\ed
\bd
\left.
\frac{1}{i}\left\{\room
\frac{\sqrt{Q}}{\alpha}\sum_\alpha\sigma_j^\alpha
\partial_x\hat{P}_\alpha
(\frac{\sqrt{Q}}{\sqrt{N}}\sum_{i\neq j}\sigma_i^\alpha\xi_i,
\frac{1}{\sqrt{N}}\sum_{i\neq j}\tau_i\xi_i)
+ \frac{1}{\alpha}\tau_j\sum_\alpha\partial_y
\hat{P}_\alpha(\frac{\sqrt{Q}}{\sqrt{N}}\sum_{i\neq j}\sigma_i^\alpha\xi_i,
\frac{1}{\sqrt{N}}\sum_{i\neq j}\tau_i\xi_i)
\right.\right.
\ed
\bd
\left.\left.~~~~~~~~~~~~~~~~~~~~~~~~~~~~~~~~~~~~~~~~~~~\room
+ u\sqrt{Q}\sigma_j^1+ v\tau_j
+\order(N^{-\frac{1}{2}})
~\right\}
\bigket_{\bxi}
\ed
so that
\bd
\cE_j[u,v]
=
-iu\sqrt{Q}\sigma_j^1\cD[u,v]
-iv\tau_j\cD[u,v]
-\frac{i}{\alpha}\sqrt{Q}\sum_\alpha \sigma_j^\alpha \cF^\alpha_1[u,v]
-\frac{i}{\alpha}\tau_j\sum_\alpha \cF^\alpha_2[u,v]
+\order(N^{-\frac{1}{2}})
\ed
\be
=
-i\sqrt{Q}\sum_\alpha\sigma_j^\alpha
\left[\frac{1}{\alpha}\cF^\alpha_1[u,v]+u\delta_{\alpha 1}\cD[u,v]\right]
-i\tau_j\sum_\alpha
\left[\frac{1}{\alpha}\cF^\alpha_2[u,v]+v\delta_{\alpha 1}\cD[u,v]\right]
+\order(N^{-\frac{1}{2}})
\label{eq:expressionforE}
\ee
Repetition/extension of this argument, by separating in the exponent
terms with two special indices $(i,j)$ rather than one, and by subsequent expansion
(whereby each term brings down a factor $N^{-\frac{1}{2}}$),
immediately shows that terms of the form
$\bra N\xi_i\xi_j ~e^{\ldots}\ket_{\bxi}$ with $i\neq j$ will be of order
$\order(N^0)$. This confirms that $\cE_{ij}[u,v]=\order(N^0)$ and
that (\ref{eq:defineanotherE}) indeed scales as indicated.
Note that the relevant combination of intensive terms in (\ref{eq:Aworkedout})
can be abbreviated as $\cL[u,v;u^\prime,v^\prime]=\frac{1}{N}\sum_j
\cE_j[u,v]\cE_j[u^\prime,v^\prime]$:
\bd
\cL[u,v;u^\prime,v^\prime]=
-Q\sum_{\alpha\beta}q_{\alpha\beta}(\{\bsigma\})
\left[\frac{1}{\alpha}\cF^\alpha_1[u,v]+u\delta_{\alpha 1}\cD[u,v]\right]
\left[\frac{1}{\alpha}\cF^\beta_1[u^\prime,v^\prime]+u^\prime\delta_{\beta 1}\cD[u^\prime,v^\prime]\right]
\ed
\bd
-R\sum_{\alpha\beta}
\left[\frac{1}{\alpha}\cF^\alpha_1[u,v]+u\delta_{\alpha 1}\cD[u,v]\right]
\left[\frac{1}{\alpha}\cF^\beta_2[u^\prime,v^\prime]+v^\prime\delta_{\beta 1}\cD[u^\prime,v^\prime]\right]
\ed
\bd
-R\sum_{\alpha\beta}
\left[\frac{1}{\alpha}\cF^\alpha_1[u^\prime,v^\prime]+u^\prime\delta_{\alpha 1}\cD[u^\prime,v^\prime]\right]
\left[\frac{1}{\alpha}\cF^\beta_2[u,v]+v\delta_{\beta 1}\cD[u,v]\right]
\ed
\be
-\sum_{\alpha\beta}
\left[\frac{1}{\alpha}\cF^\alpha_2[u,v]+v\delta_{\alpha 1}\cD[u,v]\right]
\left[\frac{1}{\alpha}\cF^\beta_2[u^\prime,v^\prime]+v^\prime\delta_{\beta 1}\cD[u^\prime,v^\prime]\right]
+\order(N^{-\frac{1}{2}})
\label{eq:nastyterm}
\ee
where we have used the built-in properties
$\frac{1}{N}\btau\inn\bsigma^\alpha=R/\sqrt{Q}$ and $\btau^2=N$,
 and in which we find the spin-glass order parameters
\be
q_{\alpha\beta}(\{\bsigma\})=
\frac{1}{N}\sum_i\sigma_i^\alpha \sigma_i^\beta
\label{eq:spinglassops}
\ee
\vsp

Let us finally work out further the remaining fundamental objects
$\cD[\ldots]$ and $\cF^{\alpha}_{1,2}[\ldots]$.
The basic property to be used is that for large $N$
the $n\plus 1$ quantities
$\{x_\alpha=\bsigma^\alpha\!\cdot\!\bxi/\sqrt{N},~
y=\btau\!\cdot\!\bxi/\sqrt{N}\}$ inside averages of the form
$\bra\ldots\ket_{\bxi}$ will become (zero average but correlated)
Gaussian variables,
with probability distribution
\bd
P(x_1,\ldots,x_n,y)=
\frac{{\rm det}^{\frac{1}{2}}\bA}{(2\pi)^{(n+1)/2}}~
e^{-\frac{1}{2}
\left(\!\!
\begin{array}{c}x_1\\[-1mm] \vdots \\[-1mm] x_n \\[-1mm] y\end{array}
\!\!\right)\cdot\bA
\left(
\!\!\begin{array}{c}x_1\\[-1mm] \vdots \\[-1mm] x_n \\[-1mm]
y\end{array}
\!\!\right)
}
~~~~~~~~
\bA^{-1}=\left(
\!\!\begin{array}{cccc} q_{11} & \!\cdots\! & q_{1n} & R/\sqrt{Q}\\
\vdots && \vdots & \vdots \\
q_{n1} & \!\cdots\! & q_{nn} & R/\sqrt{Q} \\
R/\sqrt{Q} & \!\cdots\! & R/\sqrt{Q} & 1
\end{array}\!\!\right)
\ed
This allows us to write
\be
\vspace*{-2mm}
\cD[u,v]=
\frac{{\rm det}^{\frac{1}{2}}\bA}{(2\pi)^{(n+1)/2}}~
\int\!d\bx dy~
e^{-\frac{1}{2}
\left(\!\!
\begin{array}{c}x_1\\[-1mm] \vdots \\[-1mm] x_n \\[-1mm] y\end{array}
\!\!\right)\cdot\bA
\left(
\!\!
\begin{array}{c}x_1\\[-1mm] \vdots \\[-1mm] x_n \\[-1mm] y\end{array}
\!\!\right)
-\frac{i}{\alpha}\sum_\alpha
\hat{P}_\alpha(\sqrt{Q}x_\alpha,y)
-i[u\sqrt{Q}x_1 +vy]}
\label{eq:Dasintegral}
\ee
\be
\vspace*{-2mm}
\cF_{1,2}^\alpha[u,v]=
\frac{{\rm det}^{\frac{1}{2}}\bA}{(2\pi)^{(n+1)/2}}~
\int\!d\bx dy~
\partial_{1,2}\hat{P}_\alpha
(\sqrt{Q}x_\alpha,y)
~
e^{-\frac{1}{2}
\left(
\!\!\begin{array}{c}x_1\\[-1mm] \vdots \\[-1mm] x_n \\[-1mm]
y\end{array}\!\!\right)
\cdot\bA
\left(
\!\!\begin{array}{c}x_1\\[-1mm] \vdots \\[-1mm] x_n \\[-1mm] y\end{array}\!\!\right)
-\frac{i}{\alpha}\sum_\alpha
\hat{P}_\alpha(\sqrt{Q}x_\alpha,y)
-i[u\sqrt{Q}x_1 +vy]}
\label{eq:Fasintegral}
\ee
Note that these quantities depend on the
microscopic variables $\bsigma^\alpha$ only through the macroscopic
observables $q_{\alpha\beta}(\{\bsigma\})$.

\subsection{Derivation of Saddle-Point Equations}

We will now combine the results
(\ref{eq:Pworkedout},\ref{eq:Aworkedout},\ref{eq:Bworkedout},\ref{eq:Cworkedout})
and
(\ref{eq:nastyterm}) with
the expressions
(\ref{eq:intermediateP},\ref{eq:intermediateAB},\ref{eq:intermediateC}).
We use integral representations for the remaining delta
functions, and isolate the observables $q_{\alpha\beta}$, by inserting
\bd
1=\int\frac{d\bq d\hbq d\hbQ d\hbR}{(2\pi)^{n^2+2n}}
~e^{iN[\sum_\alpha(\hQ_\alpha+\hR_\alpha R/\sqrt{Q})+\sum_{\alpha\beta}\hat{q}_{\alpha\beta}q_{\alpha\beta}]
-i\sum_i\sum_\alpha[\hQ_\alpha(\sigma_i^\alpha)^2+\hR_\alpha\tau_i\sigma_i^\alpha]-i\sum_{\alpha\beta}\hat{q}_{\alpha\beta}\sigma_i^\alpha
\sigma_i^\beta}
\ed
We hereby achieve a full factorisation over sites in the relevant
quantities (note: the objects $\cD[\ldots]$ and $\cL[\ldots]$ depend
on the microscopic variables only via $q_{\alpha\beta}(\{\bsigma\})$):
\bd
\A[x,y;x^\prime,y^\prime]
=\int\!\frac{d\hat{x}d\hat{x}^\prime d\hat{y}
d\hat{y}^\prime}{(2\pi)^4}e^{i[x\hat{x}+x^\prime\hat{x}^\prime+y\hat{y}+y\hat{y}^\prime]}
\lim_{n\to 0}\lim_{N\to\infty}
\int\! d\bq d\hbq d\hbQ d\hbR \prod_{\alpha x^\pprime y^\pprime}d\hat{P}_\alpha(x^\pprime,y^\pprime)~
~~~~~~~~~~~~~~~~
\ed
\bd
e^{iN[\sum_\alpha(\hQ_\alpha+\hR_\alpha
R/\sqrt{Q})+\sum_{\alpha\beta}\hat{q}_{\alpha\beta}q_{\alpha\beta}
+\sum_\alpha\int dx^\pprime
dy^\pprime~\hat{P}_\alpha(x^\pprime,y^\pprime)P[x^\pprime,y^\pprime]]
+\alpha N\log \cD[0,0]}~
\ed
\bd
\prod_i\left\{
\int\!d\bsigma
~
e^{-i\sum_\alpha[\hQ_\alpha(\sigma_\alpha)^2+\hR_\alpha\tau_i\sigma_\alpha]
-i\sum_{\alpha\beta}\hat{q}_{\alpha\beta}\sigma_\alpha\sigma_\beta}
\right\}
\frac{\cL[\hat{x},\hat{y};\hat{x}^\prime,\hat{y}^\prime]}
{\cD^{2}[0,0]}
\ed
and
\bd
P[x,y]
=\int\!\frac{d\hat{x} d\hat{y}}{(2\pi)^2}e^{i[x\hat{x}+y\hat{y}]}
\lim_{n\to 0}\lim_{N\to\infty}
\int\! d\bq d\hbq d\hbQ d\hbR
\prod_{\alpha x^\pprime y^\pprime}d\hat{P}_\alpha(x^\pprime,y^\pprime)~
~~~~~~~~~~~~~~~~~~~~~~~~
\ed
\bd
e^{iN[\sum_\alpha(\hQ_\alpha+\hR_\alpha
R/\sqrt{Q})+\sum_{\alpha\beta}\hat{q}_{\alpha\beta}q_{\alpha\beta}
+\sum_\alpha\int dx^\pprime
dy^\pprime~\hat{P}_\alpha(x^\pprime,y^\pprime)P[x^\pprime,y^\pprime]]
+\alpha N\log \cD[0,0]}
\ed
\bd
\prod_i
\left\{
\int\!d\bsigma
~
e^{-i\sum_\alpha[\hQ_\alpha(\sigma_\alpha)^2+\hR_\alpha\tau_i\sigma_\alpha]
-i\sum_{\alpha\beta}\hat{q}_{\alpha\beta}\sigma_\alpha\sigma_\beta}
\right\}
\frac{\cD[\hat{x},\hat{y}]}
{\cD[0,0]}
\ed
Both can be written in the form of an integral dominated by
saddle-points:
\bd
\A[x,y;x^\prime,y^\prime]
=\int\!\frac{d\hat{x}d\hat{x}^\prime d\hat{y}
d\hat{y}^\prime}{(2\pi)^4}e^{i[x\hat{x}+x^\prime\hat{x}^\prime+y\hat{y}+y\hat{y}^\prime]}
~~~~~~~~~~~~~~~~~~~~~~~~~~~~~~~~~~~~~~~~~~~~~~~~~~~~~
\ed
\bd
\lim_{n\to 0}
\lim_{N\to\infty}
\int\! d\bq d\hbq d\hbQ d\hbR \prod_{\alpha x^\pprime
y^\pprime}d\hat{P}_\alpha(x^\pprime,y^\pprime)~
e^{N\Psi[\bq,\hbq,\hbQ,\hbR,\{\hat{P}\}]}
\frac{\cL[\hat{x},\hat{y};\hat{x}^\prime,\hat{y}^\prime]}
{\cD^{2}[0,0]}
\ed
and
\bd
P[x,y]
=\int\!\frac{d\hat{x} d\hat{y}}{(2\pi)^2}e^{i[x\hat{x}+y\hat{y}]}
\lim_{n\to 0}\lim_{N\to\infty}
\int\! d\bq d\hbq d\hbQ d\hbR
\prod_{\alpha x^\pprime
y^\pprime}d\hat{P}_\alpha(x^\pprime,y^\pprime)~
e^{N\Psi[\bq,\hbq,\hbQ,\hbR,\{\hat{P}\}]}
\frac{\cD[\hat{x},\hat{y}]}
{\cD[0,0]}
\ed
with
\bd
\Psi[\ldots]=
i\sum_\alpha(\hQ_\alpha+\hR_\alpha
R/\sqrt{Q})+i\sum_{\alpha\beta}\hat{q}_{\alpha\beta}q_{\alpha\beta}
+i\sum_\alpha\int dx^\pprime
dy^\pprime~\hat{P}_\alpha(x^\pprime,y^\pprime)P[x^\pprime,y^\pprime]
\ed
\bd
+\alpha \log \cD[0,0]
+\lim_{N\to\infty}\frac{1}{N}\sum_i
\log \int\!d\bsigma
~
e^{-i\sum_\alpha[\hQ_\alpha\sigma_\alpha^2+\hR_\alpha\tau_i\sigma_\alpha]
-i\sum_{\alpha\beta}\hat{q}_{\alpha\beta}\sigma_\alpha\sigma_\beta}
\ed
Finally we use that fact
that the above expressions will be given by the intensive parts
evaluated in the dominating saddle-point of $\Psi$. We can use the
expression for $P[x,y]$ and its property $\int\!dxdy~P[x,y]=1$ to verify
that all expressions are properly normalised (no overall prefactors
are to be taken into account). We perform a simple transformation on
some of our integration variables:
\bd
\hat{q}_{\alpha\beta}\to\hat{q}_{\alpha\beta}-\hat{Q}_\alpha
\delta_{\alpha\beta}
~~~~~~~~~~~~~
\hat{R}_\alpha \to \sqrt{Q}\hat{R}_\alpha
\ed
 and finally
we get
\be
\A[x,y;x^\prime,y^\prime]
=\int\!\frac{d\hat{x}d\hat{x}^\prime d\hat{y}
d\hat{y}^\prime}{(2\pi)^4}e^{i[x\hat{x}+x^\prime\hat{x}^\prime+y\hat{y}+y\hat{y}^\prime]}
\lim_{n\to 0}
\frac{\cL[\hat{x},\hat{y};\hat{x}^\prime,\hat{y}^\prime]}
{\cD^{2}[0,0]}
\label{eq:finalA}
\ee
\be
P[x,y]=\int\!\frac{d\hat{x}
d\hat{y}}{(2\pi)^2}e^{i[x\hat{x}+y\hat{y}]}
\lim_{n\to 0} \frac{\cD[\hat{x},\hat{y}]}
{\cD[0,0]}
\label{eq:finalP}
\ee
in which all functions are to be evaluated upon choosing for the order
parameters the appropriate saddle-points of $\Psi$ (variation with
respect to $\bq,~\hbq,~\hbQ,~\hbR$ and $\{\hat{P}\}$), which itself
takes the
form:
\bd
\Psi[\ldots]=
i\sum_\alpha\hQ_\alpha(1\minus q_{\alpha\alpha})+iR\sum_\alpha \hR_\alpha
+i\sum_{\alpha\beta}\hat{q}_{\alpha\beta}q_{\alpha\beta}
+i\sum_\alpha\int dx^\pprime
dy^\pprime~\hat{P}_\alpha(x^\pprime,y^\pprime)P[x^\pprime,y^\pprime]
\ed
\be
+\alpha \log \cD[0,0]
+\lim_{N\to\infty}\frac{1}{N}\sum_i
\log \int\!d\bsigma
~
e^{-i\tau_i\sqrt{Q}\sum_\alpha \hR_\alpha\sigma_\alpha
-i\sum_{\alpha\beta}\hat{q}_{\alpha\beta}\sigma_\alpha\sigma_\beta}
\label{eq:finalpsi}
\ee
With $\cD[\ldots]$ given by (\ref{eq:Dasintegral}), which depends on
the variational parameters $\{\hat{P}\}$ and $q_{\alpha\beta}$ only.
The function $\cL[\ldots]$ is given by (\ref{eq:nastyterm}).
The order parameters $q_{\alpha\beta}$ have the usual
interpretation in terms of the average probability density for finding
a mutual overlap $q$ of two independently evolving weight vectors
$(\bJ^a,\bJ^b)$, in two
systems $a$ and $b$ with the same realization of the training set (see e.g.
\cite{Mezardetal}):
\be
\bigbra
P(q)\bigket_{\!\rmsets}=
\bigbra~\bigbras\delta\left[q-\frac{\bJ^a\inn\bJ^b}{|\bJ^a||\bJ^b|}
\right]\bigkets~\bigket_{\!\rmsets}=\lim_{n\to 0}\frac{1}{n(n\minus
1)}\sum_{\alpha\neq \beta}\delta[q-q_{\alpha\beta}]
\label{eq:qabmeaning}
\ee
Note that upon applying the above procedure to the functions
$\B[\ldots]$ and
$\C[\ldots]$ in
(\ref{eq:intermediateAB},\ref{eq:intermediateC}) we find again
integrals dominated by the dominant saddle-point of $\Psi$; here, in view of
(\ref{eq:Bworkedout}) and (\ref{eq:Cworkedout}), the intensive parts
are zero, and thus
\be
\B[x,y;x^\prime,y^\prime]=
\C[x,y;x^\prime,y^\prime;x^\pprime,y^\pprime]=0
\label{eq:nodiffusionproof}
\ee
as anticipated earlier.

\subsection{Replica-Symmetric Saddle-Points}

We now make the replica symmetric (RS) ansatz in the extremisation
problem, which according to (\ref{eq:qabmeaning}) is equivalent to
assuming ergodicity. With a modest amount of foresight we put
\bd
q_{\alpha\beta}=q_0\delta_{\alpha\beta}+q[1\minus
\delta_{\alpha\beta}],
~~~~~~~\hat{q}_{\alpha\beta}=\frac{1}{2}i[r\minus r_0\delta_{\alpha\beta}],
~~~~~~~\hat{R}_\alpha=i\rho,
~~~~~~~\hat{Q}_\alpha=i\phi,
~~~~~~~\hat{P}_\alpha[u,v]=i\chi[u,v]
\ed
This converts the quantity $\Psi$ of equation (\ref{eq:finalpsi}) for
small $n$ into
\bd
\lim_{n\to 0}\frac{1}{n}\Psi[\ldots]=
-\phi(1\minus q_0)-\rho R
+\frac{1}{2}qr-\frac{1}{2}q_0(r\minus r_0)
-\int dx^\pprime
dy^\pprime~\chi[x^\pprime,y^\pprime]P[x^\pprime,y^\pprime]
\ed
\bd
+\lim_{n\to 0}\frac{\alpha}{n} \log \cD[0,0]
+\lim_{n\to 0}\lim_{N\to\infty}\frac{1}{Nn}\sum_i
\log \int\!Dz\int\!d\bsigma
~
e^{\tau_i\rho\sqrt{Q}\sum_\alpha \sigma_\alpha
-\frac{1}{2}r_0\sum_{\alpha}\sigma^2_\alpha
+z\sqrt{r}\sum_{\alpha}\sigma_\alpha}
\ed
with the abbreviation $Dz=(2\pi)^{-\frac{1}{2}}e^{-\frac{1}{2}z^2}dz$.
We do the Gaussian integral in the last term, and expand the result
for small $n$:
\bd
\lim_{n\to 0}\frac{1}{n}\Psi[\ldots]
=
-\phi(1\minus q_0)-\rho R
+\frac{1}{2}qr-\frac{1}{2}q_0(r\minus r_0)
-\frac{1}{2}\log r_0 + \frac{1}{2r_0}(r \plus \rho^2 Q)
\ed
\be
-\int dx^\pprime
dy^\pprime~\chi[x^\pprime,y^\pprime]P[x^\pprime,y^\pprime]
+\lim_{n\to 0}\frac{\alpha}{n} \log \cD[0,0]
+ ~{\rm const}
\label{eq:RSsaddle}
\ee
Note that `const' refers to terms which do not depend on the order
parameters to be varied, and will thus not show up in saddle-point
equations; such terms can, however, depend on time via quantities such
as $(Q,R)$.
At this stage it is useful to work out four of our saddle-point
equations:
\bd
\frac{\partial\Psi}{\partial\phi}=\frac{\partial\Psi}{\partial r}=
\frac{\partial\Psi}{\partial \rho}=\frac{\partial\Psi}{\partial
r_0}=0:
~~~~~~~q_0=1,
~~~~~~~r_0=\frac{1}{1\minus q},
~~~~~~~\rho=\frac{R}{Q(1\minus q)},
~~~~~~~r=\frac{qQ\minus R^2}{Q(1\minus q)^2}
\ed
These allow us to eliminate most variational parameters, leaving a
saddle-point problem involving only the function $\chi[x,y]$
and the scalar $q$:
\be
\lim_{n\to 0}\frac{1}{n}\Psi[q,\{\chi\}]
=
\frac{1\minus R^2/Q}{2(1\minus q)} +\frac{1}{2}\log(1\minus q)
-\int dx^\prime
dy^\prime~\chi[x^\prime,y^\prime]P[x^\prime,y^\prime]
+\lim_{n\to 0}\frac{\alpha}{n} \log \cD[0,0;q,\{\chi\}]
+ ~{\rm const}
\label{eq:simpleRSsaddle}
\ee
Finally we have to work out the RS version of $\cD[u,v;q,\{\chi\}]$:
\be
\cD[u,v;\chi,q,1]=
\frac{{\rm det}^{\frac{1}{2}}\bA}{(2\pi)^{(n+1)/2}}~
\int\!d\bx dy~
e^{-\frac{1}{2}
\left(
\!\!\begin{array}{c}x_1\\[-1mm] \vdots \\[-1mm] x_n \\[-1mm]
y\end{array}\!\!\right)
\cdot\bA
\left(
\!\!\begin{array}{c}x_1\\[-1mm] \vdots \\[-1mm] x_n \\[-1mm] y\end{array}\!\!\right)
+\frac{1}{\alpha}\sum_\alpha
\chi(\sqrt{Q}x_\alpha,y)
-i[u\sqrt{Q}x_1 +vy]}
\label{eq:RSD}
\ee
with
\bd
\bA^{-1}=\left(
\!\!\begin{array}{cccc} 1 & \!\cdots\! & q & R/\sqrt{Q}\\
\vdots && \vdots & \vdots \\
q & \!\cdots\! & 1 & R/\sqrt{Q} \\
R/\sqrt{Q} & \!\cdots\! & R/\sqrt{Q} & 1
\end{array}\!\!\right)
\ed
The inverse of the above matrix is found to be
\bd
\bA=\left(
\!\!\begin{array}{cccc} C_{11} & \!\cdots\! & C_{1n} & \gamma\\
\vdots && \vdots & \vdots \\
C_{n1} & \!\cdots\! & C_{nn} & \gamma \\
\gamma & \!\cdots\! & \gamma & b
\end{array}\!\!\right)
~~~~~~~~~~
C_{\alpha\beta}=\frac{\delta_{\alpha\beta}}{1\minus q}-d
~~~~~~~~~~
\begin{array}{l}
\gamma=-\frac{R\sqrt{Q}}{Q(1- q)}+ \order(n)\\[2mm]
b=1+\order(n)\\[2mm]
d=\frac{q-R^2/Q}{(1\minus q)^2}+\order(n)
\end{array}
\ed
With this expression, and upon linearising the terms in the exponents
which are quadratic in $\bx$ in the usual manner with Gaussian integrals,  we obtain
\bd
\cD[u,v;q,\{\chi\}]=
\frac{
\int\!d\bx dy~
e^{-\frac{1}{2}\bx\cdot\bC\bx-\frac{1}{2}by^2 -\gamma y\sum_{\alpha=1}^n x_\alpha
+\frac{1}{\alpha}\sum_\alpha\chi[\sqrt{Q}x_\alpha,y]-i[u\sqrt{Q}x_1 +vy]}}
{\int\!d\bx dy~
e^{-\frac{1}{2}\bx\cdot\bC\bx-\frac{1}{2}by^2 -\gamma y\sum_{\alpha=1}^n x_\alpha}}
\ed
\be
=
\frac{\int\!DzDy e^{-ivy/\sqrt{b}}\left[
\int\!dx~
e^{
-\frac{x^2}{2(1-q)}+ [z\sqrt{d}-\gamma \frac{y}{\sqrt{b}}
]x
+\frac{1}{\alpha}\chi[\sqrt{Q}x,\frac{y}{\sqrt{b}}]}
\right]^{n-1}
\!\int\!dx~
e^{
-\frac{x^2}{2(1-q)} + [z\sqrt{d}-\gamma\frac{y}{\sqrt{b}}]x
+\frac{1}{\alpha}\chi[\sqrt{Q}x,\frac{y}{\sqrt{b}}]-iu\sqrt{Q}x}}
{\int\!DzDy \left[
\int\!dx~
e^{-\frac{1}{2(1-q)}x^2 +
[z\sqrt{d}-\gamma\frac{y}{\sqrt{b}}]x}\right]^n}
\label{eq:Duv}
\ee
For the saddle-point problem we only need to calculate
$\lim_{n\to 0}\frac{\alpha}{n}\log \cD[0,0;q,\{\chi\}]$:
\bd
\lim_{n\to 0}\frac{\alpha}{n}\log \cD[0,0;q,\{\chi\}]
=
\lim_{n\to 0}\frac{\alpha}{n}\left\{
\log\int\!DzDy \left[
\int\!dx~
e^{
-\frac{x^2}{2(1-q)}+ [z\sqrt{d}-\gamma y/\sqrt{b}]x
+\frac{1}{\alpha}\chi[\sqrt{Q}x,y/\sqrt{b}]}
\right]^{n}
\right.
\ed
\bd
\left.~~~~~~~~~~~~~~~~~~~~~~~~~~~~~~~~~~~~~~~~~~~~~
-\log \int\!DzDy \left[\int\!dx~
e^{-\frac{1}{2(1-q)}x^2 + [z\sqrt{d}-\gamma y/\sqrt{b}]x}\right]^n
\right\}
\ed
\bd
=
\alpha\int\!DzDy \log \left\{
\frac{\int\!dx~
e^{
-\frac{x^2}{2Q(1-q)}+ x[z\sqrt{d}-\gamma y]/\sqrt{Q}
+\frac{1}{\alpha}\chi[x,y]}}
{\int\!dx~
e^{-\frac{x^2}{2Q(1-q)} + x[z\sqrt{d}-\gamma y]/\sqrt{Q}}}
\right\}
\ed
with $\gamma$ and $d$ evaluated in the limit $n\to 0$.
Equivalently we can define
\be
A=R/Q(1\minus q)~~~~~~~~~~B=\sqrt{qQ\minus R^2}/Q(1\minus q)
\label{eq:AandB}
\ee
which gives
\bd
\lim_{n\to 0}\frac{\alpha}{n}\log \cD[0,0;q,\{\chi\}]
=
\alpha\int\!DzDy \log \left\{
\frac{\int\!dx~
e^{
-\frac{x^2}{2Q(1-q)}+ x[Ay+Bz]
+\frac{1}{\alpha}\chi[x,y]}}
{\int\!dx~
e^{-\frac{x^2}{2Q(1-q)} + x[Ay+Bz]}}
\right\}
\ed
Upon doing the $x$-integration in the denominator of this expression
we can write the explicit expression for the surface $\Psi$ to be
extremised with respect to $q$ and the function $\chi[x,y]$, apart from
irrelevant constants, in the surprisingly simple form (with the
short-hand (\ref{eq:AandB})):
\bd
\lim_{n\to 0}\frac{1}{n}\Psi[q,\{\chi\}]
=
\frac{1\minus\alpha\minus R^2/Q}{2(1\minus q)} +\frac{1}{2}(1\minus \alpha)\log(1\minus q)
-\int dx
dy~\chi[x,y]P[x,y]
\ed
\be
+~\alpha
\int\!DzDy ~\log \int\!dx~
e^{
-\frac{x^2}{2Q(1-q)}+ x[Ay+Bz]
+\frac{1}{\alpha}\chi[x,y]}
\label{eq:simplestRSsaddle}
\ee
Note that (\ref{eq:simplestRSsaddle}) is to be
{\em minimised}, both with
respect to $q$ (which originated as an $n(n\minus 1)$ fold entry in a
matrix, leading to curvature sign change for $n<1$) and with respect
to the function $\chi[x,y]$ (obtained from the $n$-fold occurrence of
the original function
$\hat{P}$, multiplied by $i$, which also leads to curvature
sign change).

The remaining saddle point equations are obtained by
variation of (\ref{eq:simplestRSsaddle}) with respect to $\chi$ and
$q$. Functional variation with respect to $\chi$ gives:
\be
{\rm for~all}~x,y:~~~~~~~~~~
P[x,y]=\frac{e^{-\frac{1}{2}y^2}}{\sqrt{2\pi}}
\int\!Dz \left\{
\frac{ e^{-\frac{x^2}{2Q(1-q)}+
x[Ay+Bz]+\frac{1}{\alpha}\chi[x,y]}}
{\int\!dx^\prime~ e^{-\frac{x^{\prime 2}}{2Q(1-q)}+
x^\prime[Ay+Bz]+\frac{1}{\alpha}\chi[x^\prime,y]}}
\right\}
\label{eq:RSsaddle1}
\ee
Note that $P[x,y]=P[x|y]P[y]$ with
$P[y]=(2\pi)^{-\frac{1}{2}}e^{-\frac{1}{2}y^2}$, as could have been
expected.
Next we vary $q$, and use (\ref{eq:RSsaddle1}) wherever possible:
\bd
\frac{1\minus\alpha\minus R^2/Q}{2(1\minus q)^2} -\frac{1\minus \alpha}{2(1\minus q)}
= \alpha\int\!DzDy \left\{
\frac{\int\!dx~ e^{-\frac{x^2}{2Q(1-q)}+ x[Ay+Bz]
+\frac{1}{\alpha}\chi[x,y]}\left[\frac{x^2}{2Q(1-q)^2}-
x[y\frac{\partial A}{\partial q}+z\frac{\partial B}{\partial q}]\right]}
{\int\!dx~ e^{
-\frac{x^2}{2Q(1-q)}+ x[Ay+Bz]
+\frac{1}{\alpha}\chi[x,y]}}\right\}
\ed
giving
\bd
\int\!dxdy~P[x,y](x\minus Ry)^2
+(R^2\minus qQ)(\frac{1}{\alpha}\minus 1)
~~~~~~~~~~~~~~~~~~~~~~~~~~~~~~~~~~~~~~~~
\ed
\be
~~~~~~~~~~~~~~~~~~~~
=\left[2\sqrt{qQ\minus R^2}
+\frac{Q(1\minus q)}{\sqrt{qQ\minus R^2}}\right]
\int\!DzDy~z\left[
\frac{\int\!dx~ e^{-\frac{x^2}{2Q(1-q)}+ x[Ay+Bz]
+\frac{1}{\alpha}\chi[x,y]}x}
{\int\!dx~ e^{
-\frac{x^2}{2Q(1-q)}+ x[Ay+Bz]
+\frac{1}{\alpha}\chi[x,y]}}\right]
\label{eq:RSsaddle2}
\ee

\subsection{Explicit Expression for the Green's Function}

In order to work out the Green's function (\ref{eq:finalA}) we need
the function $\cL[u,v;u^\prime,v^\prime]$ as defined in
(\ref{eq:nastyterm}) which, in turn, is given in terms of the
integrals (\ref{eq:Dasintegral},\ref{eq:Fasintegral}).
First we calculate the $n\to 0$ limit of
$D[u,v;q,\{\chi\}]$ (\ref{eq:Duv}), and simplify the result with the saddle-point
equation (\ref{eq:RSsaddle1}):
\bd
\lim_{n\to 0}\cD[u,v;q,\{\chi\}]
=\int\!DzDy ~e^{-ivy}
\frac{\int\!dx~ e^{
-\frac{x^2}{2Q(1-q)}+ x[Ay+Bz]
+\frac{1}{\alpha}\chi[x,y] -iux}}
{\int\!dx~ e^{
-\frac{x^2}{2Q(1-q)}+ x[Ay+Bz]
+\frac{1}{\alpha}\chi[x,y]}}
\ed
\be
=\int\!dxdy P[x,y]e^{-ivy-iux}
\label{eq:finalDuv}
\ee
Next we work out the quantities $F_{1,2}^\alpha[u,v]$ of equation
(\ref{eq:Fasintegral}) in RS ansatz, using Gaussian linearizations:
\bd
\lim_{n\to 0}\cF_{1,2}^\alpha[u,v]=~i\lim_{n\to 0}~
\frac{\int\!d\bx dy~
\partial_{1,2}\chi[\sqrt{Q}x_\alpha,y]
~
e^{-\frac{1}{2}
\left(
\!\!\begin{array}{c}x_1\\[-1mm] \vdots \\[-1mm] x_n \\[-1mm]
y\end{array}
\!\!\right)\cdot\bA
\left(\!\!
\begin{array}{c}x_1\\[-1mm] \vdots \\[-1mm] x_n \\[-1mm] y\end{array}
\!\!\right)
+\frac{1}{\alpha}\sum_\alpha
\chi[\sqrt{Q}x_\alpha,y]
-i[u\sqrt{Q}x_1 +vy]}}
{\int\!d\bx dy~
~e^{-\frac{1}{2}
\left(\!\!\begin{array}{c}x_1\\[-1mm] \vdots \\[-1mm] x_n \\[-1mm]
y\end{array}\!\!\right)\cdot\bA
\left(\!\!\begin{array}{c}x_1\\[-1mm] \vdots \\[-1mm] x_n \\[-1mm]
y\end{array}\!\!\right)}}
\ed
\bd
=~i\lim_{n\to 0}~
\int\!
DyDz~e^{-ivy}\int\!d\bx~e^{\sum_\beta
\left[-\frac{1}{2}\frac{x^2_\beta}{1\minus q}+[z\sqrt{d}-\gamma y]x_\beta
+\frac{1}{\alpha}\chi[\sqrt{Q}x_\beta,y]\right] -iux_1\sqrt{Q}}
\partial_\lambda\chi[\sqrt{Q}x_\alpha,y]
\ed
The replica permutation symmetries of this expression allow us to
conclude
\be
\lim_{n\to 0}\cF_\lambda^\alpha[u,v]=\delta_{\alpha
1}F_\lambda^1[u,v]+(1\minus \delta_{\alpha 1})F_\lambda^2[u,v]
\label{eq:bigF}
\ee
where
\be
F_{1,2}^1[u,v]
=i\int\!dxdy~P[x,y]e^{-ivy-iux}\partial_{1,2}\chi[x,y]
\label{eq:Fsame}
\ee
\be
F_{1,2}^2[u,v]=
i\int\!
DyDz~e^{-ivy}
\frac{
\left[\int\!dx~e^{-\frac{x^2}{2Q(1\minus
q)}+x[Ay+Bz]+\frac{1}{\alpha}\chi[x,y]}
\partial_{1,2}\chi[x,y]\right]
\left[\int\!dx~e^{-\frac{x^2}{2Q(1\minus
q)}+x[Ay+Bz]+\frac{1}{\alpha}\chi[x,y]-iux}
\right]}
{\left[\int\!dx~e^{-\frac{x^2}{2Q(1\minus
q)}+x[Ay+Bz]+\frac{1}{\alpha}\chi[x,y]}\right]^2}
\label{eq:Fdiff}
\ee
We can now proceed to the calculation of (\ref{eq:nastyterm}).
First we note that the basic building blocks of (\ref{eq:nastyterm})
are most easily expressed in terms of the functions
\be
G_1[u,v]=\frac{1}{\alpha}\cF^1_1[u,v]+u\cD[u,v]~~~~~~~~~~
\tilde{G}_1[u,v]=\frac{1}{\alpha}\cF^2_1[u,v]
\label{eq:G1}
\ee
\be
G_2[u,v]=\frac{1}{\alpha}\cF^1_2[u,v]+v\cD[u,v]~~~~~~~~~~
\tilde{G}_2[u,v]=\frac{1}{\alpha}\cF^2_2[u,v]
\label{eq:G2}
\ee
With these short-hands we obtain, upon performing the summations over replica
indices in (\ref{eq:nastyterm}):
\bd
\cL[u,v;u^\prime,v^\prime]=
-Q(1\minus q)G_1[u,v]G_1[u^\prime,v^\prime]
-Q(1\minus q)(n\minus 1)\tilde{G}_1[u,v]\tilde{G}_1[u^\prime,v^\prime]
\ed
\bd
-Q q
\left[G_1[u,v]+(n\minus 1)\tilde{G}_1[u,v]\right]
\left[G_1[u^\prime,v^\prime]+(n\minus 1)\tilde{G}_1[u^\prime,v^\prime]\right]
\ed
\bd
-R
\left[G_1[u,v]+(n\minus 1)\tilde{G}_1[u,v]\right]
\left[G_2[u^\prime,v^\prime]+(n\minus 1)\tilde{G}_2[u^\prime,v^\prime]\right]
\ed
\bd
-R
\left[G_1[u^\prime,v^\prime]+(n\minus 1)\tilde{G}_1[u^\prime,v^\prime]\right]
\left[G_2[u,v]+(n\minus 1)\tilde{G}_2[u,v]\right]
\ed
\bd
-
\left[G_2[u,v]+(n\minus 1)\tilde{G}_2[u,v]\right]
\left[G_2[u^\prime,v^\prime]+(n\minus 1)\tilde{G}_2[u^\prime,v^\prime]\right]
\ed
and so
\bd
\lim_{n\to 0}\cL[u,v;u^\prime,v^\prime]=
-Q(1\minus q)\left[G_1[u,v]G_1[u^\prime,v^\prime]-\tilde{G}_1[u,v]\tilde{G}_1[u^\prime,v^\prime]\right]
\ed
\bd
-Q q
\left[G_1[u,v]-\tilde{G}_1[u,v]\right]
\left[G_1[u^\prime,v^\prime]-\tilde{G}_1[u^\prime,v^\prime]\right]
\ed
\bd
-R
\left[G_1[u,v]- \tilde{G}_1[u,v]\right]
\left[G_2[u^\prime,v^\prime]-\tilde{G}_2[u^\prime,v^\prime]\right]
-R
\left[G_1[u^\prime,v^\prime]-\tilde{G}_1[u^\prime,v^\prime]\right]
\left[G_2[u,v]-\tilde{G}_2[u,v]\right]
\ed
\bd
-
\left[G_2[u,v]-\tilde{G}_2[u,v]\right]
\left[G_2[u^\prime,v^\prime]-\tilde{G}_2[u^\prime,v^\prime]\right]
\ed
With the Fourier transforms of the functions $G[\ldots]$, given by
\be
\hat{G}_1[\hat{u},\hat{v}]=
\int\!\frac{dudv}{(2\pi)^2}e^{iu\hat{u}+iv\hat{v}}
\left[\frac{1}{\alpha}\cF^1_1[u,v]+u\cD[u,v]\right]~~~~~~~~~~
\overline{G}_1[\hat{u},\hat{v}]=\frac{1}{\alpha}
\int\!\frac{dudv}{(2\pi)^2}e^{iu\hat{u}+iv\hat{v}}\cF^2_1[u,v]
\label{eq:G1fourier}
\ee
\be
\hat{G}_2[\hat{u},\hat{v}]=
\int\!\frac{dudv}{(2\pi)^2}e^{iu\hat{u}+iv\hat{v}}
\left[\frac{1}{\alpha}\cF^1_2[u,v]+v\cD[u,v]\right]~~~~~~~~~~
\overline{G}_2[\hat{u},\hat{v}]=\frac{1}{\alpha}
\int\!\frac{dudv}{(2\pi)^2}e^{iu\hat{u}+iv\hat{v}}\cF^2_2[u,v]
\label{eq:G2fourier}
\ee
the Green's function $\A[x,y;x^\prime,y^\prime]$ (\ref{eq:finalA})
can now be written in explicit form as
\bd
\A[x,y;x^\prime,y^\prime]
=
-Q(1\minus q)\left[\hat{G}_1[x,y]\hat{G}_1[x^\prime,y^\prime]
-\overline{G}_1[x,y]\overline{G}_1[x^\prime,y^\prime]\right]
\ed
\bd
-Q q
\left[\hat{G}_1[x,y]-\overline{G}_1[x,y]\right]
\left[\hat{G}_1[x^\prime,y^\prime]-\overline{G}_1[x^\prime,y^\prime]\right]
\ed
\bd
-R
\left[\hat{G}_1[x,y]- \overline{G}_1[x,y]\right]
\left[\hat{G}_2[x^\prime,y^\prime]-\overline{G}_2[x^\prime,y^\prime]\right]
-R
\left[\hat{G}_1[x^\prime,y^\prime]-\overline{G}_1[x^\prime,y^\prime]\right]
\left[\hat{G}_2[x,y]-\overline{G}_2[x,y]\right]
\ed
\be
-
\left[\hat{G}_2[x,y]-\overline{G}_2[x,y]\right]
\left[\hat{G}_2[x^\prime,y^\prime]-\overline{G}_2[x^\prime,y^\prime]\right]
\label{eq:RSA}
\ee
Finally, working out the four relevant Fourier transforms,
using (\ref{eq:finalDuv},\ref{eq:Fsame},\ref{eq:Fdiff}),  gives:
\be
\hat{G}_1[x,y]=
iP[x,y]\left[\frac{1}{\alpha}\frac{\partial}{\partial x}\chi[x,y]
-\frac{\partial}{\partial x}\log P[x,y] \right]
\ee
\be
\hat{G}_2[x,y]=
iP[x,y]\left[\frac{1}{\alpha}\frac{\partial}{\partial y}\chi[x,y]
-\frac{\partial}{\partial y}\log P[x,y] \right]
\ee
\be
\overline{G}_1[x,y]=\frac{i}{\alpha}P[y]
\int\!Dz\frac{
\left[\int\!dx^\prime~e^{-\frac{x^{\prime 2}}{2Q(1\minus
q)}+x^\prime[Ay+Bz]+\frac{1}{\alpha}\chi[x^\prime,y]}
\partial_1\chi[x^\prime,y]\right]
e^{-\frac{x^2}{2Q(1\minus
q)}+x[Ay+Bz]+\frac{1}{\alpha}\chi[x,y]}}
{\left[\int\!dx^\prime~e^{-\frac{x^{\prime 2}}{2Q(1\minus
q)}+x^\prime[Ay+Bz]+\frac{1}{\alpha}\chi[x^\prime,y]}\right]^2}
\ee
\be
\overline{G}_2[x,y]=\frac{i}{\alpha}P[y]
\int\!Dz\frac{
\left[\int\!dx^\prime~e^{-\frac{x^{\prime 2}}{2Q(1\minus
q)}+x^\prime[Ay+Bz]+\frac{1}{\alpha}\chi[x^\prime,y]}
\partial_2\chi[x^\prime,y]\right]
e^{-\frac{x^2}{2Q(1\minus
q)}+x[Ay+Bz]+\frac{1}{\alpha}\chi[x,y]}
}
{\left[\int\!dx^\prime~e^{-\frac{x^{\prime 2}}{2Q(1\minus
q)}+x^\prime[Ay+Bz]+\frac{1}{\alpha}\chi[x^\prime,y]}\right]^2}
\ee
with $P[y]=(2\pi)^{-\frac{1}{2}}e^{-\frac{1}{2}y^2}$.
\vsp

Since the distribution $P[x,y]$  obeys $P[x,y]=P[x|y]P[y]$ with
$P[y]=(2\pi)^{-\frac{1}{2}}e^{-\frac{1}{2}y^2}$, our equations can be
simplified by choosing as our order parameter function
the conditional distribution $P[x|y]$. We also replace the conjugate
order parameter function $\chi[x,y]$ by the effective measure
$M[x,y]$, and we introduce a compact notation for the relevant averages in our
problem:
\be
M[x,y]=e^{-\frac{x^2}{2Q(1-q)}+Axy +\frac{1}{\alpha}\chi[x,y]}
~~~~~~~~~~~~~~~~
\bra f[x,y,z]\ket_\star =
 \frac{\int\!dx~M[x,y]e^{Bxz}f[x,y,z]}
{\int\!dx~M[x,y]e^{Bxz}}
\label{eq:chi_M}
\ee
Instead of the original Green's function $\cA[x,y;x^\prime,y^\prime]$
we turn to the transformed Green's function
$\tilde{\cA}[x,y;x^\prime,y^\prime]$, defined as
\bd
\cA[x,y;x^\prime,y^\prime]=P[x,y]\tilde{\cA}[x,y;x^\prime,y^\prime]P[x^\prime,y^\prime]
\ed
With these notational conventions one finds that (\ref{eq:RSA})
translates into the following expression:
\bd
\tilde{\A}[x,y;x^\prime,y^\prime]
=
Q(1\minus q)\left[J_1[x,y]J_1[x^\prime,y^\prime]\minus
\tilde{J}_1[x,y]
\tilde{J}_1[x^\prime,y^\prime]\right]
+Q q
\left[J_1[x,y]\minus \tilde{J}_1[x,y]\right]
\left[J_1[x^\prime,y^\prime]\minus \tilde{J}_1[x^\prime,y^\prime]\right]
\ed
\be
+R
\left[J_1[x,y]\minus  \tilde{J}_1[x,y]\right]J_2[x^\prime,y^\prime]
+R
\left[J_1[x^\prime,y^\prime]\minus \tilde{J}_1[x^\prime,y^\prime]\right]
J_2[x,y]
+J_2[x,y]J_2[x^\prime,y^\prime]
\label{eq:RSAwithJs}
\ee
with
\bd
J_1[X,Y]=
\frac{\partial}{\partial X}\log\frac{M[X,Y]}{P[X|Y]}
+\frac{X\minus RY}{Q(1\minus q)}
\ed
\bd
\tilde{J}_1[X,Y]=P[X|Y]^{-1}\int\!Dz
\bra
\frac{\partial}{\partial x}\log M[x,Y]+\frac{x\minus RY}{Q(1\minus q)}
\ket_\star
\bra \delta[X\minus x]\ket_\star
\ed
\bd
J_2[X,Y]=\frac{\partial}{\partial Y}\log \frac{M[X,Y]}{P[X|Y]}
-\frac{RX}{Q(1\minus q)}+Y
-
P[X|Y]^{-1}\!\int\!Dz
\bra
\frac{\partial}{\partial Y}\log M[x,Y]-\frac{Rx}{Q(1\minus q)}
\ket_\star
\bra \delta[X\minus x]\ket_\star
\ed
It turns out that significant simplification of the result
(\ref{eq:RSAwithJs}) is possible, upon using the following two identities
to rewrite the functions $J_1[\ldots]$, $\tilde{J}_1[\ldots]$ and
$J_2[\ldots]$:
\be
\bra \frac{\partial}{\partial x}\log M[x,y]\ket_\star= -Bz
\label{eq:staridentity1}
\ee
\be
\bra \frac{\partial}{\partial y}\log M[x,y]\ket_\star=
\frac{\partial}{\partial y}\log \int\!dx~e^{Bxz}M[x,y]
\label{eq:staridentity2}
\ee
Identity (\ref{eq:staridentity1})
results upon integrating by parts with respect to $x$, whereas
identity (\ref{eq:staridentity2}) is a direct consequence of $y$
dependencies occurring  in $M[x,y]$ only.
Note that $B=\sqrt{qQ\minus R^2}/Q(1\minus q)$.
To achieve the desired simplification of
$\tilde{\cA}[x,y;x^\prime,y^\prime]$ we define the following object:
\be
\Phi[X,y]=
\left\{\room Q(1\minus q)P[X|y]\right\}^{-1} \int\!Dz
\bra X\minus x \ket_\star
\bra \delta[X\minus x]\ket_\star
\label{eq:W}
\ee
We can now, after additional integration by parts with respect to $z$,
simplify the above expressions for $J_1[\ldots]$, $\tilde{J}_1[\ldots]$ and
$J_2[\ldots]$ to
\bd
J_1[X,Y]=\frac{X\minus RY}{Q(1\minus q)}
 -\frac{qQ\minus R^2}{Q(1\minus q)} \Phi[X,Y]
~~~~~~~~~~~~~~
\tilde{J}_1[X,Y]=J_1[X,Y]-\Phi[X,Y]
\ed
\bd
J_2[X,Y]=Y -R\Phi[X,Y]
\ed
and consequently
\be
\cA[x,y;x^\prime,y^\prime]=P[x,y]\tilde{\cA}[x,y;x^\prime,y^\prime]P[x^\prime,y^\prime]
\label{eq:greenandtransformedgreen}
\ee
\be
\room
\tilde{\cA}[x,y;x^\prime,y^\prime]
=
yy^\prime
+(x\minus Ry)\Phi[x^\prime,y^\prime]
+(x^\prime\minus Ry^\prime)\Phi[x,y]
-(Q\minus R^2)\Phi[x,y]\Phi[x^\prime,y^\prime]
\label{eq:simplifiedGreen}
\ee
with $\Phi[x,y]$ as given in (\ref{eq:W}).

\end{document}